\documentstyle{article}

\def\noheaderplainsetup{%

\topmargin=0pt \headheight=0pt \headsep=0pt  \oddsidemargin=0pt \evensidemargin=0pt  \textheight=8.9truein \textwidth=6.5truein}

\noheaderplainsetup

\newcommand{\kn}{\mbox{$\sqcap\hspace{-6.5pt}\raisebox{-0.11cm}{--}$}}
\newcommand{\mov}{\mbox{\em Nonrep}}
\newcommand{\rep}{\mbox{\em Rep}}
\newcommand{\predell}{\mbox{\bf CL4}}
\newcommand{\clt}{\mbox{\bf CL2}}
\newcommand{\al}{\mbox{\bf AL}}
\newcommand{\valid}{\mbox{$\vdash\hspace{-4pt}\vdash$}}
\newcommand{\uvalid}{\mbox{$\vdash\hspace{-5pt}\vdash\hspace{-5pt}\vdash$}}
\newcommand{\chess}{\mbox{\em Chess}}
\newcommand{\checkers}{\mbox{\em Checkers}}
\newcommand{\rneg}{\neg}               
\newcommand{\pneg}{\neg}               
\newcommand{\emptyrun}{\langle\rangle} 
\newcommand{\oo}{\bot}            
\newcommand{\pp}{\top}            
\newcommand{\xx}{\wp}               
\newcommand{\s}[1]{\underline{#1}} 
\newcommand{\blu}[1]{\overline{#1}} 
\newcommand{\yel}[1]{\underline{#1}} 
\newcommand{\cont}[1]{\underline{\overline{#1}}} 
\newcommand{\legal}[2]{\mbox{\bf Lr}^{#1}_{#2}} 
\newcommand{\Legal}[1]{\mbox{\bf LR}^{#1}} 
\newcommand{\win}[2]{\mbox{\bf Wn}^{#1}_{#2}} 
\newcommand{\tree}[2]{\mbox{\em Tree}^{#1}{#2}} 
\newcommand{\seq}[1]{\langle #1 \rangle}           
\newcommand{\col}[1]{\mbox{$#1$:}}

\newcommand{\sst}{\mbox{\raisebox{-0.07cm}{\scriptsize $-$}\hspace{-0.2cm}$\pst$}}
\newcommand{\scost}{\mbox{\raisebox{0.20cm}{\scriptsize $-$}\hspace{-0.2cm}$\pcost$}}
\newcommand{\sqc}{\mbox{\small \raisebox{0.0cm}{$\bigtriangleup$}}}
\newcommand{\sqd}{\mbox{\small \raisebox{0.049cm}{$\bigtriangledown$}}}

\newcommand{\mla}{\mbox{{\Large $\wedge$}}}
\newcommand{\mlai}{\mbox{{$\wedge$}}}
\newcommand{\mle}{\mbox{{\Large $\vee$}}}
\newcommand{\mlei}{\mbox{{$\vee$}}}
\newcommand{\pst}{\mbox{\raisebox{-0.01cm}{\scriptsize $\wedge$}\hspace{-4pt}\raisebox{0.16cm}{\tiny $\mid$}\hspace{2pt}}}
\newcommand{\gneg}{\neg}                  
\newcommand{\intimpl}{\mbox{\hspace{2pt}$\circ$\hspace{-0.14cm} \raisebox{-0.043cm}{\Large --}\hspace{2pt}}}
\newcommand{\pintimpl}{\mbox{\hspace{2pt}\raisebox{0.033cm}{\tiny $>$}\hspace{-0.18cm} \raisebox{-0.043cm}{\large --}\hspace{2pt}}}
\newcommand{\sintimpl}{\mbox{\hspace{2pt}\raisebox{0.033cm}{\tiny $ | \hspace{-4pt} >$}\hspace{-0.14cm} \raisebox{-0.039cm}{\large --}\hspace{2pt}}}
\newcommand{\mli}{\rightarrow}                     
\newcommand{\mleq}{\hspace{2pt}\leftrightarrow\hspace{2pt}}   
\newcommand{\cla}{\mbox{\large $\forall$}}      
\newcommand{\cle}{\mbox{\large $\exists$}}        
\newcommand{\mld}{\vee}    
\newcommand{\mlc}{\wedge}  
\newcommand{\ade}{\mbox{\Large $\sqcup$}}      
\newcommand{\ada}{\mbox{\Large $\sqcap$}}      
\newcommand{\add}{\sqcup}                      
\newcommand{\adc}{\sqcap}                      
\newcommand{\clai}{\forall}     
\newcommand{\clei}{\exists}        
\newcommand{\adei}{\mbox{$\sqcup$}}      
\newcommand{\adai}{\mbox{$\sqcap$}}      
\newcommand{\sti}{\mbox{\raisebox{-0.02cm}
{\scriptsize $\circ$}\hspace{-0.121cm}\raisebox{0.08cm}{\tiny $.$}\hspace{-0.079cm}\raisebox{0.10cm}
{\tiny $.$}\hspace{-0.079cm}\raisebox{0.12cm}{\tiny $.$}\hspace{-0.085cm}\raisebox{0.14cm}
{\tiny $.$}\hspace{-0.079cm}\raisebox{0.16cm}{\tiny $.$}\hspace{1pt}}}

\newcommand{\psti}{\mbox{\raisebox{-0.02cm}
{\tiny $\wedge$}\hspace{-0.121cm}\raisebox{0.08cm}{\tiny $.$}\hspace{-0.079cm}\raisebox{0.10cm}
{\tiny $.$}\hspace{-0.079cm}\raisebox{0.12cm}{\tiny $.$}\hspace{-0.085cm}\raisebox{0.14cm}
{\tiny $.$}\hspace{-0.079cm}\raisebox{0.16cm}{\tiny $.$}\hspace{1pt}}}

\newcommand{\costi}{\mbox{\raisebox{0.08cm}
{\scriptsize $\circ$}\hspace{-0.121cm}\raisebox{-0.01cm}{\tiny $.$}\hspace{-0.079cm}\raisebox{0.01cm}
{\tiny $.$}\hspace{-0.079cm}\raisebox{0.03cm}{\tiny $.$}\hspace{-0.085cm}\raisebox{0.05cm}
{\tiny $.$}\hspace{-0.079cm}\raisebox{0.07cm}{\tiny $.$}\hspace{1pt}}}

\newcommand{\pcosti}{\mbox{\raisebox{0.08cm}
{\tiny $\vee$}\hspace{-0.121cm}\raisebox{-0.01cm}{\tiny $.$}\hspace{-0.079cm}\raisebox{0.01cm}
{\tiny $.$}\hspace{-0.079cm}\raisebox{0.03cm}{\tiny $.$}\hspace{-0.085cm}\raisebox{0.05cm}
{\tiny $.$}\hspace{-0.079cm}\raisebox{0.07cm}{\tiny $.$}\hspace{1pt}}}

\newcommand{\tlg}{\bot}               
\newcommand{\twg}{\top}               
\newcommand{\st}{\mbox{\raisebox{-0.05cm}{$\circ$}\hspace{-0.13cm}\raisebox{0.16cm}{\tiny $\mid$}\hspace{2pt}}}
\newcommand{\cost}{\mbox{\raisebox{0.12cm}{$\circ$}\hspace{-0.13cm}\raisebox{0.02cm}{\tiny $\mid$}\hspace{2pt}}}
\newcommand{\pcost}{\mbox{\raisebox{0.12cm}{\scriptsize $\vee$}\hspace{-4pt}\raisebox{0.02cm}{\tiny $\mid$}\hspace{2pt}}}

\newtheorem{theoremm}{Theorem}[section]
\newtheorem{factt}[theoremm]{Fact}
\newtheorem{definitionn}[theoremm]{Definition}
\newtheorem{lemmaa}[theoremm]{Lemma}
\newtheorem{conventionn}[theoremm]{Convention}
\newtheorem{claimm}[theoremm]{Claim}
\newtheorem{examplee}[theoremm]{Example}
\newtheorem{exercisee}[theoremm]{Exercise}

\newenvironment{definition}{\begin{definitionn} \em}{ \end{definitionn}}
\newenvironment{theorem}{\begin{theoremm}}{\end{theoremm}}
\newenvironment{lemma}{\begin{lemmaa}}{\end{lemmaa}}

\newenvironment{example}{\begin{examplee}}{\end{examplee}}
\newenvironment{exercise}{\begin{exercisee} \em}{\end{exercisee}}

\newenvironment{proof}{ {\bf Proof.} }{\  $\Box$ \vspace{.1in} }

\makeindex

\begin{document}
\title{In the beginning was game semantics\vspace{-5pt}}
\author{Giorgi Japaridze\thanks{This material is based upon work supported by the National Science Foundation under Grant No. 0208816}}
\date{\vspace{-25pt}}
\maketitle

\begin{abstract} This chapter presents an overview of {\em computability logic} --- the game-semantically constructed  
logic of interactive computational tasks and resources. There is only one non-overview, technical section in it, devoted to a proof of the soundness of affine logic with respect to the semantics of computability logic.

\end{abstract}

Invited book chapter to appear in {\bf Games: Unifying Logic, Language and Philosophy.} O. Majer, A.-V. Pietarinen and T. Tulenheimo, eds., Springer.  

\tableofcontents

\section{Introduction}\label{intr}

In the beginning was Semantics, and Semantics was Game Semantics, and Game Semantics was Logic.\footnote{`In the beginning was the Word, and the Word was with God, 
and the Word was God... Through him all things were made; without him nothing was made that has been made.' --- John's Gospel.} Through it all concepts were conceived; for it all axioms are written, and to it all deductive systems should serve... 

This is not an evangelical story, but the story and philosophy of {\em computability logic} (CL),\index{CL (Computability Logic)}\label{0CL (Computability Logic)} the recently introduced \cite{Jap03} mini-religion within logic. 
According to its philosophy, {\em syntax}\index{syntax}\label{0syntax} --- the study of axiomatizations or any  
other, deductive or nondeductive string-manipulation systems --- exclusively owes its right on existence to 
{\em semantics},\index{semantics}\label{0semantics}  and is thus secondary to it. CL believes that logic is meant to be the most basic, 
general-purpose formal tool potentially usable by intelligent agents in successfully navigating the real life. And it is semantics 
that establishes that ultimate real-life meaning of logic. Syntax is important, yet it is so not in its own 
right but only as much as it serves a meaningful semantics, allowing us to realize the potential of that semantics in some systematic and perhaps convenient or efficient way. Not passing the test for soundness with respect to the underlying semantics would fully disqualify any syntax, no matter how otherwise appealing it is. Note --- disqualify the syntax and not the semantics.
Why this is so hardly requires any explanation: relying on an unsound syntax might result in wrong beliefs, misdiagnosed patients or crashed spaceships.

Unlike soundness, completeness is a desirable but not necessary condition. Sometimes --- as, say, in the case of pure second-order logic, or first-order applied number theory with $+$ and $\times$ --- completeness is impossible to achieve in principle. In such cases 
we may still benefit from continuing working with various reasonably strong syntactic constructions. A good example of such a ``reasonable" yet incomplete syntax is Peano arithmetic. Another example, as we are going to see later, is 
affine logic, which turns out to be sound but incomplete with respect to the semantics of CL.
And even when complete axiomatizations are known, it is not fully unusual for them to be sometimes artificially downsized and made incomplete for efficiency, simplicity, convenience or even esthetic considerations. Ample examples of this can be found 
in applied computer science. But again, while there might be meaningful trade-offs between (the degrees of) completeness, efficiency and other desirable-but-not-necessary properties of a syntax, the underlying semantics remains untouchable, and the condition 
of soundness unnegotiable. It is that very untouchable core that should be the point of departure for logic as a fundamental science.  

A separate question, of course, is what counts as a semantics. The model example of a semantics with a capital `S'\index{capital `S' semantics}\label{0capital `S' semantics} is that of classical logic. But in the logical literature this term often has a more generous meaning than 
what CL is ready to settle for. As pointed out, CL views logic as a universal-utility tool. So, a capital-`S'-semantics should be non-specific enough, and applicable to the world in general rather than some very special and artificially selected (worse yet, artificially created) fragment of it. Often what is called a semantics is just a special-purpose apparatus designed to help analyze a given syntactic construction rather than understand and navigate the outside world. The usage of Kripke models\index{Kripke model}\label{0Kripke model} as a derivability test for intuitionistic formulas, or as a validity criterion in various systems of modal logic is an example. An attempt to see more than a technical, syntax-serving instrument (which, as such, may be indeed  very important and useful) in this type of lowercase `s' 
semantics\index{lowercase `s' semantics}\label{0lowercase `s' semantics} might create a vicious circle: a deductive system $L$ under question is ``right" because it derives exactly the formulas that are valid in a such and such Kripke semantics; and then it turns out that the reason why we are considering the such and such Kripke semantics is that ... it validates exactly what $L$ derives.  

This was about why {\em in the beginning was Semantics}. Now a few words about why {\em Semantics was Game Semantics}. 
For CL, game is not just a game. It is a foundational mathematical concept on which a powerful enough logic (=semantics) should be based. This is so because, as noted, CL sees logic as a ``real-life navigational tool", 
and it is games that appear to offer the most comprehensive, coherent, natural, adequate and convenient mathematical models for the very essence of all ``navigational" activities of agents: their interactions with the surrounding world. An {\em agent} and its {\em environment}\index{environment}\label{0environment} translate into game-theoretic terms as two {\em players}; their {\em actions} as {\em moves};
{\em situations} arising in the course of interaction as {\em positions}; and {\em success} or {\em failure} as {\em wins} or {\em losses}. 

It is natural to require that the interaction strategies of the party that we have referred to as an ``agent" be limited 
to {\em algorithmic} ones, allowing us to henceforth call that player a {\em machine}.\index{machine}\label{0machine}   
This is a minimum condition that any non-esoteric game semantics would have to satisfy. On the other hand, no restrictions can or should be imposed on the environment, who represents `the blind forces of nature, or the devil himself' (\cite{Jap03}).
Algorithmic activities being synonymous to {\em computations}, games thus represent {\em computational problems}\index{computational problem}\label{0computational problem} --- interactive tasks performed by computing agents, with {\em computability} meaning {\em winnability}, i.e. existence of a machine that wins the game against any possible (behavior of the) environment. 

In the 1930s  mankind came up with what has been perceived as an ultimate mathematical 
definition of the precise meaning of algorithmic solvability. Curiously or not, such a definition was set forth and embraced before really having attempted to answer the seemingly more basic question about what {\em computational problems} are --- the very entities that may or may not have algorithmic solutions in the first place. The tradition established since then in theoretical computer science by computability simply means Turing computability of {\em functions}, as the task performed by every Turing machine is nothing but receiving an input $x$ and generating the output $f(x)$ for some function $f$. Turing\index{Turing}\label{0Turing} \cite{Tur36} himself, however, was more cautious about making overly broad philosophical conclusions, acknowledging that  
not everything one would potentially call a computational problem might necessarily be a function, or reducible to such.
Most tasks that computers and computer networks perform are interactive. And nowadays more and more voices are being 
heard 
\cite{Gol04,Japic,Mil93,Weg98} pointing out that true interaction might be going beyond what functions and hence ordinary Turing machines are meant to capture.  

Two main concepts on which the semantics of CL is based are those of {\em static games}\index{static game}\label{0static game1} and their {\em winnability}\index{winnability}\label{0winnability}
(defined later in Sections \ref{ss5} and \ref{icp}). 
Correspondingly, the philosophy of CL relies on two beliefs that, together, present what can be considered an interactive version of the Church-Turing thesis:\index{Church-Turing thesis}\label{0Church-Turing thesis}\vspace{10pt} 

{\bf Belief 1.} {\em 
The concept of static games is an adequate formal counterpart of our intuition of $($``pure", speed-independent$)$  interactive computational problems.}

{\bf Belief 2.} {\em 
The concept of winnability is an adequate formal counterpart of our intuition of algorithmic solvability of such problems.}\vspace{10pt}

As will be seen later, one of the main features distinguishing the CL games from more traditional game models 
is the absence of {\em procedural rules}\index{Procedural rule}\label{0Procdural rule}  (\cite{Ben01}) --- 
rules strictly regulating which player is to move in any given situation. Here, in a general case, either player is free to move. It is exactly this feature that makes players' strategies no longer definable as functions (functions from positions to moves). And it is this highly relaxed nature that makes the CL games apparently most flexible and general of all two-player, two-outcome games. 

Trying to understand strategies as functions would not be a good idea even if the type of games we consider naturally allowed us to do so. Because, when it comes to long or infinite games, functional strategies would be disastrously inefficient, making it hardly possible to develop any reasonable complexity theory 
for interactive computation (the next important frontier for CL or theoretical computer science in general). To understand this, it would be sufficient to just reflect on the behavior of one's personal computer. The job of your computer is to play one long --- potentially infinite --- game against you. Now, have you noticed your faithful servant getting slower every time you use it? Probably not. That is because the computer is smart enough to follow a non-functional strategy in this game. If its strategy was a function from positions (interaction histories) to moves, the response time would inevitably keep worsening due to the need to read the entire --- continuously lengthening and, in fact, practically infinite --- interaction history every time before responding. Defining strategies as functions of only the latest moves (rather than entire interaction histories) in Abramsky and Jagadeesan's \cite{Abr94} tradition  is also not a way out, as typically more than just the last move matters. Back to your personal computer, its actions certainly depend on more than 
your last keystroke.

Computability in the traditional Church-Turing sense is a special case of winnability --- winnability restricted to two-step (input/output, question/answer) interactive problems. So is the classical concept of truth, which is nothing but winnability restricted to propositions, viewed by CL as zero-step problems, i.e. games with no moves that are automatically won or lost depending on whether they are true or false. This way, the semantics of CL is a generalization, refinement and conservative extension of that of classical logic.

Thinking of a human user in the role of the environment, computational problems are synonymous to computational tasks 
 --- tasks performed by a machine for the user/environment. 
What is a task for a machine is then a resource for the environment, and vice versa. So the CL games, at the same time, formalize our intuition of {\em computational resources}.\index{computational resource}\label{0computational resource}    Logical operators are understood as operations on such  tasks/ resources/games, atoms as variables ranging over tasks/resources/games, and validity of a logical formula as being ``always winnable", i.e. as existence --- under every particular interpretation of atoms --- of a machine that successfully accomplishes/provides/wins the corresponding task/resource/game no matter how the 
environment behaves.  With this semantics, `computability logic is a formal theory of computability in the same sense as classical logic is a formal theory of truth' (\cite{CL1}).   Furthermore, as mentioned, the classical concept of truth is a special case of
winnability,   
 which eventually translates into classical logic's being nothing but a special fragment of computability logic.  

CL is a semantically constructed logic and, at this young age, its syntax is only just starting to develop, with open problems and unverified conjecture prevailing over answered questions. In a sense, this situation is opposite to the case with some other non-classical traditions such as intuitionistic or linear logics  where, as most logicians would probably agree, ``in the beginning was Syntax", and really good formal semantics convincingly justifying the proposed syntactic constructions are still being looked for. In fact, the semantics of CL can be seen to be providing such a justification, although, at least for linear logic, this is only in a limited sense 
explained below. 

The set of valid formulas in a certain fragment of the otherwise more expressive language of CL forms a logic that is similar to but by no means  the same as linear logic.\index{linear logic}\label{0linear logic} The two logics typically agree on short and simple formulas (perhaps with the exception for those involving exponentials, where disagreements may start already on some rather short formulas). For instance, both logics reject  $P\mli P\mlc P$  and accept $P\mli P\adc P$, with classical-shape propositional connectives here and later understood as the corresponding multiplicative operators of linear logic, and square-shape operators as additives ($\adc$=``with", $\add$=``plus"). Similarly, both logics reject $P\add\gneg P$ and accept $P\mld \gneg P$. On the other hand, CL disagrees with linear logic on many more evolved formulas. E.g., CL validates the following two principles rejected even by {\em affine logic 
$\al$}\index{affine logic}\label{0affine logic} --- linear logic with the weakening rule: 

\[((P\mlc Q)\mld(R\mlc S))\mli((P\mld R)\mlc(Q\mld S));\] 
\[(P\mlc(R\adc S))\adc(Q\mlc(R\adc S))\adc((P\adc Q)\mlc R)\adc ((P\adc Q)\mlc S) \mli (P\adc Q)\mlc(R\adc S).\]

Neither the similarities nor the discrepancies are a surprise. The philosophies of CL and linear logic overlap in their striving to develop a logic of resources. But the ways this philosophy is materialized are rather different. CL starts with a mathematically strict and intuitively convincing semantics, and only after that, as a natural second step, asks what the corresponding logic and its possible axiomatizations are. On the other hand, it would be accurate to say   
that linear logic started directly from the second step. Even though certain companion semantics were provided 
for it from the very beginning, those are not quite what we earlier agreed to call capital-`S'. As a formal theory of resources (rather than that of phases or coherent spaces), linear logic has been motivated and introduced  syntactically
rather than semantically, essentially by taking classical sequent calculus and deleting the rules that seemed unacceptable from a certain intuitive, naive resource point of view. Hence,  in the absence of a clear formal concept of resource-semantical truth or validity, the question about whether the resulting system was complete could not even be meaningfully asked. In this process of syntactically rewriting classical logic some innocent, deeply hidden principles could have easily gotten victimized. CL believes that this is exactly what happened, with the above-displayed formulas separating it from linear logic --- and more such formulas to be seen later --- viewed as babies thrown out with the bath water. Of course, many retroactive attempts have been made to find semantical (often game-semantical) justifications for linear logic. Technically it is always possible to come up with some sort of a formal semantics that matches a given target syntactic construction, but the whole question is how natural and meaningful such a semantics is in its own rights, and how adequately it corresponds to the logic's underlying philosophy and ambitions. `Unless, by good luck, the target system really {\em is} ``the right logic", the chances of a decisive success when following the odd scheme {\em from syntax to semantics} could be rather slim' (\cite{Jap03}). The natural scheme is {\em from semantics to syntax}. It matches the way classical logic evolved and climaxed in G\"{o}del's completeness theorem. And, as we now know, this is exactly the scheme that computability logic, too, follows. 
  
Intuitionistic logic\index{intuitionistic logic}\label{0intuitionistic logic} is another example of a syntactically conceived logic. Despite decades of efforts, no fully convincing semantics has been found for it. Lorenzen's game semantics\index{Lorenzen's game semantics}\label{0Lorenzen's game semantics}
\cite{Fel85,Lor59}, which has a concept of validity without having a concept of truth, has been perceived as a technical supplement to the existing syntax rather than as having independent importance. Some other semantics, such as Kleene's realizability\index{Kleene's realizability}\label{0Kleene's realizability} \cite{Kle52} or G\"{o}del's Dialectica interpretation\index{G\"{o}del's Dialectica interpretation}\label{0Dialectica interpretation} \cite{God58}, are closer to what we might qualify as capital-`S'. But, unfortunately, they validate certain principles unnegotiably rejected by intuitionistic logic.
From our perspective, the situation here is much better than with linear logic though. In \cite{int1}, Heyting's first-order intuitionistic 
calculus\index{Heyting's intuitionistic calculus}\label{0Heyting's intuitionistic calculus} has been shown to be sound with respect to the CL semantics. And the propositional fragment of Heyting's calculus has also been shown to be complete (\cite{Japjsl1,Japapal,Japjsl2,Ver}). This signifies success 
--- at least at the propositional level --- in semantically justifying intuitionistic logic, and a materialization of 
Kolmogorov's\index{Kolmogorov's thesis}\label{0Kolmogorov's thesis} \cite{Kol32} well known yet so far rather abstract thesis according to which intuitionistic logic is a logic of problems. Just as the resource philosophy of CL overlaps with that of linear logic, so does its algorithmic philosophy with the constructivistic philosophy of intuitionism. The difference, again, is in the ways this philosophy is materialized.  Intuitionistic logic has come up with a ``constructive syntax" without having an adequate underlying formal semantics, such as a clear concept of truth in some constructive sense. This sort of a syntax was essentially obtained from the classical one by banning the offending law of  the excluded middle.\index{excluded middle}\label{0excluded middle} But, as in the case of linear logic, the critical question immediately springs out: where is a guarantee that together with excluded middle some innocent principles would not be expelled as well? The constructivistic claims of CL, on the other hand, are based on the fact that it defines truth as algorithmic solvability. The philosophy of CL does not find the term {\em constructive syntax} meaningful unless it is understood as soundness with respect to some {\em constructive semantics}, for only a semantics may or may not be constructive in a reasonable sense. The reason for the failure of $P\add\gneg P$ in CL is not that this principle ... is not included in its axioms. Rather, the failure of this principle is exactly the reason why this principle, or anything else entailing it, would not be among the axioms of a sound system for CL. Here ``failure'' has 
a precise semantical meaning. It is non-validity, i.e. existence of a problem $A$ for which $A\add\gneg A$ is not algorithmically solvable.  

It is also worth noting that, while intuitionistic logic irreconcilably defies classical logic, computability logic comes up with a peaceful solution acceptable for everyone. The key is the expressiveness of its language, that has (at least) two versions for each traditionally controversial logical operator, and particularly the two versions $\mld$ and $\add$ of disjunction. As will be seen later, the semantical meaning of $\mld$ conservatively extends --- from moveless games to all games --- its classical meaning, and the principle $P\mld\gneg P$ survives as it represents  an always-algorithmically-solvable combination of problems, even if solvable in a sense that some constructivistically-minded might fail --- or pretend to fail --- to understand. And the semantics of $\add$, on the other hand, formalizes and conservatively extends a different, stronger meaning which apparently every constructivist associates with disjunction. As expected, then $P\add\gneg P$ turns out to be semantically invalid. CL's
proposal for settlement between classical and constructivistic logics then reads: `If you are open (=classically) minded, take advantage of the full expressive power of CL; and if you are constructivistically minded, just identify a collection of the operators whose meanings seem constructive enough to you, mechanically disregard everything containing the other operators, and put an end to those fruitless fights about what deductive methods or principles should be considered right and what should be deemed wrong' (\cite{Jap03}).
  
Back to linear --- more precisely, affine --- logic. As mentioned, $\al$ is sound with respect to the CL semantics, a proof of which is the main new technical contribution of the present paper. This is definitely good news from the ``better something than nothing" standpoint. $\al$ is simple and, even though incomplete, still  reasonably strong. 
What is worth noting is that our soundness proof for $\al$, just as all other soundness proofs known so far in CL, including that for the intuitionistic fragment (\cite{int1}), or the {\bf CL4} fragment (\cite{CL4}) that will be discussed in Section \ref{ss9}, is {\em constructive}. This is     
in the sense that, whenever a formula $F$ is provable in a given deductive system, an algorithmic solution for the problem(s) represented by $F$ can be automatically extracted from the proof of $F$. The persistence of this phenomenon for various fragments of CL carries another piece of good news: CL provides a systematic answer not only to the theoretical  
question ``{\em what} can be computed?" but, as it happens, also to the more terrestrial  question ``{\em how} can be computed?".
  
The main practical import of the constructive soundness result for $\al$ (just as for any other sublogic of CL) is related to the potential of basing applied theories or knowledge base systems on that logic, the latter being a reasonable, computationally meaningful 
alternative to classical logic. The non-logical axioms --- or knowledge base --- of an $\al$-based applied system/theory would be any collection of (formulas representing) problems whose algorithmic solutions are known. Then our soundness result for $\al$ guarantees that every theorem $T$ of the theory also has an algorithmic solution and that, furthermore, such a solution, itself, can be effectively constructed from a proof of $T$. This makes $\al$ a systematic problem-solving tool: finding a solution for a given problem reduces to finding a proof of that problem in an $\al$-based theory. The incompleteness of $\al$ only means that, in its language, this logic is not as perfect/strong as 
as a formal tool could possibly be, and that, depending on needs, it makes sense to continue looking for further sound extensions (up to a complete one) of it. As pointed out earlier, 
when it comes to applications, unlike soundness, completeness is a desirable but not necessary condition. 

With the two logics in a sense competing for the same market, the main --- or perhaps only --- advantage of linear logic over CL is its having a nice and simple syntax. In fact, linear logic 
{\em is} (rather than {\em has}) a beautiful syntax; and computability logic {\em is} (rather than {\em has}) a meaningful semantics. At this point it is not clear what a CL-semantically complete extension of $\al$ would look like syntactically. As a matter of fact, while the set of  valid formulas of the exponential-free fragment of the language of linear logic has been shown  to be decidable \mbox{(\cite{CL4}),} so far it is not even known whether that set in the full language is recursively enumerable. If it is, finding a complete axiomatization for it would most likely require a substantially new syntactic approach, going far beyond the traditional sequent-calculus framework within which linear logic is constructed (a possible candidate here is cirquent calculus, briefly discussed at the end of this section). And, in any case, such an axiomatization would hardly be as simple as that of $\al$, so the syntactic simplicity advantage of linear logic will always remain. Well, CL has one thing to say: simplicity is good, yet, if it is most important, then nothing can ever beat ... the empty logic. 

The rest of this paper is organized as follows. Sections \ref{ss2}-\ref{ss8} provide a detailed introduction to the basic semantical concepts of computability logic: games and operations on them, two equivalent models of interactive computation (algorithmic strategies), and validity. The coverage of most of these concepts is more detailed here than in the earlier survey-style papers \cite{Jap03,Japic} on CL, and is supported with ample examples and illustrations. Section \ref{ss9} provides an overview, without a proof, of the strongest 
technical result obtained so far in computability logic, specifically, the soundness and completeness of system $\predell$, whose logical vocabulary contains negation $\gneg$, parallel (``multiplicative'') connectives $\mlc,\mld,\mli$, choice (``additive'') connectives $\adc,\add$ with their quantifier counterparts $\ada,\ade$, and blind (``classical'') quantifiers $\cla,\cle$. Section   \ref{applc} outlines potential applications of computability logic in 
knowledge base systems, systems for planning and action, and constructive applied theories. There the survey part of the paper ends, and the following two sections  are devoted to a formulation (Section \ref{ss11}) and proof (Section \ref{snew}) of the new result --- the soundness of affine logic with respect to the semantics of CL. The final Section \ref{ssconcl} outlines some possible future developments in the area.

This paper gives an overview of most results known in computability logic
as of the end of 2005, by the time when the main body of the text was written. The present paragraph is a last-minute addition made at the beginning of 2008. Below is a list of the most important developments that, due to being very recent, have received no coverage in this chapter:
\begin{itemize} 
\item As already mentioned, the earlier conjecture about the completeness of Heyting's propositional intuitionistic calculus with respect to the semantics of CL has been resolved positively. A completeness proof for the implicative fragment of intuitionistic logic was given in \cite{Japjsl1}, and that proof was later extended to the full propositional intuitionistic calculus in \cite{Japapal}. With a couple of months' delay, Vereshchagin \cite{Ver} came up with an alternative proof of the same result. 
\item In \cite{Japjsl2}, the implicative fragment of affine logic has been proven to be complete with respect to the semantics of computability logic. The former is nothing but implicative intuitionistic logic without the rule of contraction. Thus, both the implication of intuitionistic logic and the implication of affine logic have adequate interpretations in CL --- specifically, as the operations $\intimpl$ and $\mli$, respectively. Intuitively, as will be shown later in Section \ref{ss4}, these are two natural versions of the operation of reduction, with the difference between $A\intimpl B$ and $A\mli B$ being that in the former $A$ can be ``reused'' while in the latter it cannot. 
\cite{Japjsl2} also introduced a series of intermediate-strength natural versions of reduction operations. 
\item Section \ref{ss4.6} briefly mentions {\em sequential operations}. The recent paper \cite{Japseq} has provided an elaboration of this new group of operations (formal definitions, associated computational intuitions, motivations, etc.), making them full-fledged citizens of computability logic. It has also constructed a sound and complete axiomatization of the fragment of computability logic whose logical vocabulary, together with negation, contains three --- parallel, choice and sequential --- sorts of conjunction and disjunction. 
\item  Probably the most significant of the relevant recent developments is the invention of {\em cirquent calculus}\index{cirquent calculus}\label{0cirquent calculus} in \cite{Japcirq,Japjlc2}. Roughly, this is a deductive approach based on circuits instead of formulas or sequents. It can be seen as a refinement of Gentzen's\index{Gentzen}\label{0Gentzen} methodology, and correspondingly the methodology of linear logic based on  the latter, achieved by allowing shared resources between different parts of sequents and proof trees. Thanks to the sharing mechanism, cirquent calculus, being more general and flexible than sequent calculus,\index{sequent calculus}\label{0sequent calculus} appears to be the only reasonable proof-theoretic approach capable of syntactically taming the otherwise wild computability logic. Sharing also makes it possible to achieve exponential-magnitude compressions of formulas and proofs, whether it be in computability logic or the kind old classical logic. 
\end{itemize}

\section{Constant games}\label{ss2}

The symbolic names used in CL for the two players {\em machine} and {\em environment} are 
$\twg$\index{$\twg$ (as a player)}\label{0twg (as a player)} and 
$\tlg$,\index{$\tlg$ (as a player)}\label{0tlg (as a player)}
 respectively.  $\xx$\index{$\xx$}\label{0xx} is always  a variable ranging over $\{\twg,\tlg\}$, with 
\[\pneg \xx\index{$\pneg$ (as an operation on players)}\label{0pneg (as an operation on players)}\] meaning $\xx$'s adversary, i.e. the player which is not $\xx$. Even though it is often a human user who acts in the role of $\oo$, our sympathies are with $\twg$ rather than $\tlg$, 
and by just saying ``won" or ``lost" without specifying a player we always mean won or lost by $\twg$. 

 The reason why I should be a fan of the machine even --- in fact especially --- when it is playing against me is that the machine is a tool, and what makes it valuable as such is 
exactly its winning the game, i.e. its not being malfunctional (it is precisely losing by a machine the game that it was supposed to win what in everyday language is called {\em malfunctioning}). 
Let me imagine myself using a computer for computing the ``$28\%$ of $x$" function in the process of preparing my federal tax return. This is a game where the first move is mine, consisting in inputting a number $m$ and meaning asking $\pp$ the question ``what is $28\%$ of $m$?". The machine wins iff it answers by the move/output $n$ such that $n=0.28m$. Of course, I do not want the machine to tell me that $27,000$ is $28\%$ of $100,000$. In other words, I do not want to win against the machine. For then I could lose the more important game against Uncle Sam. 

Before getting to a formal definition of games, let us agree without loss of generality that a 
{\bf move}\index{move}\label{0move} is always a string over the standard keyboard alphabet. One of the non-numeric and non-punctuation symbols of this alphabet, denoted $\spadesuit$,\index{$\spadesuit$}\label{0spadesuit} is designated as a special-status move, intuitively meaning a move that is always illegal to make. 
A {\bf labeled move}\index{labeled move (labmove)}\label{0labeled move (labmove)} ({\bf labmove}) is a move prefixed with $\pp$ or $\oo$, with its prefix ({\bf label})\index{label}\label{0label} indicating which player has made the move. 
A {\bf run}\index{run}\label{0run} is a (finite or infinite) sequence of labmoves, and a 
{\bf position}\index{position}\label{0position} is a finite run.

We will be exclusively using the letters $\Gamma,\Delta,\Theta,\Phi,\Psi,\Upsilon,\Lambda,\Sigma,\Pi$  for runs, $\alpha,\beta,\gamma,\delta$ for moves, and $\lambda$ for labmoves. 
Runs will be often delimited by ``$\langle$" and ``$\rangle$", with $\emptyrun$\index{$\emptyrun$}\label{0emptyrun} thus denoting the {\bf empty run}.\index{empty run}\label{0empty run} The meaning of an expression such as $\seq{\Phi,\xx\alpha,\Gamma}$ must be clear: this is the result of appending to position $\Phi$ 
the labmove $\xx\alpha$ and then the run $\Gamma$. We write \[\rneg\Gamma\index{$\rneg$ (as an operation on runs)}\label{0rneg (as an operation on runs)}\]  for the result of simultaneously replacing every label $\xx$ in every labmove of $\Gamma$ by $\gneg\xx$.

Our ultimate definition of games will be given later in terms of the simpler and more basic class of games called {\em constant}. The following is a formal definition of constant games combined with some less formal conventions regarding the usage of certain terminology.

\begin{definition}\label{game}
 A {\bf constant game}\index{constant game}\label{0constant game} is a pair $A=(\legal{A}{},\win{A}{})$, where:\vspace{10pt}

1. $\legal{A}{}$,\index{$\legal{}{}$}\label{0legalruns} called the {\bf structure}\index{structure (of a game)}\label{0structure (of a game)} of $A$, is a set of runs not containing (whatever-labeled) move $\spadesuit$, satisfying the condition that a finite or infinite run is in $\legal{A}{}$ iff all of its nonempty finite --- not necessarily proper --- initial
segments are in $\legal{A}{}$ (notice that this implies $\emptyrun\in\legal{A}{}$). The elements of $\legal{A}{}$ are
said to be {\bf legal runs}\index{legal run}\label{0legal run} of $A$, and all other runs are said to be {\bf illegal runs}\index{illegal run}\label{0illegal run} of $A$. We say that $\alpha$ is a {\bf legal move}\index{legal move}\label{0legal move} for $\xx$ in a position $\Phi$ of $A$ iff $\seq{\Phi,\xx\alpha}\in\legal{A}{}$; otherwise 
$\alpha$ is an {\bf illegal move}.\index{illegal move}\label{0illegal move} When the last move of the shortest illegal initial segment of $\Gamma$  is $\xx$-labeled, we say that $\Gamma$ is a {\bf $\xx$-illegal run}\index{$\xx$-illegal run}\label{0xx-illegal run} of $A$.\vspace{5pt} 

2. $\win{A}{}$,\index{$\win{}{}$}\label{0winA} called the {\bf content}\index{content (of a game)}\label{0content (of a game)} of $A$,  is a function that sends every run $\Gamma$ to one of the players $\pp$ or $\oo$, satisfying the condition that if $\Gamma$ is a $\xx$-illegal run of $A$, then $\win{A}{}\seq{\Gamma}\not=\xx$. When $\win{A}{}\seq{\Gamma}=\xx$, we say that $\Gamma$ is a {\bf $\xx$-won\index{won run}\label{0won run}} (or {\bf won by $\xx$}) run of $A$; otherwise $\Gamma$ is {\bf lost by $\xx$}.\index{lost run}\label{0lost run} Thus, an illegal run is always lost by the player who has made the first illegal move in it.  
\end{definition}

Let $A$ be a constant game. $A$ is said to be {\bf finite-depth}\index{finite-depth game}\label{0finite-depth game} iff there is a (smallest) integer $d$, called the {\bf depth}\index{depth (of a game)}\label{0depth (of a game)} of $A$, such that the length of every legal run of $A$ is $\leq d$. And $A$ is 
{\bf perifinite-depth}\index{perifinite-depth game}\label{0perifinite-depth game} iff every legal run of it is finite, even if there are arbitrarily long legal runs. \cite{Jap03} defines the depths of perifinite-depth games in terms of ordinal numbers, which are finite for finite-depth games and transfinite for all other perifinite-depth games.  Let us call a legal run $\Gamma$ of $A$ {\bf maximal}\index{maximal run}\label{0maximal run} iff $\Gamma$ is not a proper initial segment of any other legal run of $A$. Then we say that $A$ is 
{\bf finite-breadth}\index{finite-breadth game}\label{0finite-breadth game} if the total number of maximal legal runs of $A$, called the {\bf breadth}\index{breadth (of a game)}\label{0breadth (of a game)} of $A$, is finite. Note that, in a general case, the breadth of a game may be not only infinite, but even uncountable. $A$ is said to be (simply) {\bf finite}\index{finite game}\label{0finite game} iff it only has a finite number of legal runs. Of course, $A$ is finite only if it is finite-breadth, and when $A$ is finite-breadth, it is finite iff it is finite-depth iff it is perifinite-depth.  

The structure component of a constant game can be visualized as a tree whose arcs are labeled with labmoves, as shown in Figure 1. Every branch of such a tree represents a legal run, specifically, the sequence of the labels of the arcs of that branch in the top-down direction starting from the root. For instance, the rightmost branch (in its full length) of the 
tree of Figure 1 corresponds to the run $\seq{\oo\gamma,\pp\gamma,\pp\alpha}$. Thus the nodes of a tree, identified with the (sub)branches that end in those nodes, represent legal positions;  the root stands for the empty position, and leaves for maximal positions.

\begin{center}
\begin{picture}(322,200)

\put(159,192){\circle{16}}
\put(159,184){\line(-3,-1){104}}
\put(80,163){{\tiny $\pp$}{\footnotesize $\alpha$}}
\put(159,184){\line(0,-1){34}}
\put(161,163){{\tiny $\oo$}{\footnotesize $\beta$}}
\put(159,184){\line(3,-1){104}}
\put(228,163){{\tiny $\oo$}{\footnotesize $\gamma$}}

\put(52,142){\circle{16}}
\put(52,134){\line(-3,-2){51}}
\put(8,113){{\tiny $\oo$}{\footnotesize $\beta$}}
\put(52,134){\line(3,-2){51}}
\put(87,113){{\tiny $\oo$}{\footnotesize $\gamma$}}
\put(159,142){\circle{16}}
\put(159,134){\line(0,-1){34}}
\put(161,113){{\tiny $\pp$}{\footnotesize $\alpha$}}
\put(266,142){\circle{16}}
\put(266,134){\line(-3,-2){51}}
\put(221,113){{\tiny $\pp$}{\footnotesize $\alpha$}}
\put(266,134){\line(0,-1){34}}
\put(268,113){{\tiny $\pp$}{\footnotesize $\beta$}}
\put(266,134){\line(3,-2){51}}
\put(299,113){{\tiny $\pp$}{\footnotesize $\gamma$}}

\put(0,92){\circle{16}}
\put(104,92){\circle{16}}
\put(104,84){\line(-1,-3){11}}
\put(85,63){{\tiny $\pp$}{\footnotesize $\beta$}}
\put(104,84){\line(1,-3){11}}
\put(112,63){{\tiny $\pp$}{\footnotesize $\gamma$}}
\put(159,92){\circle{16}}
\put(214,92){\circle{16}}
\put(214,84){\line(-1,-3){11}}
\put(195,63){{\tiny $\pp$}{\footnotesize $\beta$}}
\put(222,63){{\tiny $\pp$}{\footnotesize $\gamma$}}
\put(214,84){\line(1,-3){11}}
\put(266,92){\circle{16}}
\put(266,84){\line(0,-1){34}}
\put(268,63){{\tiny $\pp$}{\footnotesize $\alpha$}}
\put(318,92){\circle{16}}
\put(318,84){\line(0,-1){34}}
\put(320,63){{\tiny $\pp$}{\footnotesize $\alpha$}}

\put(92,42){\circle{16}}
\put(116,42){\circle{16}}
\put(202,42){\circle{16}}
\put(226,42){\circle{16}}
\put(266,42){\circle{16}}
\put(318,42){\circle{16}}

\put(105,10){{\bf Figure 1:} A structure}
\end{picture}
\end{center}

Notice the relaxed nature of our games. In the empty position of the above-depicted structure, both players have legal moves. This can be seen from the two ($\pp$-labeled and $\oo$-labeled) sorts of labmoves on the outgoing arcs of the root. Let us call such positions/nodes {\bf heterogenous}.\index{heterogeneous position}\label{0heterogeneous position} 
Generally any non-leaf nodes can be heterogenous, even though in our particular example only the root is so.  As we are going to see later, in heterogenous positions indeed either player is free to move. Based on this liberal attitude, our games can be called {\bf free},\index{free game}\label{0free game} as opposed to {\em strict} games where, in every situation, at most one of the players is allowed to move. Of course, strict games can be considered special cases of our free games --- the cases with no heterogenous nodes. Even though not having legal moves does not formally preclude the ``wrong" player to move in a given position, such an action, as we remember, results in an immediate loss for that player and hence amounts to not being permitted to move. 
 There are good reasons in favor of the free-game approach. Hardly many tasks that humans, computers or robots perform in real life are strict. Imagine you are playing chess over the Internet on two boards against two independent adversaries that, together, form the (one) environment for you. Let us say you play white on both boards. Certainly the initial position of this game is not heterogenous. However, once you make your first move --- say, on board \#1 --- the picture changes. Now both you and the environment have legal moves, and who will be the next to move depends on who can or wants to act sooner. Namely, you are free to make another opening move on board \#2, while the environment --- adversary \#1 --- can make a reply move on board \#1. A strict-game approach would have to impose some not-very-adequate supplemental conditions uniquely determining the next player to move, such as not allowing you to move again until receiving a response to your previous move. Let alone that this is not how the real two-board game would proceed, such regulations defeat the very purpose of the idea of parallel/distributed computations with all the known benefits it offers. 

While the above discussion used the term ``strict game'' in a perhaps somewhat more general sense, let us agree that from now on we will stick to the following meaning of that term: 

\begin{definition}\label{strict}
A constant game \ $A$ \ is said to be \ {\bf strict}\index{strict game}\label{0strict game} \ iff, \ for every \ legal position \ $\Phi$ \ of \ $A$, \  we have \  
$\{\alpha\ |\ \seq{\Phi,\pp\alpha}\in\legal{A}{}\}=\emptyset$ \ or \ $\{\alpha\ |\ \seq{\Phi,\oo\alpha}\in\legal{A}{}\}=\emptyset.$
\end{definition}

Figure 2 adds a content to the structure of Figure 1, thus turning it into a constant game:

\begin{center}
\begin{picture}(322,200)

\put(159,192){\circle{16}}
\put(155,188){$\oo$}
\put(159,184){\line(-3,-1){104}}
\put(80,163){{\tiny $\pp$}{\footnotesize $\alpha$}}
\put(159,184){\line(0,-1){34}}
\put(161,163){{\tiny $\oo$}{\footnotesize $\beta$}}
\put(159,184){\line(3,-1){104}}
\put(228,163){{\tiny $\oo$}{\footnotesize $\gamma$}}

\put(52,142){\circle{16}}
\put(48,138){$\pp$}
\put(52,134){\line(-3,-2){51}}
\put(8,113){{\tiny $\oo$}{\footnotesize $\beta$}}
\put(52,134){\line(3,-2){51}}
\put(87,113){{\tiny $\oo$}{\footnotesize $\gamma$}}
\put(159,142){\circle{16}}
\put(155,138){$\pp$}
\put(159,134){\line(0,-1){34}}
\put(161,113){{\tiny $\pp$}{\footnotesize $\alpha$}}
\put(266,142){\circle{16}}
\put(262,138){$\oo$}
\put(266,134){\line(-3,-2){51}}
\put(221,113){{\tiny $\pp$}{\footnotesize $\alpha$}}
\put(266,134){\line(0,-1){34}}
\put(268,113){{\tiny $\pp$}{\footnotesize $\beta$}}
\put(266,134){\line(3,-2){51}}
\put(299,113){{\tiny $\pp$}{\footnotesize $\gamma$}}

\put(0,92){\circle{16}}
\put(-4,88){$\pp$}
\put(104,92){\circle{16}}
\put(100,88){$\oo$}
\put(104,84){\line(-1,-3){11}}
\put(85,63){{\tiny $\pp$}{\footnotesize $\beta$}}
\put(104,84){\line(1,-3){11}}
\put(112,63){{\tiny $\pp$}{\footnotesize $\gamma$}}
\put(159,92){\circle{16}}
\put(155,88){$\pp$}
\put(214,92){\circle{16}}
\put(210,88){$\oo$}
\put(214,84){\line(-1,-3){11}}
\put(195,63){{\tiny $\pp$}{\footnotesize $\beta$}}
\put(222,63){{\tiny $\pp$}{\footnotesize $\gamma$}}
\put(214,84){\line(1,-3){11}}
\put(266,92){\circle{16}}
\put(262,88){$\pp$}
\put(266,84){\line(0,-1){34}}
\put(268,63){{\tiny $\pp$}{\footnotesize $\alpha$}}
\put(318,92){\circle{16}}
\put(314,88){$\oo$}
\put(318,84){\line(0,-1){34}}
\put(320,63){{\tiny $\pp$}{\footnotesize $\alpha$}}

\put(92,42){\circle{16}}
\put(88,38){$\pp$}
\put(116,42){\circle{16}}
\put(112,38){$\oo$}
\put(202,42){\circle{16}}
\put(198,38){$\pp$}
\put(226,42){\circle{16}}
\put(222,38){$\oo$}
\put(266,42){\circle{16}}
\put(262,38){$\pp$}
\put(318,42){\circle{16}}
\put(314,38){$\oo$}

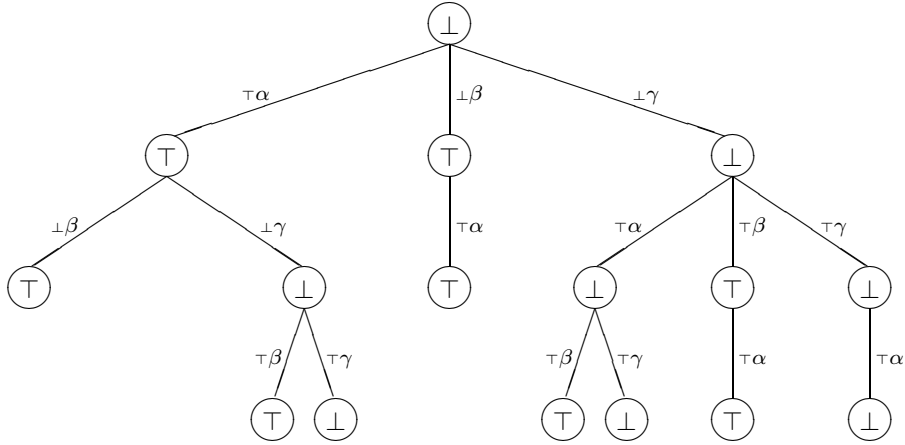
\put(40,10){{\bf Figure 2:} Constant game = structure + content}
\end{picture}
\end{center}

Here the label of each node indicates the winner in the corresponding position. E.g., we see that the empty run is won by $\oo$, and the run $\seq{\pp\alpha,\oo\gamma,\pp\beta}$ won by $\pp$. There is no need to indicate winners for illegal runs: as we remember, such runs are lost by the player responsible for making them illegal, so we can tell at once that, say, 
$\seq{\pp\alpha,\oo\gamma,\pp\alpha,\pp\beta,\oo\gamma}$ is lost by $\pp$ because the offending third move of it is $\pp$-labeled. Generally, every perifinite-depth constant game can be fully represented in the style of Figure 2 by labeling the nodes of the corresponding structure tree. To capture a non-perifinite-depth game, we will need some additional way to indicate the winners in infinite branches, for no particular (end)nodes represent such branches.
 
The traditional, strict-game approach usually defines a player $\xx$'s strategy\index{strategy}\label{0strategy} as a function that sends every position in which $\xx$ has legal moves to one of those moves. As pointed out earlier, such a functional view is no longer applicable in the case of properly free games. Indeed, if $f_{\pp}$ and $f_{\oo}$ are the two players' functional strategies for the game of Figure 2 with $f_{\pp}(\emptyrun)=\alpha$ and $f_{\oo}(\emptyrun)=\beta$, then it is not clear whether the first move of the corresponding run will be $\pp\alpha$ or $\oo\beta$. Yet, even if not functional, 
$\pp$ does have a winning strategy for that game. What, exactly, a {\em strategy} means will be explained in Section \ref{icp}. For now, in our currently available ad hoc terms, one of $\pp$'s winning strategies sounds as follows: ``Regardless of what the adversary is doing or has done, go ahead and make move $\alpha$; make $\beta$ as your second move if and when you see that the adversary has made move $\gamma$, no matter whether this happened before or after your first move''. Which of the runs consistent with this strategy will become the actual one depends on how (and how fast) $\oo$ acts, yet every such run will be a success for $\pp$. It is left as an exercise for the reader to see that there are exactly five possible legal runs consistent with $\pp$'s above strategy, all won by $\pp$:  $\seq{\pp\alpha}$, 
$\seq{\pp\alpha,\oo\beta}$, $\seq{\pp\alpha,\oo\gamma,\pp\beta}$, $\seq{\oo\beta,\pp\alpha}$ and $\seq{\oo\gamma,\pp\alpha,\pp\beta}$. As for illegal runs consistent with that strategy,
it can be seen that every such run would be   $\oo$-illegal and hence, again, won by $\pp$.  

Below comes our first formal definition of a game operation. This operation, called 
{\bf prefixation},\index{prefixation}\label{0prefixation} is somewhat reminiscent of the modal operator(s) of dynamic logic. It takes two arguments: 
a (here constant) game $A$ and a legal position $\Phi$ of $A$, and generates the game $\seq{\Phi}A$ that, with  $A$ visualized as a tree in the style of Figure 2, is nothing but the subtree rooted at 
the node corresponding to position $\Phi$. This operation is undefined when $\Phi$ is an illegal position of $A$.

\begin{definition}\label{prfx}
Let $A$ be a constant game, and $\Phi$ a legal position of $A$. The game 
$\seq{\Phi}A$\index{$\seq{\Phi}A$}\label{0seqPhiA} is defined by: 
\begin{itemize}
\item $\legal{\seq{\Phi}A}{}=\{\Gamma\ |\ \seq{\Phi,\Gamma}\in\legal{A}{}\}$.
\item $\win{\seq{\Phi}A}{}\seq{\Gamma}=\win{A}{}\seq{\Phi,\Gamma}$.
\end{itemize}
\end{definition}  

Intuitively, $\seq{\Phi}A$ is the game playing which means playing $A$ starting (continuing) from position $\Phi$. 
That is, $\seq{\Phi}A$ is the game to which $A$ {\bf evolves} (will be ``{\bf brought down}") after the moves of $\Phi$ have been made. 

\section{Games in general, and nothing but games}\label{ss3}

Computational problems in the traditional, Church-Turing sense can be seen as strict, depth-$2$ games of the special type shown in Figure 3. The first-level arcs of such a game represent inputs, i.e. $\oo$'s moves; and the second-level arcs represent outputs, i.e. $\pp$'s moves. 
The root of this sort of a game is always $\pp$-labeled as it corresponds to the situation when there was no input, in which case the machine is considered the winner because the absence of an input removes any further responsibility from it.  All second-level nodes, on the other hand, are $\oo$-labeled, for they represent the situations when there was an input but the machine failed to generate any output. Finally, each group of siblings of the third-level nodes has exactly one $\pp$-labeled member. This is so because traditional problems are about computing functions, meaning that 
there is exactly one ``right'' output per given input. What particular nodes of those groups will have label $\pp$ --- and only this part of the game tree --- depends on what particular function is the one under question. The game of Figure 3 is about computing the successor function.

\begin{center}
\begin{picture}(322,160)

\put(175,142){\circle{16}}
\put(-10,109){\scriptsize Input}
\put(171,138){$\pp$}
\put(175,134){\line(-3,-1){106}}
\put(83,109){\tiny $\oo 1$}
\put(175,134){\line(0,-1){34}}
\put(163,109){\tiny $\oo 2$}
\put(175,134){\line(3,-1){106}}
\put(226,109){\tiny $\oo 3$}
\put(175,134){\line(6,-1){119}}
\put(285,109){\Huge ...}

\put(-10,59){\scriptsize Output}
\put(65,92){\circle{16}}
\put(61,88){$\oo$}
\put(30,59){\tiny $\pp 1$}
\put(65,84){\line(-1,-3){11}}
\put(46,59){\tiny $\pp 2$}
\put(65,84){\line(-1,-1){34}}
\put(62,59){\tiny $\pp 3$}
\put(65,84){\line(1,-3){11}}
\put(77,59){\tiny $\pp 4$}
\put(65,84){\line(1,-1){34}}
\put(95,59){\large ...}
\put(65,84){\line(3,-2){34}}

\put(29,42){\circle{16}}
\put(25,38){$\oo$}
\put(53,42){\circle{16}}
\put(49,38){$\pp$}
\put(77,42){\circle{16}}
\put(73,38){$\oo$}
\put(101,42){\circle{16}}
\put(97,38){$\oo$}

\put(175,92){\circle{16}}
\put(171,88){$\oo$}
\put(140,59){\tiny $\pp 1$}
\put(175,84){\line(-1,-3){11}}
\put(156,59){\tiny $\pp 2$}
\put(175,84){\line(-1,-1){34}}
\put(172,59){\tiny $\pp 3$}
\put(175,84){\line(1,-3){11}}
\put(187,59){\tiny $\pp 4$}
\put(175,84){\line(1,-1){34}}
\put(205,59){\large ...}
\put(175,84){\line(3,-2){34}}

\put(139,42){\circle{16}}
\put(135,38){$\oo$}
\put(163,42){\circle{16}}
\put(159,38){$\oo$}
\put(187,42){\circle{16}}
\put(183,38){$\pp$}
\put(211,42){\circle{16}}
\put(207,38){$\oo$}

\put(285,92){\circle{16}}
\put(281,88){$\oo$}
\put(250,59){\tiny $\pp 1$}
\put(285,84){\line(-1,-3){11}}
\put(266,59){\tiny $\pp 2$}
\put(285,84){\line(-1,-1){34}}
\put(282,59){\tiny $\pp 3$}
\put(285,84){\line(1,-3){11}}
\put(297,59){\tiny $\pp 4$}
\put(285,84){\line(1,-1){34}}
\put(315,59){\large ...}
\put(285,84){\line(3,-2){34}}

\put(249,42){\circle{16}}
\put(245,38){$\oo$}
\put(273,42){\circle{16}}
\put(269,38){$\oo$}
\put(297,42){\circle{16}}
\put(293,38){$\oo$}
\put(321,42){\circle{16}}
\put(317,38){$\pp$}

\put(55,10){{\bf Figure 3:} The problem of computing $n+1$}
\end{picture}
\end{center}

Once we agree that computational problems are nothing but games, the difference in the degrees of generality and flexibility between the traditional approach to computational problems and our approach becomes apparent and appreciable. 
What we see in Figure 3 is indeed a very special sort of games, and there is no good call for confining ourselves to its limits. In fact, staying within those limits 
would seriously retard any more or less advanced and systematic study of computability. First of all, one would want to get rid of the ``one $\pp$-labeled node per sibling group'' restriction for the third-level nodes. Many natural problems, such as 
the problem of finding a prime integer between $n$ and $2n$, or finding an integral root of $x^2-2n=0$, may have more than one as well as less than one solution. That is, there can be 
more than one as well as less than one ``right'' output on a given input $n$. 
And why not further get rid of any remaining restrictions on the labels of whatever-level nodes and whatever-level arcs.
One can easily think of natural situations when, say, some inputs do not obligate the machine to generate an output and thus the corresponding second-level nodes should be $\pp$-labeled. An example would be the case when the machine is computing a partially-defined function $f$ and receives an input $n$ on which $f$ is undefined. So far we have been talking about generalizations within the depth-2 restriction, corresponding to viewing computational problems as very short dialogues between the machine and its environment. Permitting longer-than-2 or even infinitely long branches would allow us to capture 
problems with arbitrarily high degrees of interactivity and arbitrarily complex interaction protocols.  The task performed by a network server is a tangible example 
of an infinite dialogue between the server and its environment --- the collection of clients, or let us just say the rest of the network. Notice that such a dialogue is usually a properly free
game with a much more sophisticated interface between the interacting parties than the simple input/output interface offered by the ordinary Turing machine model, where the whole story starts 
by the environment asking a question (input) and ends by the machine generating an answer (output), with no interaction whatsoever inbetween these two steps.  

Removing restrictions on depths yields a meaningful generalization not only in the upward, but in the downward direction as well: it does make perfect sense to consider ``dialogues'' of lengths 
less than $2$. Constant games of depth $0$ we call {\bf elementary}.\index{elementary game}\label{0elementary game} There are exactly two elementary constant games, for which we use the same symbols $\twg$\index{$\twg$ (as a game)}\label{0twg (as a game)} and $\tlg$\index{$\tlg$ (as a game)}\label{0tlg (as a game)} as for the two players: 

\begin{center}
\begin{picture}(182,72)
\put(30,62){\em game $\twg$}
\put(48,42){\circle{16}}
\put(44,38){$\pp$}

\put(124,62){\em game $\tlg$}
\put(142,42){\circle{16}}
\put(138,38){$\oo$}

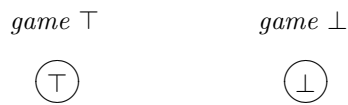
\put(0,10){{\bf Figure 4:} Elementary constant games}
\end{picture}
\end{center}

We identify these with the two propositions of classical logic: $\twg$ ({\em true}) and $\tlg$ ({\em false}). ``Snow is white'' is thus a moveless game automatically won by the machine, while
``Snow is black'' is automatically lost. So, not only traditional computational problems are special cases of our games, but traditional propositions as well. This is exactly what eventually makes classical logic a natural --- elementary --- fragment of computability logic. 

As we know, however, propositions 
are not sufficient to build a reasonably expressive logic. 
For higher expressiveness, classical logic generalizes propositions to predicates. Let us fix two infinite sets of expressions: the set 
$\{v_1,v_2,\ldots\}$ of {\bf variables}\index{variable}\label{0variable1} and the set $\{1,2,,\ldots\}$ of 
{\bf constants}.\index{constant1}\label{0constant1} Without loss of generality here we assume that this collection of constants is exactly the universe of discourse in all cases that we consider.  By a {\bf valuation}\index{valuation}\label{0valuation} we mean 
a function $e$ that sends each variable $x$ to a constant $e(x)$. In these terms, a classical {\bf predicate}\index{predicate}\label{0predicate} $p$ can be understood as 
a function that sends each valuation $e$ to either $\twg$ (meaning that $p$ is true at $e$) or $\tlg$ (meaning that $p$ is false at $e$). 
Propositions can thus be thought of as special, {\em constant} cases of predicates --- predicates that return the same proposition for every valuation.

The concept of games that we define below generalizes constant games in exactly the same sense as the above classical concept of predicates generalizes propositions:

\begin{definition}\label{ngame}
A {\bf game}\index{game}\label{0game} is a function from valuations to constant games. 

We write $e[A]$\index{$e[A]$}\label{0eee} (rather than $A(e)$) to denote the constant game returned by game $A$ for valuation $e$. Such a constant game $e[A]$ is said to be an {\bf instance}\index{instance of a game}\label{0instance of a game} of $A$. 

We also typically write $\legal{A}{e}$\label{0lr}\index{$\legal{A}{e}$} and $\win{A}{e}$\label{0wn}\index{$\wn{A}{e}$} instead of $\legal{e[A]}{}$ and $\win{e[A]}{}$. 
\end{definition}

Throughout this paper, $x,y,z$ will be usually used as metavariables for variables, $c$ for constants, and $e$ for valuations. 

Just as this is the case with propositions versus predicates, we think of constant games in the sense of Definition \ref{game} 
as special, {\em constant} cases of games in the sense of Definition \ref{ngame}. In particular, each constant game $A'$ is the game $A$ such that, for every valuation $e$,
$e[A]=A'$. From now on we will no longer distinguish between such $A$ and $A'$, so that, if $A$ is a constant game,
it is its own instance, with $A=e[A]$ for every $e$.

The notion of elementary game that we defined for constant games naturally generalizes to all games by stipulating that a given game is {\bf elementary}\index{elementary game}\label{0elementary game2} iff all of its instances are so. Hence, just as we identified classical propositions with constant elementary games, classical predicates from now on will be identified with elementary games. For instance, $Even(x)$ is the elementary game such that $e[Even(x)]$ is the game $\twg$ if $e(x)$ is even, and the game $\tlg$ if $e(x)$ is odd. Many other concepts 
originally defined only for constant games --- including the properties {\em strict}, {\em finite}, {\em (peri)finite-depth} and {\em finite-breadth} --- can be extended to all games in a similar way. 

We say that a game $A$ {\bf depends on}\index{depend (a game on a variable)}\label{0depend (a game on a variable)} a variable $x$ iff there are two valuations $e_1,e_2$ which agree on all variables except $x$ such that $e_1[A]\not=e_2[A]$. Constant games thus do not depend on any variables. $A$ is said to be {\bf finitary}\index{finitary game}\label{0finitary game} iff there is a finite set $\vec{x}$ of variables such that, for every two valuations $e_1$ and $e_2$ that agree on all variables 
of $\vec{x}$, we have $e_1[A]=e_2[A]$. The cardinality of (the smallest) such $\vec{x}$ is said to be the {\bf arity}\index{arity of a game}\label{0arity of a game} of $A$. So, ``constant game'' and ``$0$-ary game'' are synonyms.

To generalize the standard operation of substitution of variables to games, let us agree that by a {\bf term}\index{term}\label{0term1} we mean either 
a variable or a constant. The domain of each valuation $e$ is extended to all terms by stipulating that, 
\[\mbox{\em for any constant $c$, $e(c)=c$.}\vspace{5pt}\] 

\begin{definition}\label{sov}
Let $A$ be a game, $x_1,\ldots,x_n$ pairwise distinct variables, and $t_1,\ldots,t_n$ any (not necessarily distinct) terms.
The result of {\bf substituting $x_1,\ldots,x_n$ by $t_1,\ldots,t_n$ in $A$},\index{substitution of variables}\label{0substitution of variables} denoted $A(x_1/t_1,\ldots,x_n/t_n)$,\index{$A(x_1/t_1,\ldots,x_n/t_n)$}\label{0suv} is defined by stipulating that, for every valuation $e$, $e[A(x_1/t_1,\ldots,x_n/t_n)]=e'[A]$, where $e'$ is the valuation for which we have:\vspace{3pt}

$\begin{array}{l}
1.\ \ e'(x_1)=e(t_1),\ \ldots,\ e'(x_n)=e(t_n);\\
2.\  \ \mbox{for every variable $y\not\in\{x_1,\ldots,x_n\}$, $e'(y)=e(y)$.} 
\end{array}$
\end{definition}

Intuitively $A(x_1/t_1,\ldots,x_n/t_n)$ is $A$ with $x_1,\ldots,x_n$ remapped to $t_1,$ $\ldots,t_n$, respectively. 
For instance, if $A$ is the predicate/elementary game $x<y$, then $A(x/y,y/x)$ is $y<x$, $A(x/y)$ is $y<y$, $A(y/3)$ is $x<3$, and $A(z/3)$ --- where $z$ is different from $x,y$ --- remains $x<y$ because $A$ does not depend on $z$.

Following the standard readability-improving practice established in the literature for predicates, we will often fix a tuple $(x_1,\ldots,x_n)$ of pairwise distinct variables for a game $A$ and write $A$ as $A(x_1,\ldots,x_n)$. 
It should be noted that when doing so, by no means do we imply that $x_1,\ldots,x_n$ are all of 
(or only) the variables on which $A$ depends. Representing $A$ in the form $A(x_1,\ldots,x_n)$ sets a context in which we can write $A(t_1,\ldots,t_n)$ to mean the same as the more clumsy expression $A(x_1/t_1,\ldots,x_n/t_n)$. So, if the game $x<y$ is represented as $A(x)$, then $A(3)$ will mean $3<y$ and $A(y)$ mean 
$y<y$. And if the same game is represented as $A(y,z)$ (where $z\not=x,y$), then $A(z,3)$ means $x<z$ while $A(y,3)$ again means $x<y$.

The entities that in common language we call games are at least as often non-constant as constant. Chess  
is a classical example of a constant game. On the other hand, many of the card games --- including solitaire games where only 
one player is active --- are more naturally represented as non-constant games: each session/instance of such a game is set 
by a particular permutation of the card deck, and thus the game can be understood as a game that depends on a variable $x$
 ranging over the possible settings of the deck. Even the game of checkers --- another ``classical example" of a constant game --- has a natural non-constant generalization 
$\checkers\hspace{1pt}(x)$ (with $x$ ranging over $\{8,10,12,14,\ldots\}$), meaning a play on the board of size $x\times x$ where,
in the initial position,  
the first $\frac{3}{2}x$ black cells are filled with white pieces and the last $\frac{3}{2}x$ black cells with black pieces.  Then the 
ordinary checkers can be written as $\checkers\hspace{1pt}(8)$. Furthermore, the numbers of 
pieces of either color also can be made variable, getting an even more general game $\checkers\hspace{1pt}(x,y,z)$, with the ordinary checkers being 
the instance $\checkers\hspace{1pt}(8,12,12)$ of it. By further allowing rectangular- (rather than just square-) shape boards, we would get a game that depends on four variables, etc. Computability theory texts also often appeal to non-constant games to illustrate 
certain complexity-theory concepts such as alternating computation or PSPACE-completeness. The {\em Formula Game} or 
{\em Generalized Geography} \mbox{(\cite{Sip06},} Section 8.3) are typical examples. Both can be understood as games that depend on a variable $x$, with $x$ ranging over quantified Boolean formulas in Formula Game and over directed graphs in Generalized Geography. 

A game $A$ is said to be {\bf unistructural in} a variable $x$ --- or simply {\bf $x$-unistructural} --- iff, for every two valuations $e_1$ and $e_2$ that agree on all variables except $x$, we have $\legal{A}{e_1}=\legal{A}{e_2}$. And $A$ is (simply) {\bf unistructural}\index{unistructural game}\label{0unistructural game} iff 
$\legal{A}{e_1}=\legal{A}{e_2}$ for any two valuations $e_1$ and $e_2$. A unistructural game is thus a game whose every instance has the same structure (the $\legal{}{}$ component). And $A$ is unistructural in $x$ iff the structure 
of any instance $e[A]$ of $A$ does not depend on how $e$ evaluates the variable $x$. Of course, every constant or elementary game is unistructural, and every unistructural game is unistructural in all variables.   While natural examples of non-unistructural games exist such as the games mentioned in the above paragraph, all examples of particular games discussed elsewhere in the present paper are unistructural. In fact, every non-unistructural game can be 
rather easily rewritten into an equivalent (in a certain reasonable sense) unistructural game. One of the standard ways to 
convert a non-unistructural game $A$ into a corresponding unistructural game $A'$ is to take the  
union  (or anything bigger) $U$ of the structures of all instances of $A$ to be the common-for-all-instances structure of $A'$, and then extend the (relevant part of the) $\win{}{}$ function of each instance $e[A]$ of $A$ to $U$ by stipulating that, if $\Gamma\in(U- \legal{A}{e})$, then the player who made the first illegal (in the sense of $e[A]$) move is the loser in $e[A']$. So, say, in the unistructural version of generalized checkers, an attempt by a player to move to a non-existing cell would result in a loss for that player but otherwise considered a legal move. 
The class of naturally emerging unistructural games is very wide. All elementary games are trivially there, and Theorem 14.1 of \cite{Jap03} establishes that all of the game operations studied in CL preserve the unistructural property of games. In view of these remarks, if the reader feels more comfortable this way, without much loss of generality (s)he can always understand ``game" as ``unistructural game". 

What makes unistructural games nice is that, even when non-constant, they can still be visualized in the style of Figures 2 and 3. The difference will be that whereas the nodes of a game tree of a constant game are always labeled by propositions ($\twg$ or $\tlg$), now such labels can be any predicates. The constant game of Figure 3 was about the problem of computing $n+1$. We can generalize it to the problem of computing $n+z$, where 
$z$ is a (the only) variable on which the game depends. The corresponding non-constant game then can be drawn by modifying the labels of the bottom-level nodes of Figure 3 as follows:

  \begin{center}
\begin{picture}(311,160)

\put(175,142){\circle{16}}
\put(171,138){$\pp$}
\put(175,134){\line(-3,-1){106}}
\put(87,112){\scriptsize $\oo 1$}
\put(175,134){\line(2,-1){69}}
\put(193,112){\scriptsize $\oo 2$}
\put(175,134){\line(6,-1){110}}
\put(285,112){\Huge ...}

\put(65,92){\circle{16}}
\put(61,88){$\oo$}
\put(18,62){\scriptsize $\pp 1$}
\put(65,84){\line(0,-3){34}}
\put(65,84){\line(-3,-2){50}}
\put(52,62){\scriptsize $\pp 2$}
\put(78,62){\scriptsize $\pp 3$}
\put(65,84){\line(3,-2){50}}
\put(115,62){\large ...}
\put(65,84){\line(5,-2){50}}

\put(15,42){\oval(42,16)}
\put(-1,40){\scriptsize $1+z=1$}
\put(65,42){\oval(42,16)}
\put(49,40){\scriptsize $1+z=2$}
\put(115,42){\oval(42,16)}
\put(99,40){\scriptsize $1+z=3$}

\put(247,92){\circle{16}}
\put(243,88){$\oo$}
\put(200,62){\scriptsize $\pp 1$}
\put(247,84){\line(0,-3){34}}
\put(247,84){\line(-3,-2){50}}
\put(234,62){\scriptsize $\pp 2$}
\put(260,62){\scriptsize $\pp 3$}
\put(247,84){\line(3,-2){50}}
\put(297,62){\large ...}
\put(247,84){\line(5,-2){50}}

\put(197,42){\oval(42,16)}
\put(181,40){\scriptsize $2+z=1$}
\put(247,42){\oval(42,16)}
\put(231,40){\scriptsize $2+z=2$}
\put(297,42){\oval(42,16)}
\put(281,40){\scriptsize $2+z=3$}

\put(55,10){{\bf Figure 5:} The problem of computing $n+z$}
\end{picture}
\end{center}

Denoting the above game by $A(z)$, the game of Figure 3 becomes the instance $A(1)$ of it. The latter results from replacing $z$ by $1$ in the tree of Figure 5. This replacement turns every label 
$n+z=m$ into the constant game/proposition $n+1=m$, i.e. --- depending on its truth value --- into $\twg$ or $\tlg$. 

Let $A$ be an arbitrary game. We say that $\Gamma$ is a {\bf unilegal run}\index{unilegal run}\label{0unilegal run} (position if finite) of $A$ iff, for every valuation $e$, $\Gamma$ is a legal run of $e[A]$. The set of all unilegal runs of $A$  is denoted by $\Legal{A}$.\index{$\Legal{}$}\label{0lll} Of course, for unistructural games, ``legal'' and ``unilegal'' mean the same.   
The operation of prefixation defined in Section \ref{ss2} only for constant games naturally extends to all games. For $\seq{\Phi}A$ to be defined, $\Phi$ should be a unilegal position of $A$. Once this condition is satisfied, we define $\seq{\Phi}A$ as the unique game such that, for every valuation $e$, $e[\seq{\Phi}A]=\seq{\Phi}e[A]$. For example, where $A(z)$ is the game of 
Figure 5, $\seq{\oo 1}A(z)$ is the subtree rooted at the first (leftmost) child of the root, and $\seq{\oo 1,\pp 2}A(z)$ is the subtree rooted at the second grandchild from the first child, 
i.e. simply the predicate $1+z=2$.

Computability logic can be seen as an approach that generalizes both the traditional theory of computation and traditional logic, and unifies them on the basis of one general formal framework. 
The main objects of study of the traditional theory of computation are traditional computational problems, and the main objects of study of traditional logic are predicates. 
Both of these sorts of objects turn out to be special cases of our games. So, one can characterize classical logic as the elementary --- non-interactive --- fragment of computability logic. And 
characterize (the core of) the traditional theory of computation as
the fragment of computability logic where interaction is limited to its simplest, two-step --- input/output, or question/answer --- form.  The basic entities on which such a unifying framework needs to focus 
are thus games, and nothing but games.\vspace{20pt}  

\section{Game operations}\label{ss4}

As we already know, logical operators in CL stand for operations on games. There is an open-ended pool of operations of potential interest, and which of those to study may depend on particular needs and taste. Yet, there is a core collection of the most basic and natural game operations, to the definitions of which the present section is devoted: 
the {\bf propositional connectives}\footnote{The term ``propositional'' is not very adequate here, and we use it only by inertia from classical logic. Propositions are very special --- elementary and constant --- cases of games. On the other hand, our ``propositional'' operations are applicable to all games, and not all of them preserve the elementary property of their arguments, even though they do preserve the constant property.}
$\gneg$, $\mlc$, $\mld$, $\mli$, $\adc$, $\add$, $\pst$, $\pcost$, $\pintimpl$, $\st$, $\cost$, $\intimpl$ and the {\bf quantifiers}\index{quantifier}\label{0quantifier} $\ada$, $\ade$, $\mla$, $\mle$, $\cla$, $\cle$. Among these we see all operators of classical logic, and our choice of the classical notation for them is no accident. It was pointed out earlier that classical logic is nothing but the elementary, zero-interactivity fragment of computability logic. Indeed, after analyzing the relevant definitions, each of the classically-shaped operations, {\em when restricted to elementary games}, can be easily seen to be virtually the same as
the corresponding operator of classical logic. For instance, if $A$ and $B$ are elementary games, then so is $A\mlc B$, and the latter is exactly the classical conjunction of $A$ and $B$ understood as an (elementary) game. In a general --- not-necessarily-elementary --- case, however, $\gneg,\mlc,\mld,\mli$ 
become more reminiscent of (yet not the same as) the corresponding multiplicative operators of linear logic. Of course, here we are essentially comparing apples with oranges for, as noted earlier, linear logic is a syntax while computability logic is a semantics, and it may be not clear in what precise sense one can talk about similarities or differences. 
In the same apples and oranges style, our operations $\adc,\add,\ada,\ade$ can be perceived  
as relatives of the additive connectives and quantifiers of linear logic, $\mla,\mle$ as ``multiplicative quantifiers'', and $\pst,\pcost,\st,\cost$ as ``exponentials'', even though it is hard to guess which of the two groups --- $\pst,\pcost$ or $\st,\cost$ --- would be closer to an orthodox linear logician's heart.  The quantifiers $\cla,\cle$, on the other hand, hardly have any reasonable linear-logic counterparts.

Let us agree that in every definition of this section $x$ stands for an arbitrary variable, $A,B,A(x),A_1,A_2,\ldots$ for arbitrary games, $e$ for an arbitrary valuation, and $\Gamma$ for an arbitrary run. 
Note that it is sufficient to define the content ($\win{}{}$ component) of a given constant game only for its legal runs, for then it uniquely extends to all runs. 
Furthermore, as usually done in logic textbooks and as we already did with the operation of prefixation, propositional connectives can be initially defined just as operations on constant games; then they automatically extend to all games by stipulating that $e[\ldots]$ simply commutes with all of those operations. That is, $\gneg A$ is the unique game such that, for every $e$, $e[\gneg A]=\gneg e[A]$; $e[A_1\mlc A_2]$ is the unique game such that, for every $e$, $e[A_1\mlc A_2]=e[A_1]\mlc e[A_2]$, etc. 
With this remark in mind, in each of our definitions of propositional connectives that follow in this section, games $A,B,A_1,A_2,\ldots$  are implicitly assumed to be constant. Alternatively, this assumption can be dropped; all one needs to change in the corresponding definitions in this case is  to write $\legal{A}{e}$ and $\win{A}{e}$ instead of simply $\legal{A}{}$ and $\win{A}{}$. 

For similar reasons, it would be sufficient to define ${\cal Q}xA$ (where $\cal Q$ is a quantifier) just for $1$-ary games $A$ that only depend on $x$. Since we are lazy to explain how, exactly, ${\cal Q}x$ would then extend to all games, our definitions of quantifiers given in this section, unlike those of propositional connectives,  
neither explicitly nor implicitly do assume any conditions on the arity of $A$.\vspace{20pt}

\subsection{Negation}
Negation\index{negation}\label{0negation} $\gneg$\index{$\gneg$ (as an operation on games)}\label{0gneg (as an operation on games)} is the role-switch operation: it turns  $\pp$'s wins and legal moves into $\oo$'s wins and legal moves, and vice versa. For instance, if $\chess$ is the game of chess from the point of view of the white player, then $\gneg\chess$ is the same game as seen by the black player. Figure 6 illustrates how applying $\gneg$ to a game $A$ generates the exact ``negative image'' of $A$, with $\pp$ and $\oo$ interchanged both in the nodes and the arcs of the game tree.

  \begin{center}
\begin{picture}(311,178)

\put(49,160){$A$}
\put(53,142){\circle{16}}
\put(49,138){$\pp$}
\put(53,134){\line(-1,-1){34}}
\put(20,113){\scriptsize $\oo 1$}
\put(53,134){\line(1,-1){34}}
\put(76,113){\scriptsize $\oo 2$}

\put(17,92){\circle{16}}
\put(13,88){$\oo$}
\put(-2,62){\scriptsize $\pp 1$}
\put(17,84){\line(-1,-3){11}}
\put(17,84){\line(1,-3){11}}
\put(25,62){\scriptsize $\pp 2$}
\put(4,42){\circle{16}}
\put(0,38){$\pp$}
\put(28,42){\circle{16}}
\put(24,38){$\oo$}

\put(88,92){\circle{16}}
\put(84,88){$\oo$}
\put(69,62){\scriptsize $\pp 1$}
\put(88,84){\line(-1,-3){11}}
\put(88,84){\line(1,-3){11}}
\put(96,62){\scriptsize $\pp 2$}
\put(75,42){\circle{16}}
\put(71,38){$\oo$}
\put(99,42){\circle{16}}
\put(95,38){$\pp$}

\put(244,160){$\gneg A$}
\put(253,142){\circle{16}}
\put(249,138){$\oo$}
\put(253,134){\line(-1,-1){34}}
\put(220,113){\scriptsize $\pp 1$}
\put(253,134){\line(1,-1){34}}
\put(276,113){\scriptsize $\pp 2$}

\put(217,92){\circle{16}}
\put(213,88){$\pp$}
\put(198,62){\scriptsize $\oo 1$}
\put(217,84){\line(-1,-3){11}}
\put(217,84){\line(1,-3){11}}
\put(225,62){\scriptsize $\oo 2$}
\put(204,42){\circle{16}}
\put(200,38){$\oo$}
\put(228,42){\circle{16}}
\put(224,38){$\pp$}

\put(288,92){\circle{16}}
\put(284,88){$\pp$}
\put(269,62){\scriptsize $\oo 1$}
\put(288,84){\line(-1,-3){11}}
\put(288,84){\line(1,-3){11}}
\put(296,62){\scriptsize $\oo 2$}
\put(275,42){\circle{16}}
\put(271,38){$\pp$}
\put(299,42){\circle{16}}
\put(295,38){$\oo$}

\put(99,10){{\bf Figure 6:} Negation}
\end{picture}
\end{center}

Notice the three different meanings that we associate with symbol $\gneg$. In Section \ref{ss2} we agreed to use $\gneg$ as an operation on players (turning $\pp$ into $\oo$ and vice versa), and an operation on runs (interchanging $\pp$ with $\oo$ in every labmove). Below comes our formal definition of the third meaning of $\gneg$ as an operation on games:
  
\begin{definition}\label{neg} {\bf Negation} $\gneg A$:
\begin{itemize}
\item $\Gamma\in\legal{\gneg A}{}$ iff $\rneg{\Gamma}\in\legal{A}{}$.
\item $\win{\gneg A}{}\seq{\Gamma}=\pp$ iff $\win{A}{}\seq{\rneg\Gamma}=\oo$.
\end{itemize}
\end{definition}

Even from the informal explanation of $\gneg$ it is clear that $\gneg\gneg A$ is always the same as $A$, for interchanging in $A$ the payers' roles twice brings the players to their original roles. It would also be easy to show that we always have $\gneg(\seq{\Phi}A)=\seq{\rneg \Phi}\gneg A$. So, say, if $\alpha$ is $\pp$'s legal move in the empty position of $A$ that brings $A$ down to $B$, then the same $\alpha$ is $\oo$'s legal move in the empty position of $\gneg A$, and it brings $\gneg A$ down to $\gneg B$. Test the game $A$ of Figure 6 to see that this is indeed so.

\subsection{Choice operations}
$\adc,\add,\ada$ and $\ade$ are called {\bf choice operations}.\index{choice operations}\label{0choice operations} $A_1\adc A_2$ is the game where, in the initial position, $\oo$ has two legal moves (choices): $1$ and $2$. Once such a choice $i$ is made, the game continues as the chosen component $A_i$, meaning that $\seq{\oo i}(A_1\adc A_2)=A_i$; if a choice is never made, $\oo$ loses. $A_1\add A_2$ is similar/symmetric, with $\pp$ and $\oo$ interchanged; that is, in $A_1\add A_2$ it is $\pp$ who makes an initial choice and who loses if such a choice is never made. Figure 7 helps us visualize the way $\adc$ and $\add$ combine two games $A$ and $B$:

  \begin{center}
\begin{picture}(226,123)

\put(26,105){$A\adc B$}
\put(41,87){\circle{16}}
\put(37,83){$\pp$}
\put(41,79){\line(-1,-1){34}}
\put(7,58){\scriptsize $\oo 1$}
\put(41,79){\line(1,-1){34}}
\put(64,58){\scriptsize $\oo 2$}

\put(0,34){$A$}
\put(70,34){$B$}

\put(176,105){$A\add B$}
\put(191,87){\circle{16}}
\put(187,83){$\oo$}
\put(191,79){\line(-1,-1){34}}
\put(157,58){\scriptsize $\pp 1$}
\put(191,79){\line(1,-1){34}}
\put(214,58){\scriptsize $\pp 2$}

\put(150,34){$A$}
\put(220,34){$B$}

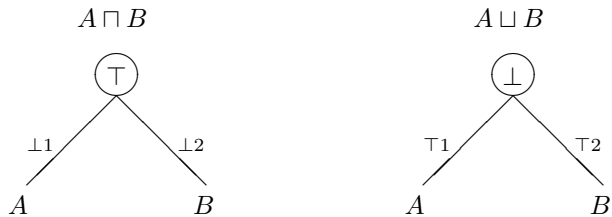
\put(10,10){{\bf Figure 7:} Choice propositional connectives}
\end{picture}
\end{center}

The game $A$ of Figure 6 can now be easily seen to be $(\twg\add\tlg)\adc(\tlg\add\twg)$, and its negation be $(\tlg\adc\twg)\add(\twg\adc\tlg)$. The symmetry/duality familiar from classical logic persists: we always have $\gneg(A\adc B)=\gneg A\add\gneg B$ and $\gneg(A\add B)=\gneg A\adc \gneg B$. Similarly for the quantifier counterparts $\ada$ and $\ade$ of $\adc$ and $\add$. 
We might have already guessed that $\ada xA(x)$ is nothing but the infinite $\adc$-conjunction $A(1)\adc A(2)\adc A(3)\adc\ldots$ \ and \ $\ade xA(x)$ \ is \ $A(1)\add A(2)\add A(3)\add\ldots$, \  as can be seen from Figure 8.
 
 \begin{center}
\begin{picture}(226,125)

\put(28,107){$\ada xA(x)$}
\put(46,89){\circle{16}}
\put(42,85){$\pp$}
\put(46,81){\line(-1,-1){34}}
\put(11,58){\scriptsize $\oo 1$}
\put(46,81){\line(1,-1){34}}
\put(54,58){\scriptsize $\oo 3$}
\put(46,81){\line(0,-1){33}}
\put(34,58){\scriptsize $\oo 2$}
\put(46,81){\line(3,-2){30}}
\put(76,58){\bf \ldots}

\put(0,34){$A(1)$}
\put(35,34){$A(2)$}
\put(70,34){$A(3)$}

\put(178,107){$\ade xA(x)$}
\put(196,89){\circle{16}}
\put(192,85){$\oo$}
\put(196,81){\line(-1,-1){34}}
\put(161,58){\scriptsize $\pp 1$}
\put(196,81){\line(1,-1){34}}
\put(204,58){\scriptsize $\pp 3$}
\put(196,81){\line(0,-1){33}}
\put(184,58){\scriptsize $\pp 2$}
\put(196,81){\line(3,-2){30}}
\put(226,58){\bf \ldots}

\put(150,34){$A(1)$}
\put(185,34){$A(2)$}
\put(220,34){$A(3)$}

\put(52,10){{\bf Figure 8:} Choice quantifiers}
\end{picture}
\end{center}

So, we always have $\seq{\oo c}\ada xA(x)=A(c)$ and $\seq{\pp c}\ade xA(x)=A(c)$. The meaning of such a labmove $\xx c$
can be characterized as that player $\xx$ selects/specifies the particular value $c$ for $x$, after which the game continues --- and the winner is determined --- according to the rules of $A(c)$.  

Now we are already able to express traditional computational problems using formulas. Traditional problems come in two forms: the problem of computing a function $f(x)$, or the problem of deciding a predicate $p(x)$. The former can be captured by $\ada x\ade y(f(x)=y)$, and the latter (which, of course, can be seen as a special case of the former) by $\ada x \bigl(p(x)\add\gneg p(x)\bigr)$. So, the game of Figure 3 will be written as $\ada x\ade y(x+1=y)$, and the game of \mbox{Figure\hspace{-1pt} 5} as $\ada x\ade y(x+z=y)$. 

The following Definition \ref{ch} summarizes the above-said, and generalizes $\adc,\add$ from binary to any $\geq 2$-ary operations. Note the perfect symmetry in it: the definition of each choice operation can be obtained from that of its dual by just interchanging $\pp$  with $\oo$.

\begin{definition}\label{ch} \ In clauses 1 and 2, $n$ is $2$ or any greater integer.\vspace{10pt} 

\noindent 1. {\bf Choice conjunction}\index{choice conjunction}\label{0choice conjunction}\index{$\adc$}\label{0adc} $A_1\adc\ldots\adc A_n$:
\begin{itemize}
\item $\legal{A_1\adc\ldots\adc A_n}{}=\{\emptyrun\}\cup\{\seq{\oo i,\Gamma} \ |\ i\in\{1,\ldots,n\}, \ \Gamma\in\legal{A_i}{}\}$.
\item $\win{A_1\adc\ldots\adc A_n}{}\emptyrun=\pp$; \\
where $i\in\{1,\ldots,n\}$, $\win{A_1\adc\ldots\adc A_n}{}\seq{\oo i,\Gamma}=\win{A_i}{}\seq{\Gamma}$.\vspace{7pt}  
\end{itemize}

\noindent 2. {\bf Choice disjunction}\index{choice disjunction}\label{0choice disjunction}\index{$\add$}\label{0add} $A_1\add\ldots \add A_n$:
\begin{itemize}
\item $\legal{A_1\add\ldots\add A_n}{}=\{\emptyrun\}\cup\{\seq{\pp i,\Gamma} \ |\ i\in\{1,\ldots,n\}, \ \Gamma\in\legal{A_i}{}\}$.
\item $\win{A_1\add \ldots\add A_n}{}\emptyrun=\oo$; \\
where $i\in\{1,\ldots,n\}$, $\win{A_1\add \ldots\add A_n}{}\seq{\pp i,\Gamma}=\win{A_i}{}\seq{\Gamma}$.\vspace{7pt}  
\end{itemize}

\noindent 3. {\bf Choice universal quantification}\index{choice universal quantification}\label{0choice universal quantification}\index{$\ada$}\label{0ada} $\ada xA(x)$:
\begin{itemize}
\item $\legal{\adai xA(x)}{e}=\{\emptyrun\}\cup\{\seq{\oo c,\Gamma} \ |\ c\in\{1,2,3,\ldots\}, \ \Gamma\in\legal{A(c)}{e}\}$.
\item $\win{\adai xA(x)}{e}\emptyrun=\pp$; \\
where $c\in\{1,2,3,\ldots\}$, $\win{\adai xA(x)}{e}\seq{\oo c,\Gamma}=\win{A(c)}{e}\seq{\Gamma}$.\vspace{7pt}  
\end{itemize}

\noindent 4. {\bf Choice existential quantification}\index{choice existential quantification}\label{0choice existential quantification}\index{$\ade$}\label{0ade} $\ade xA(x)$:
\begin{itemize}
\item $\legal{\adei xA(x)}{e}=\{\emptyrun\}\cup\{\seq{\pp c,\Gamma} \ |\ c\in\{1,2,3,\ldots\}, \ \Gamma\in\legal{A(c)}{e}\}$.
\item $\win{\adei xA(x)}{e}\emptyrun=\oo$; \\ 
where $c\in\{1,2,3,\ldots\}$, $\win{\adei xA(x)}{e}\seq{\pp c,\Gamma}=\win{A(c)}{e}\seq{\Gamma}$.\vspace{10pt}
\end{itemize}

\end{definition}

\subsection{Parallel operations}\label{ss4.3}
The operations $\mlc,\mld,\mla,\mle$ combine games in a way that corresponds to our intuition of parallel computations. For this reason we call such operations {\bf parallel}.\index{parallel operations}\label{0parallel operations} 
Playing $A_1\mlc A_2$ (resp. $A_1\mld A_2$) means playing the two games simultaneously where, in order to win, $\pp$ needs to win in both (resp. at least one) of the components $A_i$. Back to our chess example, the two-board game $\gneg\chess\mld\chess$ can be easily won by just mimicking in $\chess$ the moves that the adversary makes in $\gneg \chess$, and vice versa. This is very different from  the situation with $\gneg\chess \add \chess$, winning which is not easy at all: there $\pp$ needs to choose between $\gneg\chess$ and $\chess$ (i.e. between playing black or white), and then win the chosen one-board game. Technically, a move $\alpha$ in the $k$th $\mlc$-conjunct or $\mld$-disjunct is made by prefixing $\alpha$ with `$k.$'. For instance, in (the initial position of) $(A\add B)\mld (C\adc D)$, the move `$2.1$' is legal for $\oo$, meaning choosing the first $\adc$-conjunct in the 
second $\mld$-disjunct of the game. If such a move is made, the game will continue as $(A\add B)\mld C$. The player $\pp$, too, has initial legal moves in $(A\add B)\mld (C\adc D)$, which are `$1.1$' and `$1.2$'. 
As we may guess, $\mla xA(x)$ is nothing but $A(1)\mlc A(2)\mlc A(3)\mlc\ldots$, and $\mle xA(x)$ is nothing but $A(1)\mld A(2)\mld A(3)\mld\ldots$. 

The following formal definition summarizes this meaning of parallel operations, generalizing the arity of $\mlc,\mld$ to any $n\geq 2$. In that definition and throughout the rest of this paper, we use the important notational convention according to which, for a string/move $\alpha$, 
\[\Gamma^\alpha\index{$\Gamma^\alpha$}\label{apr2}\] means  the result of removing from $\Gamma$ all (lab)moves 
except those of the form $\xx\alpha\beta$, and then deleting the prefix\footnote{Here and later, when talking about a prefix of a labmove $\xx\gamma$, we do not count the label $\xx$ as a part of the prefix.} `$\alpha$' in the remaining moves, i.e. replacing each such $\xx \alpha\beta$ by $\xx\beta$. For example, where $\Gamma$ is the leftmost branch of the tree for  $(\twg\adc\tlg)\mld(\tlg\add\twg)$ shown in Figure 9, we have $\Gamma^{1.}=\seq{\oo 1}$ and $\Gamma^{2.}=\seq{\pp 1}$. Intuitively, we view this $\Gamma$ as consisting of two subruns, one ($\Gamma^{1.}$) being a run in the first $\mld$-disjunct of  $(\twg\adc\tlg)\mld(\tlg\add\twg)$, and the other ($\Gamma^{2.}$) being a run in the second disjunct.\vspace{5pt}

\begin{definition}\label{par} In clauses 1 and 2, $n$ is $2$ or any greater integer.\vspace{10pt} 

\noindent 1. {\bf Parallel conjunction}\index{parallel conjunction}\label{0parallel conjunction}\index{$\mlc$}\label{0mlc} $A_1\mlc\ldots\mlc A_n$:
\begin{itemize}
\item $\Gamma\in\legal{A_1\mlc\ldots\mlc A_n}{}$ iff every move of $\Gamma$ has the prefix `$i.$' for some $i\in\{1,\ldots,n\}$ and, for each such $i$, $\Gamma^{i.}\in\legal{A_i}{}$.
\item Whenever $\Gamma\in\legal{A_1\mlc\ldots\mlc A_n}{}$, $\win{A_1\mlc\ldots\mlc A_n}{}\seq{\Gamma}=\pp$ iff, for each $i\in\{1,\ldots,n\}$, $\win{A_i}{}\seq{\Gamma^{i.}}=\pp$.\vspace{7pt}  
\end{itemize}

\noindent 2. {\bf Parallel disjunction}\index{parallel disjunction}\label{0parallel disjunction}\index{$\mld$}\label{0mld} $A_1\mld\ldots\mld A_n$:
\begin{itemize}
\item $\Gamma\in\legal{A_1\mld\ldots\mld A_n}{}$ iff every move of $\Gamma$ has the prefix `$i.$' for some $i\in\{1,\ldots,n\}$ and, for each such $i$, $\Gamma^{i.}\in\legal{A_i}{}$.
\item Whenever $\Gamma\in\legal{A_1\mld\ldots\mld A_n}{}$, $\win{A_1\mld\ldots\mld A_n}{}\seq{\Gamma}=\oo$ iff, for each $i\in\{1,\ldots,n\}$, $\win{A_i}{}\seq{\Gamma^{i.}}=\oo$.\vspace{7pt}  
\end{itemize}

\noindent 3. {\bf Parallel universal quantification}\index{parallel universal quantification}\label{0parallel universal quantification}\index{$\mla$}\label{0mla} $\mla xA(x)$:
\begin{itemize}
\item $\Gamma\in\legal{\mlai xA(x)}{e}$ iff every move of $\Gamma$ has the prefix `$c.$' for some $c\in\{1,2,3,\ldots\}$ and, for each such $c$, $\Gamma^{c.}\in\legal{A(c)}{e}$.
\item Whenever $\Gamma\in\legal{\mlai xA(x)}{e}$, $\win{\mlai xA(x)}{e}\seq{\Gamma}=\pp$ iff, for each $c\in\{1,2,3,\ldots\}$, $\win{A(c)}{e}\seq{\Gamma^{c.}}=\pp$.\vspace{7pt}  
\end{itemize}

\noindent 4. {\bf Parallel existential quantification}\index{parallel existential quantification}\label{0parallel existential quantification}\index{$\mle$}\label{0mle} $\mle xA(x)$:
\begin{itemize}
\item $\Gamma\in\legal{\mlei xA(x)}{e}$ iff every move of $\Gamma$ has the prefix `$c.$' for some $c\in\{1,2,3,\ldots\}$ and, for each such $c$, $\Gamma^{c.}\in\legal{A(c)}{e}$.
\item Whenever $\Gamma\in\legal{\mlei xA(x)}{e}$, $\win{\mlei xA(x)}{e}\seq{\Gamma}=\oo$ iff, for each $c\in\{1,2,3,\ldots\}$, $\win{A(c)}{e}\seq{\Gamma^{c.}}=\oo$.
\end{itemize}

\end{definition}

As was the case with choice operations, we can see that the definition of each of the parallel operations can be obtained from the definition of its dual by just interchanging $\pp$ with $\oo$. 
Hence it is easy to verify that we always have $\gneg(A\mlc B)=\gneg A\mld\gneg B$, $\gneg(A\mld B)=\gneg A\mlc\gneg B$, $\gneg\mla xA(x)=\mle x\gneg A(x)$, $\gneg\mle xA(x)=\mla x\gneg A(x)$.

Note also that just like negation (and unlike choice operations), parallel operations preserve the elementary property of games and, when restricted to elementary games, the meanings of $\mlc$ and $\mld$ coincide with those of classical conjunction and disjunction, while the meanings of $\mla$ and $\mle$ coincide with those of classical universal quantifier and existential quantifier. The same 
conservation of classical meaning is going to be the case with the blind quantifiers $\cla,\cle$ defined later; so, at the elementary level, $\mla$ and $\mle$ are indistinguishable from $\cla$ and $\cle$. 

A strict definition of our understanding of validity --- which, as we may guess, conserves the classical meaning of this concept in the context of elementary games --- will  be given later in Section \ref{ss7}. For now, let us adopt an intuitive explanation according to which validity means 
being ``always winnable by a machine''. 
While all classical tautologies automatically remain valid when parallel operators are applied to elementary games, in the general case the class of valid principles shrinks. For example,  $\gneg P\mld (P\mlc P)$ is not valid. Proving this might require some thought, but at least we can see that the earlier ``mimicking'' (``copy-cat'') strategy successful for 
$\gneg \chess\mld \chess$ would be inapplicable to \(\gneg \mbox{\em Chess}\mld (\mbox{\em Chess}\mlc \mbox{\em Chess})\). The best that $\pp$ can do in this three-board game is to pair $\gneg \mbox{\em Chess}$ with one of the two conjuncts of $\mbox{\em Chess}\mlc \mbox{\em Chess}$. It is possible that then $\gneg \mbox{\em Chess}$ and the unmatched {\em Chess} are both lost, in which case the whole game will be lost.  
  
When $A$ and $B$ are finite (or finite-depth) games, the depth of $A\mlc B$ or $A\mld B$ is the sum of the depths of $A$ and $B$, which signifies an exponential growth of the breadth. Figure 9 illustrates this growth, suggesting that once we have reached the level of parallel operations --- let alone recurrence operations that will be defined shortly --- continuing drawing trees in the earlier style becomes no fun. Not to be disappointed though: making it possible to express large- or infinite-size game trees in a compact way is what our game operators are all about after all. 

  \begin{center}
\begin{picture}(311,178)

\put(-1,160){$\twg\adc\tlg$}
\put(14,142){\circle{16}}
\put(10,138){$\pp$}
\put(14,134){\line(-1,-4){8}}
\put(-3,113){\scriptsize $\oo 1$}
\put(14,134){\line(1,-4){8}}
\put(20,113){\scriptsize $\oo 2$}

\put(4,92){\circle{16}}
\put(0,88){$\pp$}

\put(24,92){\circle{16}}
\put(20,88){$\oo$}

\put(49,160){$\tlg\add\twg$}
\put(64,142){\circle{16}}
\put(60,138){$\oo$}
\put(64,134){\line(-1,-4){8}}
\put(47,113){\scriptsize $\pp 1$}
\put(64,134){\line(1,-4){8}}
\put(70,113){\scriptsize $\pp 2$}

\put(54,92){\circle{16}}
\put(50,88){$\oo$}

\put(74,92){\circle{16}}
\put(70,88){$\pp$}

\put(162,160){$(\twg\adc\tlg)\mld(\tlg\add\twg)$}
\put(206,142){\circle{16}}
\put(202,138){$\pp$}
\put(206,134){\line(-3,-4){26}}
\put(172,113){\scriptsize $\oo 1.2$}
\put(206,134){\line(3,-4){26}}
\put(132,113){\scriptsize $\oo 1.1$}
\put(206,134){\line(-5,-2){85}}
\put(206,134){\line(5,-2){85}}
\put(221,113){\scriptsize $\pp 2.1$}
\put(260,113){\scriptsize $\pp 2.2$}

\put(119,92){\circle{16}}
\put(115,88){$\pp$}
\put(94,62){\scriptsize $\pp 2.1$}
\put(118,84){\line(-1,-4){8}}
\put(118,84){\line(1,-4){8}}
\put(124,62){\scriptsize $\pp 2.2$}
\put(108,42){\circle{16}}
\put(104,38){$\pp$}
\put(128,42){\circle{16}}
\put(124,38){$\pp$}

\put(177,92){\circle{16}}
\put(173,88){$\oo$}
\put(153,62){\scriptsize $\pp 2.1$}
\put(177,84){\line(-1,-4){8}}
\put(177,84){\line(1,-4){8}}
\put(183,62){\scriptsize $\pp 2.2$}
\put(167,42){\circle{16}}
\put(163,38){$\oo$}
\put(187,42){\circle{16}}
\put(183,38){$\pp$}

\put(235,92){\circle{16}}
\put(231,88){$\pp$}
\put(210,62){\scriptsize $\oo 1.1$}
\put(235,84){\line(-1,-4){8}}
\put(235,84){\line(1,-4){8}}
\put(243,62){\scriptsize $\oo 1.2$}
\put(225,42){\circle{16}}
\put(221,38){$\pp$}
\put(245,42){\circle{16}}
\put(241,38){$\oo$}

\put(293,92){\circle{16}}
\put(289,88){$\pp$}
\put(268,62){\scriptsize $\oo 1.1$}
\put(293,84){\line(-1,-4){8}}
\put(293,84){\line(1,-4){8}}
\put(301,62){\scriptsize $\oo 1.2$}
\put(283,42){\circle{16}}
\put(279,38){$\pp$}
\put(303,42){\circle{16}}
\put(299,38){$\pp$}

\put(79,10){{\bf Figure 9:} Parallel disjunction}
\end{picture}
\end{center}

An alternative approach to graphically representing $A\mld B$ (or $A\mlc B$) would be to just draw two trees --- one for $A$ and one for $B$ --- next to each other rather than draw  
one tree for $A\mld B$. The legal positions of $A\mld B$ can then be visualized as pairs $(\Phi,\Psi)$, where $\Phi$ is a node of the $A$-tree and $\Psi$ a node of the $B$-tree; the ``label'' of 
each such position $(\Phi,\Psi)$ will be $\pp$ iff the label of at least one (or both if we are dealing with $A\mlc B$) of the positions/nodes $\Phi,\Psi$ in the corresponding tree is $\pp$. For instance, the root of 
the $(\twg\adc\tlg)\mld(\tlg\add\twg)$-tree of Figure 9 can just be thought of as the pair consisting of the roots of the $(\twg\adc\tlg)$- and $(\tlg\add\twg)$-trees; child \#1 of the root of the 
 $(\twg\adc\tlg)\mld(\tlg\add\twg)$-tree as the pair whose first node is the left child of the root of the $(\twg\adc\tlg)$-tree and the second node is the root of the 
$(\tlg\add\twg)$-tree, etc. It is true that, under this approach, a pair $(\Phi,\Psi)$ might correspond to more than one position of $A\mld B$. For example, grandchildren \#1 and \#5 of the root of 
the $(\twg\adc\tlg)\mld(\tlg\add\twg)$-tree, i.e. the positions $\seq{\oo 1.1,\pp 2.1}$ and $\seq{\pp 2.1,\oo 1.1}$, would become indistinguishable. This, however, is OK, because 
such two positions would always be equivalent, in the sense that \[\seq{\oo 1.1,\pp 2.1}((\twg\adc\tlg)\mld(\tlg\add\twg))=\seq{\pp 2.1,\oo 1.1}((\twg\adc\tlg)\mld(\tlg\add\twg)).\]

Whether trees are or are not helpful in visualizing parallel combinations of games, prefixation is still very much so if we think of each (uni)legal position $\Phi$ of $A$ as the game 
$\seq{\Phi}A$. This way, every (uni)legal run $\Gamma$ of $A$ becomes a sequence of games.

\begin{example}\label{exp1} {\em To the legal run $\Gamma=\seq{\oo 2.7,\pp 1.7,\oo 1.49,\pp 2.49}$  of 
game $A=\ade x\ada y(y\not=x^2)\mld \ada x\ade y(y=x^2)$ corresponds the following sequence, showing how things evolve as $\Gamma$ runs, i.e. how the moves of $\Gamma$ affect/modify the game that is being  played:\vspace{8pt}

\noindent\hspace{-4pt}$\begin{array}{lll}
A_0\hspace{-3pt}: & \ade x\ada y(y\not=x^2)\mld \ada x\ade y(y=x^2),\hspace{-5pt} &  \mbox{i.e. $A$,}\\
 & &  \mbox{i.e. $\emptyrun A$;}\vspace{4pt}\\
A_1\hspace{-3pt}: & \ade x\ada y(y\not=x^2)\mld \ade y(y=7^2), &  \mbox{i.e. $\seq{\oo 2.7}A_0$,}\\
 & &  \mbox{i.e. $\seq{\oo 2.7}A$;}\vspace{4pt}\\
A_2\hspace{-3pt}: & \ada y(y\not=7^2)\mld \ade y(y=7^2), &  \mbox{i.e. $\seq{\pp 1.7}A_1$,}\\
 & &  \mbox{i.e. $\seq{\oo 2.7,\pp 1.7}A$;}\vspace{4pt}\\
A_3\hspace{-3pt}: & 49\not=7^2\mld \ade y(y=7^2), &  \mbox{i.e. $\seq{\oo 1.49}A_2$,}\\
 & &  \mbox{i.e. $\seq{\oo 2.7,\pp 1.7,\oo 1.49}A$;}\vspace{4pt}\\
A_4\hspace{-3pt}: & 49\not=7^2\mld 49=7^2, &  \mbox{i.e. $\seq{\pp 2.49}A_3$,}\\
 & &  \mbox{i.e. $\seq{\oo 2.7,\pp 1.7,\oo 1.49,\pp 2.49}A$.}\vspace{8pt}\\
\end{array}$

The run hits the true proposition $A_4$, and hence is won by $\pp$.} 
\end{example}

When visualizing $\mla,\mle$-games in a similar style, we are better off representing them as infinite conjunctions/disjunctions. Of course, putting infinitely many conjuncts/disjuncts on paper would be no fun. But, luckily,
in every position of $\mla xA(x)$ or $\mle xA(x)$ only a finite number of conjuncts/disjuncts would be ``activated", i.e. have a non-$A(c)$ form, so that all of the other, uniform, conjuncts can be combined into blocks and represented, say, through an ellipsis, or through expressions such as $\mla m\hspace{-2pt}\leq\hspace{-2pt} x\hspace{-2pt}\leq\hspace{-2pt} n A(x)$ or $\mla x\hspace{-2pt}\geq\hspace{-2pt} m A(x)$. 
 
\begin{example}\label{exp2}
{\em Let $\mbox{\em Odd}(x)$ be the predicate ``$x$ is odd''. 

The $\pp$-won legal run $\seq{\pp 7.1}$ of $\mle x\bigl(\mbox{\em Odd}(x)\add \gneg \mbox{\em Odd}(x)\bigr)$
will be represented as follows:\vspace{5pt}

\noindent\hspace{-4pt}$\begin{array}{l}
\mle x\geq 1\bigl(\mbox{\em Odd}(x)\add \gneg \mbox{\em Odd}(x)\bigr); \\
\mle 1\hspace{-2pt}\leq\hspace{-2pt} x\hspace{-2pt}\leq \hspace{-2pt}6\bigl(\mbox{\em Odd}(x)\add \gneg \mbox{\em Odd}(x)\bigr)\mld \mbox{\em Odd}(7)\mld 
\mle x\hspace{-2pt}\geq\hspace{-2pt} 8\bigl(\mbox{\em Odd}(x)\add \gneg \mbox{\em Odd}(x)\bigr).\vspace{5pt}
\end{array}$

And the infinite legal run $\Gamma=\seq{\pp 1.1,\pp 2.2,\pp 3.1,\pp 4.2,\pp 5.1,\pp 6.2,\ldots}$ of    
$\mla x\bigl(\mbox{\em Odd}(x)\add \gneg \mbox{\em Odd}(x)\bigr)$ will be represented as follows:\vspace{5pt}

\noindent\hspace{-4pt}$\begin{array}{l}
\mla x\hspace{-2pt}\geq\hspace{-2pt} 1\bigl(\mbox{\em Odd}(x)\add \gneg \mbox{\em Odd}(x)\bigr); \\
\mbox{\em Odd}(1) \mlc\mla x\hspace{-2pt}\geq\hspace{-2pt} 2\bigl(\mbox{\em Odd}(x)\add \gneg \mbox{\em Odd}(x)\bigr); \\ 
\mbox{\em Odd}(1) \mlc \gneg\mbox{\em Odd}(2)\mlc \mla x\hspace{-2pt}\geq\hspace{-2pt} 3\bigl(\mbox{\em Odd}(x)\add \gneg \mbox{\em Odd}(x)\bigr);\\
 \mbox{\em Odd}(1) \mlc \gneg \mbox{\em Odd}(2)\mlc \mbox{\em Odd}(3)\mlc\mla x\hspace{-2pt}\geq\hspace{-2pt} 4\bigl(\mbox{\em Odd}(x)\add \gneg \mbox{\em Odd}(x)\bigr);\\
\mbox{...etc.}\vspace{5pt} 
\end{array}$

Note that $\Gamma$ is won by $\pp$ but every finite initial segment of it is lost.
}\end{example}

\subsection{Reduction}\label{ss4.4}
What we call {\bf reduction}\index{reduction (as a game operation)}\label{0reduction (as a game operation)} $\mli$\index{$\mli$}\label{0mli} is perhaps most interesting of all operations, yet we do not introduce $\mli$ as a primitive operation as it can be formally defined by
\[B\mli A\ =\ (\gneg B)\mld A.\]
From this definition we see that, when applied to elementary games, $\mli$ has its ordinary classical meaning, because so do $\gneg$ and $\mld$. 
  
Intuitively, $B\mli A$ is (indeed) the problem of {\em reducing} 
$A$ to $B$: solving $B\mli A$ means solving $A$ while having $B$ as a {\em computational resource}.\index{computational resource}\label{0cr2} Resources are symmetric to problems: what is a problem to solve for one player is a resource that the other player can use, and vice versa. Since 
$B$ is negated in  $\gneg B\mld A$ and negation means switching the roles, $B$ appears as a resource rather than problem for 
$\pp$ in $B\mli A$. For example, the game of Example \ref{exp1} can be written as $\ada x\ade y(y=x^2)\mli \ada x\ade y(y=x^2)$.  For $\pp$, $\ada x\ade y(y=x^2)$ is the problem of computing square,
which can be seen as a task (telling the square of any given number) performed by $\pp$ for $\oo$.  But 
in the antecedent it turns into a square-computing resource --- a task performed by $\oo$ for $\pp$. In the run $\Gamma$ of Example \ref{exp1}, $\pp$ took advantage of this fact, and solved problem  $\ada x\ade y(y=x^2)$ in the consequent using $\oo$'s solution to the same problem in the antecedent. That is, $\pp$ reduced $\ada x\ade y(y=x^2)$ to $\ada x\ade y(y=x^2)$. 

To get a better appreciation of $\mli$ as a problem reduction operation, let us look a less  trivial --- already ``classical" in CL --- example. Let $A(x,y)$ be the predicate 
``Turing machine (whose code is) $x$ accepts input $y$'', and $H(x,y)$ the predicate ``Turing machine $x$ halts on input $y$''.  Note that then 
$\ada x\ada y\bigl(A(x,y)\add\gneg A(x,y)\bigr)$ expresses the acceptance problem 
as a decision problem: in order to win, $\pp$ should be able to tell which of the disjuncts --- $A(x,y)$ or $\gneg A(x,y)$ --- is true for any particular values 
for $x$ and $y$ selected by the environment. Similarly, $\ada x\ada y\bigl(H(x,y)\add\gneg H(x,y)\bigr)$ expresses the halting problem as a decision problem. 
No machine can (always) win $\ada x\ada y\bigl(A(x,y)\add\gneg A(x,y)\bigr)$ because the acceptance problem, just as the halting problem, is known to be undecidable. Yet, the acceptance problem is algorithmically reducible to the halting problem. Into our terms, this fact translates as existence of a machine that always wins the game
\begin{equation}\label{apr8}
\ada x\ada y\bigl(H(x,y)\add\gneg H(x,y)\bigr)\ \mli\ \ada x\ada y\bigl(A(x,y)\add\gneg A(x,y)\bigr). 
\end{equation}

A successful strategy for such a machine ($\pp$) is as follows. At the beginning, $\pp$ waits  till 
$\oo$ specifies some values $m$ and $n$ for $x$ and $y$ in the consequent, i.e. makes the moves `$2.m$' and `$2.n$'. Such moves, bringing the consequent down to 
$A(m,n)\add\gneg A(m,n)$, 
 can be seen as asking $\pp$ the question ``does Turing machine $m$ accept input $n$?". To this question $\pp$ replies by the counterquestion ``does $m$ halt on $n$?", i.e. makes the moves `$1.m$ and `$1.n$', bringing the antecedent down to 
$H(m,n)\add\gneg H(m,n)$. The environment
has to answer this counterquestion, or else it loses. If it answers ``no" (i.e. makes the move `$1.2$' and thus further brings the antecedent down to $\gneg H(m,n)$), $\pp$ also answers ``no" to the 
original question in the consequent (i.e. makes the move `$2.2$'), with the overall game having evolved to the true
and hence $\pp$-won proposition/elementary game $\gneg H(m,n)\mli \gneg A(m,n)$. Otherwise, if the environment's answer is ``yes" 
(move `$1.1$'), $\pp$ simulates Turing machine $m$ on input $n$ until it halts, and then makes the move `$2.1$' or `$2.2$' depending whether the simulation accepted or rejected. Of course, it is a possibility that such a simulation goes on forever and thus no moves will be made by $\pp$ in the consequent. This, however, will only happen when $H(m,n)$ --- the $\add$-disjunct selected by the environment in the antecedent --- is false, and having lied in the antecedent makes $\oo$ the loser no matter what happens in the consequent.   

Again, what the machine did in the above strategy indeed was a reduction: it used an (external) solution 
to the halting problem to solve the acceptance problem. There are various natural concepts of reduction, and a strong case can be made in favor of the thesis that  
the sort of reduction captured by $\mli$ is most basic among them, with a great variety of other reasonable concepts of reduction being expressible in terms of $\mli$. Among those is 
 {\em Turing reduction}. It will be discussed a little later when we get to recurrence operations. Another frequently used concept of reduction is {\em mapping reduction} that we are going to look at in the following paragraph. And yet some other natural concepts of reduction expressible in terms of $\mli$ may or may not have established names. For example, from the above discussion it can be seen that a certain reducibility-style relation holds between the predicates $A(x,y)$ and $H(x,y)$ in an even stronger sense than the algorithmic winnability of (\ref{apr8}). In fact, not only (\ref{apr8}) is winnable, but also the generally harder-to-win game 
\begin{equation}\label{apr8a}
\ada x\ada y\bigl(H(x,y)\add\gneg H(x,y)\mli A(x,y)\add\gneg A(x,y)\bigr).
\end{equation}
 This is so because $\pp$'s above-described strategy for (\ref{apr8}) did not use (while could have used) any values for $x$ and $y$ others than the values chosen for these variables by $\oo$ in the consequent. So, the $\ada x\ada y$ prefix 
can be just made external as this is done in (\ref{apr8a}). It will be seen later that semantically our approach treats free variables as if they were (externally) bound by $\ada$. Hence, the winnability of (\ref{apr8a}), in turn, is the same as simply the winnability of 
\[H(x,y)\add\gneg H(x,y)\mli A(x,y)\add\gneg A(x,y).\]

A predicate $p(\vec{x})$ is said to be {\bf mapping reducible}\index{mapping reducibility}\label{0mapping reducibility}\footnote{This term is adopted from \cite{Sip06}. The more common but less adequate name for what we call mapping reducibility is {\bf many-one reducibility}.} to a predicate $q(\vec{y})$ iff there is an effective function $f$ such that, for any constants $\vec{c}$, $p(\vec{c})$ is true iff
$q(f(\vec{c}))$ is so. Here $\vec{x}$ abbreviates any $n$-tuple of pairwise distinct variables, $\vec{c}$ any $n$-tuple of constants, $\vec{y}$ any $m$-tuple of pairwise distinct variables, and $f$ is a function that sends $n$-tuples of constants to $m$-tuples of constants. Using $A\mleq B$ as an abbreviation for $(A\mli B)\mlc (B\mli A)$ and $\ada\vec{z}$  
for $\ada z_1\ldots\ada z_k$ where $\vec{z}=z_1,\ldots,z_k$ (and similarly for $\ade\vec{z}$), it is not hard to see that 
mapping reducibility of $p(\vec{x})$ to $q(\vec{y})$ means nothing but existence of an algorithmic winning strategy for 
\[\ada \vec{x}\ade \vec{y}\bigl(p(\vec{x})\mleq q(\vec{y})\bigr).\]
Our acceptance predicate $A(x,y)$ can be shown to be mapping reducible to the halting predicate $H(x,y)$, i.e. the game
\[\ada x\ada y\ade x'\ade y'(A(x,y)\mleq H(x',y'))\]
shown to be winnable by a machine. An algorithmic strategy for $\pp$ is the following. After $\oo$ selects values $m$ and $n$ for $x$ and $y$, select the values $m'$ and (the same) $n$ for $x'$ and $y'$, and 
rest your case. Here 
$m'$ is the Turing machine that works exactly as $m$ does, with the only difference that whenever $m$ enters its reject state, $m'$ goes into an infinite loop instead, so that $m$ accepts if and only if $m'$ halts. Such an $m'$, of course, can be effectively constructed from $m$.

Notice that while the standard approaches only allow us to talk about (a whatever sort of) reducibility as a {\em relation} between problems, in our approach reduction becomes an {\em operation} on problems, with reducibility as a relation simply meaning computability 
of the corresponding combination (such as $\ada \vec{x}\ade \vec{y}\bigl(p(\vec{x})\mleq q(\vec{y}))$ or $A\mli B$) of games. Similarly for other relations or properties such as the property of {\em decidability}. The latter becomes the operation of {\em deciding} if we define the problem of deciding 
a predicate (or any computational problem) $p(\vec{x})$ as the game $\ada \vec{x}\bigl(p(\vec{x})\add\gneg p(\vec{x})\bigr)$. So, now we can meaningfully ask questions such as
``is the reduction of the problem of deciding $p(x)$ to the problem of deciding $q(x)$ reducible to the mapping reduction of $p(x)$ to $q(x)$?". This question would be equivalent to whether the following game has an algorithmic winning strategy:
\begin{equation}\label{e1}
\begin{array}{l}  
\Bigl(\ada x\ade y\bigl(p(x)\mleq q(y)\bigr)\Bigr)\mli\\
\Bigl(\ada x\bigr(q(x)\add\gneg q(x)\bigr)\mli
\ada x\bigr(p(x)\add\gneg p(x)\bigr)\Bigr).
\end{array}
\end{equation}
\noindent This problem is indeed algorithmically solvable no matter what particular predicates $p(x)$ and $q(x)$ are, which means that mapping reduction 
is at least as strong as reduction.  Here is a strategy for $\pp$:
Wait till $\oo$ selects a value $k$ for $x$ in the consequent of the consequent of (\ref{e1}). Then specify the same value 
$k$ for $x$ in the antecedent of (\ref{e1}), and wait till $\oo$ replies there by selecting a value $n$ for $y$. Then select 
the same value $n$ for $x$ in the antecedent of the consequent of (\ref{e1}). $\oo$ will have to respond by $1$ or 
$2$ in that component of the game. Repeat that very response in the consequent of the consequent of (\ref{e1}), and celebrate victory.

We are going to see in Section \ref{ss9} that (\ref{e1}) is a legal formula of the language of system $\predell$, which, according to Theorem \ref{main5},  is sound and complete with respect to the semantics of computability logic. So, had our ad hoc methods failed to find an answer (and this would certainly be the case if we dealt with a more complex computational problem), the existence of a successful algorithmic strategy could have been established by showing that (\ref{e1}) is provable in $\predell$. 
 Moreover, by clause (a) of Theorem \ref{main5}, after finding a $\predell$-proof of (\ref{e1}), we would not only know that an algorithmic solution for (\ref{e1}) exists, but we would also be able to constructively extract such a solution from the proof. On the other hand, the fact that reduction is not as strong as mapping reduction could be established by showing that $\predell$ does not prove 
\begin{equation}\label{e2}
\begin{array}{l}  
\Bigl(\ada x\bigr(q(x)\add\gneg q(x)\bigr)\mli
\ada x\bigr(p(x)\add\gneg p(x)\bigr)\Bigr) \mli\\
 \Bigl(\ada x\ade y\bigl(p(x)\mleq q(y)\bigr)\Bigr).
\end{array}
\end{equation}
This negative fact, too, can be established effectively as, according to Theorem \ref{dec}, the relevant fragment of 
$\predell$  is decidable.
In fact, the completeness proof for $\predell$ given in \cite{CL4} shows a way how to actually construct particular predicates --- $p(x)$ and $q(x)$ in our case --- for which 
the problem represented by a given $\predell$-unprovable formula has no algorithmic solution.

\subsection{Blind operations}
Another group of core game operations, called {\bf blind},\index{blind operations}\label{0blind operations}  comprises  $\cla$ and its dual $\cle$.   Intuitively, playing $\cla xA(x)$ or $\cle xA(x)$ means just playing 
$A(x)$ ``blindly'', without knowing the value of $x$. In $\cla xA(x)$, $\pp$ wins iff the play it generates is successful for every possible value of $x$, while in $\cle xA(x)$ being successful for just one value is sufficient. $\cla $ and $\cle$ thus essentially produce what is called games with {\em imperfect information} (see \cite{Pie01}).  This sort of a blind play is meaningful or possible --- and hence $\cla xA(x)$, $\cle xA(x)$ defined --- only when what moves are available (legal) does not depend on the unknown value of $x$; in other words, when $A(x)$ is unistructural in $x$. 

\begin{definition}\label{bq} Assume $A(x)$ is unistructural in $x$.\vspace{5pt} 

\noindent 1. {\bf Blind universal quantification}\index{blind universal quantification}\label{0blind universal quantification} $\cla x  A(x)$:\index{$\cla$}\label{0cla} 
\begin{itemize}
\item $\legal{\clai xA(x)}{e}=\legal{A(x)}{e}$.  
\item $\win{\clai xA(x)}{e}\seq{\Gamma}=\pp$ iff, for every constant $c$, $\win{A(c)}{e}\seq{\Gamma}=\pp$.\vspace{5pt} 
\end{itemize}

\noindent 2. {\bf Blind existential quantification}\index{blind existential quantification}\label{0blind existential quantification} $\cle x  A(x)$:\index{$\cle$}\label{0cle} 
\begin{itemize}
\item $\legal{\clei xA(x)}{e}=\legal{A(x)}{e}$.  
\item $\win{\clei xA(x)}{e}\seq{\Gamma}=\oo$ iff, for every constant $c$, $\win{A(c)}{e}\seq{\Gamma}=\oo$. 
\end{itemize}

\end{definition}

 As with the other pairs of quantifiers, one can see that we always have 
$\gneg \cla xA(x)=\cle x\gneg A(x)$ and $\gneg\cle xA(x)=\cla x\gneg A(x)$.

To feel the difference between $\cla$ and $\ada$, compare the games
\[\ada x\bigl(\mbox{\em Even$(x)$}\add \mbox{\em Odd$(x)$}\bigr)\] and  \[\cla x\bigl(\mbox{\em Even$(x)$}\add \mbox{\em Odd$(x)$}\bigr).\] 
Both are about telling whether a given number is even or odd; the difference is only in whether that ``given number" is specified (made as a move by $\oo$) or not. 
The first problem is an easy-to-win, depth-2 game of a structure that we have already seen. The depth of the second game, on the other hand, is 1, with only the machine to make a move --- select the ``true"  disjunct, which is hardly possible to do as the value of $x$ remains unspecified. 

Figure 10 shows trees for $\mbox{\em Even}(x)\add\mbox{\em Odd}(x)$ and $\cla x\bigl(\mbox{\em Even}(x)\add \mbox{\em Odd}(x)\bigr)$ next to each other. Notice that applying $\cla x$ does not change the structure of a (unistructural) game. What it does is that it simply prefixes every node with a $\cla x$ (we do not explicitly see such a prefix on the root because $\cla x\tlg=\tlg$). This means that we always have $\seq{\Phi}\cla xA(x)=\cla x\seq{\Phi}A(x)$. Similarly for $\cle x$.

  \begin{center}
\begin{picture}(305,122)

\put(21,107){$\mbox{\em Even}(x)\add\mbox{\em Odd}(x)$}
\put(65,92){\circle{16}}
\put(61,88){$\oo$}
\put(18,62){\scriptsize $\pp 1$}
\put(65,84){\line(-3,-2){50}}
\put(99,62){\scriptsize $\pp 2$}
\put(65,84){\line(3,-2){50}}

\put(15,42){\oval(42,16)}
\put(1,40){\scriptsize $\mbox{\em Even}(x)$}

\put(115,42){\oval(42,16)}
\put(103,40){\scriptsize $\mbox{\em Odd}(x)$}

\put(190,107){$\cla x\bigl(\mbox{\em Even}(x)\add\mbox{\em Odd}(x)\bigr)$}
\put(243,92){\circle{16}}
\put(239,88){$\oo$}
\put(196,62){\scriptsize $\pp 1$}
\put(243,84){\line(-3,-2){50}}
\put(277,62){\scriptsize $\pp 2$}
\put(243,84){\line(3,-2){50}}

\put(193,42){\oval(42,16)}
\put(174,40){\scriptsize $\clai x\mbox{\em Even}(x)$}
\put(293,42){\oval(42,16)}
\put(277,40){\scriptsize $\clai x\mbox{\em Odd}(x)$}

\put(53,10){{\bf Figure 10:} Blind universal quantification}
\end{picture}
\end{center}

Of course, not all nonelementary $\cla$-games will be unwinnable. Here is an example:\vspace{-3pt}
\[\cla x\Bigl(\mbox{\em Even$(x)$}\add \mbox{\em Odd$(x)$} \mli \ada y\bigl(\mbox{\em Even$(x \times y)$}\add
\mbox{\em Odd$(x\times y)$}\bigr)\Bigr).\vspace{-3pt}\]
Solving this problem, which means reducing the consequent to the antecedent without knowing the value of $x$, is easy: 
$\pp$ waits till $\oo$ selects  a value $n$ for $y$. If $n$ is even, then $\pp$ selects the first $\add$-disjunct in the consequent. Otherwise, if $n$ is odd, $\pp$ continues waiting until $\oo$ selects one of the $\add$-disjuncts in the antecedent
(if $\oo$ has not already done so), and then $\pp$ makes the same move 
1  or 2 in the consequent as $\oo$ made in the antecedent. One of the possible runs such a strategy can yield is 
$\seq{\oo 1.2,\oo 2.5,\pp 2.2}$, which can be visualized as the following sequence of games:\vspace{5pt}

\noindent\hspace{-3pt} $\begin{array}{l}
\cla x\Bigl(\mbox{\em Even}(x)\add \mbox{\em Odd}(x) \mli \ada y\bigl(\mbox{\em Even}(x \times y)\add
\mbox{\em Odd}(x\times y)\bigr)\Bigr);\vspace{2pt}\\
\cla x\Bigl(\mbox{\em Odd}(x) \mli \ada y\bigl(\mbox{\em Even}(x \times y)\add
\mbox{\em Odd}(x\times y)\bigr)\Bigr);\vspace{2pt}\\
\cla x\Bigl(\mbox{\em Odd}(x) \mli \mbox{\em Even}(x \times 5)\add
\mbox{\em Odd}(x\times 5)\Bigr);\vspace{2pt}\\
\cla x\Bigl(\mbox{\em Odd}(x) \mli 
\mbox{\em Odd}(x\times 5)\Bigr).\vspace{5pt}
\end{array}$

By now we have seen three --- choice, parallel and blind --- natural sorts of quantifiers. The idea of a forth --- {\em sequential}\index{sequential quantifiers}\label{0sequential quantifiers} --- sort, which we will not discuss here, was outlined in \cite{Japic}.
It is worthwhile to take a quick look at how different quantifiers relate. Both $\cla xA(x)$ and $\mla xA(x)$ can be shown to be properly stronger than $\ada xA(x)$, in the sense that 
$\cla xP(x)\mli \ada xP(x)$ and $\mla xP(x)\mli\ada xP(x)$ are valid while  $\ada xP(x)\mli \cla xP(x)$ and $\ada xP(x)\mli\mla xP(x)$ are not. On the other hand, the strengths of $\cla xP(x)$ and $\mla xP(x)$ are mutually incomparable: 
neither $\cla xP(x)\mli \mla x P(x)$ nor $\mla xP(x)\mli \cla x P(x)$ is valid. 
The big difference between $\cla$ and $\mla$ is that, while playing $\cla xA(x)$ means playing one ``common" play for 
all possible $A(c)$ and thus $\cla xA(x)$ --- just like $\ada x A(x)$ --- is a  
one-board game, $\mla xA(x)$ is an infinitely-many-board game: playing it means playing, in parallel, 
game $A(1)$ on board \#1, game $A(2)$ on board \#2, etc.
When restricted to elementary games, however, the distinction between the blind and the parallel groups of quantifiers disappears as already noted and, just like $\gneg$, $\mlc$, $\mld$, $\mli$, $\mla$, $\mle$, the blind quantifiers behave exactly in the classical way. Having this collection of operators makes CL a conservative extension of classical first-order logic: the latter is nothing but CL restricted to elementary problems and the logical vocabulary $\gneg$, $\mlc$, $\mld$, $\mli$, $\cla$ (and/or $\mla$), $\cle$ (and/or $\mle$).

\subsection{Recurrence operations}\label{ss4.6}
What is common to the members of the family of game operations called {\bf recurrence operations}\index{recurrence operations}\label{0recurrence operations} is that, when applied to $A$, they turn it into a game playing which means repeatedly playing $A$. In terms of resources, recurrence operations generate multiple ``copies'' of $A$, thus making $A$ a reusable/recyclable resource. The difference between various sorts of recurrences is how ``reusage'' is exactly understood.
Imagine a computer that has a program successfully playing $\chess$. The resource that such a computer provides is obviously something stronger than just $\chess$, for it permits to play $\chess$ as many times as the user wishes, while $\chess$, as such, only assumes one play. The simplest operating system would allow to start a session of $\chess$, then --- after finishing or abandoning and destroying it --- start a new play again, and so on. The game that such a system plays --- i.e. the resource that it supports/provides --- 
is $\sst \chess$,\index{$\sst$}\label{0sst} which assumes an unbounded number of plays of $\chess$ in a sequential fashion. We call $\sst$ {\bf sequential recurrence}.\index{sequential recurrence}\label{0sequential recurrence} A more advanced operating system, however, would not require to destroy the old sessions before starting new ones; rather, it would allow to run as many parallel sessions as the user needs. This is what is captured by $\pst\chess$, meaning nothing but the infinite parallel  
conjunction $\chess\mlc\chess\mlc\chess\mlc\ldots$. Hence $\pst$ is called {\bf parallel recurrence}. As a resource, $\pst\chess$ is obviously stronger than $\sst\chess$ as it gives the user more flexibility. But $\pst$ is still not the strongest form of reusage. A really good operating system would not only allow the user to start new sessions of $\chess$ without destroying old ones; it would also make it possible to branch/replicate each particular stage of each particular session, i.e. create any number of ``copies" of any already reached position
of the multiple parallel plays of $\chess$, thus giving the user 
the possibility to try different continuations from the same position. What corresponds to this intuition is $\st\chess$, where $\st$ is called {\bf branching recurrence}.\index{branching recurrence}\label{0branching recurrence0}\footnote{The term ``branching recurrence" and the symbol $\sti$  were established in \cite{Japic}. The earlier (foundational) paper \cite{Jap03} uses ``branching conjunction" and $!$ instead. Similarly, \cite{Jap03} uses the term ``branching disjunction'' instead of our present ``branching corecurrence'', and symbol $?$ instead of $\cost$. Finally, to our present symbol $\intimpl$ in \cite{Jap03} corresponds $\Rightarrow$.}  
As all of the operations (except $\gneg,\mli$) seen in this section, each sort of recurrence comes with its dual operation, called {\bf corecurrence}. Say, the {\bf branching corecurrence}\index{branching corecurrence}\label{0branching corecurrence0} $\cost A$ of $A$ is nothing but $\st \gneg A$ as seen by the environment, so $\cost A$ can be defined as $\gneg \st\gneg A$; similarly for {\bf parallel corecurrence} $\pcost$ and {\bf sequential corecurrence}\index{sequential corecurrence} \label{0sequential coreccurrence} $\scost$.\index{$\scost$}\label{0scost} $\pcost A$ thus means the infinite parallel disjunction $A\mld A\mld A\mld\ldots$. The sequential recurrence and sequential corecurrence of $A$, on the other hand, can be defined as
infinite {\bf sequential conjunction}\index{sequential conjunction and disjunction}\label{0sequential conjunction and disjunction} $A\sqc A\sqc A\sqc\ldots$\index{$\sqc$}\label{0sqc} and infinite {\bf sequential disjunction} $A\sqd A\sqd A\sqd\ldots$,\index{$\sqd$}\label{0sqd} respectively. An idea of the sequential versions of conjunction/disjunction, quantifiers and recurrence/corecurrence was informally outlined in a footnote of Section 8 of \cite{Japic}, and then fully elaborated in \cite{Japseq}. Out of laziness, in this paper we are not going to go any farther than the above intuitive explanation of sequential recurrence, just as we have not attempted and will not attempt to define the sequential versions of propositional connectives or quantifiers.\footnote{There are similar and even more serious reasons for not attempting to introduce blind\index{blind conjunction and disjunction}\label{0blind conjunction and disjunction} versions of conjunction and disjunction.} Only the parallel and branching sorts of recurrence will receive our full attention. 

\begin{definition}\label{pr} \ \vspace{8pt}  

\noindent 1. {\bf Parallel recurrence}\index{parallel recurrence}\label{0parallel recurrence} $\pst A$:\index{$\pst$}\label{0pst}
\begin{itemize}
\item $\Gamma\in\legal{\psti A}{}$ iff every move of $\Gamma$ has the prefix `$i.$' for some $i\in\{1,2,3,\ldots\}$ and, for each such $i$, $\Gamma^{i.}\in\legal{A}{}$.
\item Whenever $\Gamma\in\legal{\psti A}{}$, we have \\
$\win{\psti A}{}\seq{\Gamma}=\pp$ iff, for each $i\in\{1,2,3,\ldots\}$, $\win{A}{}\seq{\Gamma^{i.}}=\pp$.\vspace{5pt}  
\end{itemize}

\noindent 2. {\bf Parallel corecurrence}\index{parallel corecurrence}\label{0parallel corecurrence} $\pcost A$:\index{$\pcost$}\label{0pcost}
\begin{itemize}
\item $\Gamma\in\legal{\pcosti A}{}$ iff every move of $\Gamma$ has the prefix `$i.$' for some $i\in\{1,2,3,\ldots\}$ and, for each such $i$, $\Gamma^{i.}\in\legal{A}{}$.
\item Whenever $\Gamma\in\legal{\pcosti A}{}$, we have \\
$\win{\pcosti A}{}\seq{\Gamma}=\oo$ iff, for each $i\in\{1,2,3,\ldots\}$, $\win{A}{}\seq{\Gamma^{i.}}=\oo$.  
\end{itemize}
\end{definition}

Thus, from the machine's perspective,  $\pst \ada x\ade y(y=x^2)$ is the problem of computing the square function, and 
--- unlike the case with $\ada x\ade y(y=x^2)$ --- doing so repeatedly, i.e. as many times as the environment asks a question ``what is the square of $m$?''.  In the style 
of  Example \ref{exp2}, a unilegal position of $\pst A$ (resp. $\pcost A$) can be represented as an infinite parallel conjunction (resp. disjunction), with the infinite contiguous block of 
``not-yet-activated'' conjuncts (resp. disjuncts) starting from item \#n combined together and written as 
$\pst n A$ (resp. $\pcost n A$). Below is an illustration. 

\begin{example}\label{apr18}
{\em The $\pp$-won run $\seq{\oo 1.3,\pp 1.9,\oo 2.1,\pp 2.1}$ of the game $\pst \ada x\ade y(y=x^2)$ generates the following sequence:\vspace{5pt}

\noindent\hspace{-3pt} $\begin{array}{l}
\pst \ada x\ade y(y=x^2) \ \ \ \mbox{ (or $\pst 1 \ada x\ade y(y=x^2)$)};\\
\ade y(y=3^2) \mlc \pst 2 \ada x\ade y(y=x^2);\\
9=3^2 \mlc \pst 2 \ada x\ade y(y=x^2);\\
9=3^2 \mlc \ade y(y=1^2)\mlc \pst 3 \ada x\ade y(y=x^2);\\
9=3^2 \mlc 1=1^2\mlc \pst 3 \ada x\ade y(y=x^2).\vspace{5pt}
\end{array}$
}\end{example}

Among the best known and most natural concepts of reducibility in traditional computability theory is that of {\bf Turing reducibility}\index{Turing reducibility}\label{0Turing reducibility} of a problem $A$  
to a problem $B$, meaning existence of a Turing machine that solves $A$ with an oracle for $B$. In this definition ``problem'', of course, is understood in the traditional sense, meaning a two-step, question-answer problem such as computing a function or deciding a predicate. This is so because both the oracle and the Turing machine offer a simple, question-answer 
interface, unable to handle problems with higher degrees or more sophisticated forms of interactivity.  Our approach allows us to get rid of the ``amateurish'' concept of an oracle, and reformulate the above definition of Turing reducibility of $A$ to $B$ as computability of 
$B\pintimpl A$, where $\pintimpl$\index{$\pintimpl$}\label{0pintimpl} is defined by 
\[B\pintimpl A\ =\ (\pst B)\mli A.\]  This newborn concept of \mbox{$\pintimpl${\bf -reducibility}}\index{$\pintimpl$-reducibility}\label{0pintimpl-reducibility} then not only adequately rephrases Turing reducibility, but also generalizes it, for $\pintimpl$-reducibility is defined for all games $A$ and $B$ rather than only those representing traditional computational problems. 

To get a better feel of $\pintimpl$ and appreciate the difference between it and the ordinary reduction $\mli$, remember our example of $\mli$-reducing the acceptance problem 
to the halting problem.
The reduction that $\pp$ used in its successful strategy for 
\[\ada x\ada y (H(x,y)\add\gneg H(x,y))\mli \ada x\ada y (A(x,y)\add\gneg A(x,y))\] was in fact a Turing reduction, as $\pp$'s moves $1.m$ and $1.n$ can be seen 
as querying an oracle (with the environment acting in the role of such) regarding whether $m$  halts on $n$. The potential usage of an ``oracle'', however, was limited there, as it could be employed only once. If, for some reason, $\pp$ needed to repeat the same question with some different parameters $m'$ and $n'$, it would not be able to do so, for this would require having two ``copies'' of the resource $\ada x\ada y (H(x,y)\add\gneg H(x,y))$ in the antecedent, i.e. having 
\[\ada x\ada y (H(x,y)\add\gneg H(x,y))\mlc \ada x\ada y (H(x,y)\add\gneg H(x,y))\] rather than $\ada x\ada y (H(x,y)\add\gneg H(x,y))$ there. On the other hand, Turing reduction 
assumes an unlimited oracle-querying capability. Such a capability is precisely accounted for by prefixing the antecedent with a $\pst$, i.e. replacing $\mli$ with $\pintimpl$. 
As an example of a problem $\pintimpl$-reducible but not $\mli$-reducible to the halting problem, consider the relative Kolmogorov complexity problem.\index{Kolmogorov complexity} \label{0kc} It can be expressed as 
$\ada x\ada y\ade zK(x,y,z)$, where $K(x,y,z)$ is the predicate ``$z$ is the Kolmogorov complexity of $x$ relative to $y$'', i.e. ``$z$ is the smallest (code of a) Turing machine that returns $x$ on input $y$''. 
The problem of Turing-reducing the relative Kolmogorov complexity problem to the halting problem translates into our terms as 
\[\pst \ada x\ada y (H(x,y)\add\gneg H(x,y))\mli \ada x\ada y\ade z K(x,y,z).\]
Seeing the antecedent as an infinite $\mlc$-conjunction, here is $\pp$'s algorithmic winning strategy for the above game. $\pp$ waits till $\oo$ selects some values $m$ and $n$ for $x$ and $y$   in the consequent, signifying asking $\pp$ about the Kolmogorov complexity of $m$ relative to $n$. 
Then, starting from $i=1$, $\pp$ does the following. In the $i$th $\mlc$-conjunct  of the antecedent, it makes two consecutive  moves
 by specifying $x$ and $y$ as $i$ and $n$, respectively, thus asking $\oo$ whether machine $i$ halts on input $n$. If $\oo$ responds there by ``no'', 
$\pp$ increments $i$ by one and repeats the step. Otherwise, if $\oo$ responds by ``yes'', $\pp$ simulates machine $i$ on input $n$ until it halts; if the simulation shows that 
machine $i$ on input $n$ returns $m$, $\pp$ makes the move $i$ in the consequent, thus saying that $i$ is the Kolmogorov complexity of $m$ relative to $n$; otherwise, $\pp$ increments $i$ by one and repeats the step.

Turing reducibility has well-justified claims to be a formalization of our weakest intuition of algorithmic reducibility of one traditional problem to another, \ and $\pintimpl$-reducibility, \  as we now know, \ conservatively extends Turing reducibility to all games. \ This may suggest that \mbox{$\pintimpl$-reducibility could be} an adequate formal counterpart of our weakest intuition of algorithmic reducibility of one 
interactive problem to another. Such a guess would be wrong though. As claimed earlier, it is $\st A$ rather than $\pst A$ that corresponds to our strongest intuition of using/reusing $A$. This automatically translates into another claim: it is $\intimpl$-reducibility rather than $\pintimpl$-reducibility that (in full interactive generality) captures the weakest form of reducibility.
Here $\intimpl$\index{$\intimpl$}\label{0intimpl} is defined by \[B\intimpl A\ =\ (\st B)\mli A.\index{$\sintimpl$}\label{0sintimpl}\footnote{Now we may guess that, if it ever comes to studying the sequential-recurrence-based reduction $(\sst B)\mli A$, the symbol for it would be 
$\sintimpl$.}\]
It was mentioned in Section \ref{intr} that Heyting's intuitionistic calculus is sound and, in the propositional case, also complete  with respect to the semantics of computability logic. This is so when intuitionistic implication is understood as $\intimpl$, and intuitionistic conjunction, disjunction, universal quantifier and existential quantifier as 
$\adc,\add,\ada$ and $\ade$, respectively. With intuitionistic implication read as $\pintimpl$,  intuitionistic calculus is unsound as, for example,  it proves  
\[(P\pintimpl R)\adc (Q\pintimpl R)\pintimpl (P\add Q\pintimpl R)\]  which fails to be a valid principle of computability.

$\pintimpl$-reducibility and $\intimpl$-reducibility, while being substantially different in the general case, turn out to be equivalent when restricted to certain special sorts of problems with low degrees of interactivity such as what we have been referring to as ``traditional problems'', examples of which being the halting, acceptance or relative Kolmogorov complexity problems. For this reason, 
both   $\pintimpl$-reducibility and $\intimpl$-reducibility are equally adequate as (conservative) generalizations of the traditional concept of Turing reducibility.

It is now time to get down to a formal definition of branching recurrence $\st$. This is not just as easy as defining $\pst$, and requires a number of auxiliary concepts and conventions. Let us 
start with a closer look at the associated intuitions.  
One of the ways to view both $\pst A$ and $\st A$ is to think of them as games where $\oo$ is allowed to restart $A$ an unlimited number of times without terminating the already-in-progress runs of $A$, creating, this way, more and more parallel plays of $A$ with the possibility to try different strategies in them and become the winner as long as one of those strategies succeeds. What makes $\st A$ stronger (as a resource) than $\pst A$, however, is that, as noted earlier, in $\st A$, $\oo$ does not have to really restart $A$ from the very beginning every time it ``restarts" it; rather, it can select to continue $A$ from any of the previous positions, thus depriving $\pp$ of the possibility to reconsider the moves it has already made. A little more precisely, at any time $\oo$ is allowed to replicate (backup) any of the currently reached parallel positions of $A$ before further modifying it, which gives it the possibility to come back later and continue playing $A$ from the backed-up position. This way, we get a tree of labmoves (each branch spelling a legal run of $A$) rather than just multiple parallel sequences of labmoves. Then $\pst A$ can be thought of as a weak version of $\st A$ where only the empty position of $A$ can be replicated, that is, where branching in the tree only happens at its root. 
A discussion of how $\st$ relates to Blass's\index{Blass}\label{0Blass0} \cite{Bla72,Bla92} repetition operator is given in Section 13 of \cite{Jap03}.

To visualize the scheme that lies under our definition of $\st$, consider a play over $\st\mbox{\em Chess}$. The play takes place between a computer ($\pp$) and a user ($\oo$), and its positions are displayed on the screen. In accordance with the earlier elaborated intuitions, we think of each such position $\Phi$ as the game $\seq{\Phi}\chess$, and vice versa. At the beginning, there is a window on the screen --- call it Window $\epsilon$ --- that displays the initial position of  \chess:

\begin{center}
\begin{picture}(45,36)
\put(0,24){Window $\epsilon$}
\put(6,10){\framebox{$\chess$}}
\end{picture}
\end{center}

 We denote this position by $\chess$, but the designer would probably make the window show  a colorful image of a chess board with 32 chess pieces in their initial locations. The play starts and proceeds in an ordinary fashion: the players are making legal moves of $\chess$, which correspondingly update the position displayed in the window. At some (any) point, 
when the current position in the window is $\seq{\Phi}\chess$, \ $\oo$ may decide to replicate the position, perhaps because he wants to try different continuations in different copies of it.  This splits Window $\epsilon$ into two children windows named $0$ and $1$, each containing the same position $\seq{\Phi}\chess$ as the mother window contained at the time of split. The mother window disappears, and the picture on the screen becomes 

\begin{center}\begin{picture}(175,36)
\put(126,24){Window $1$}
\put(125,10){\framebox{$\seq{\Phi}\chess$}}
\put(1,24){Window $0$}
\put(0,10){\framebox{$\seq{\Phi}\chess$}}
\end{picture}
\end{center}

\noindent(again, let us try to imagine a real chess position colorfully depicted in the two windows instead of the bleak expression ``$\seq{\Phi}\chess$").

From now on the play continues on two boards / in two windows. Either player can make a legal move in either window. After some time, when the game in Window $0$ has evolved to $\seq{\Phi,\Psi}\chess$\  and in \mbox{Window $1$} to $\seq{\Phi,\Theta}\chess$, $\oo$ can, again, decide to split one of the windows --- say, Window $0$. The mother window $0$ will be replaced by two children windows: $00$ and $01$, each having the same content as their mother had at the moment of split, so that now the screen will be showing three windows:

\begin{center}
\begin{picture}(242,36)
\put(185,24){Window $1$}
\put(176,10){\framebox{$\seq{\Phi,\Theta}\chess$}}
\put(5,24){Window $00$}
\put(0,10){\framebox{$\seq{\Phi,\Psi}\chess$}}
\put(94,24){Window $01$}
\put(88,10){\framebox{$\seq{\Phi,\Psi}\chess$}}
\end{picture}
\end{center}

If and when, at some later moment, $\oo$ decides to make a third split --- say, in Window $01$ --- the picture on the screen becomes

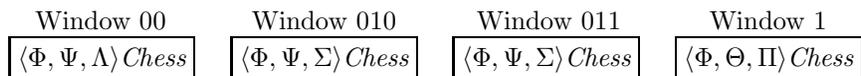
\begin{center}
\noindent\begin{picture}(327,36)
\put(265,24){Window $1$}
\put(252,10){\framebox{$\seq{\Phi,\Theta,\Pi}\chess$}}
\put(10,24){Window $00$}
\put(0,10){\framebox{$\seq{\Phi,\Psi,\Lambda}\chess$}}
\put(92,24){Window $010$}
\put(84,10){\framebox{$\seq{\Phi,\Psi,\Sigma}\chess$}}
\put(177,24){Window $011$}
\put(168,10){\framebox{$\seq{\Phi,\Psi,\Sigma}\chess$}}
\end{picture}
\end{center}

\noindent etc.  At the end, the game will be won by $\pp$ if and only if each of the windows contains a winning position of $\chess$.

The above four-window position can also be represented as the following binary tree, where the name of each window is uniquely determined by its location in the tree:

\begin{center}
\begin{picture}(253,143)
\put(215,35){\framebox{$\seq{\Phi,\Theta,\Pi}$\chess}}
\put(-39,35){\framebox{$\seq{\Phi,\Psi,\Lambda}\chess$}}
\put(45,35){\framebox{$\seq{\Phi,\Psi,\Sigma}\chess$}}
\put(130,35){\framebox{$\seq{\Phi,\Psi,\Sigma}\chess$}}
\put(43,10){{\bf Figure 11:} A position of \ $\st\chess$}

\put(120,132){\line(3,-2){128}}
\put(120,132){\line(-3,-2){128}}
\put(80,105){\line(3,-2){40}}
\put(120,78){\line(4,-3){42}}
\put(120,78){\line(-4,-3){42}}

\put(91,117){0}
\put(143,117){1}
\put(50,89){0}
\put(105,89){1}
\put(89,59){0}
\put(146,59){1}
\end{picture}
\end{center}

\noindent Window names  will be used by the players to indicate in which of the windows they are making a move.
Specifically, `$w.\alpha$' is the move meaning making move $\alpha$ in Window $w$; and the move (by $\oo$) that splits/replicates Window $w$ 
is `$\col{w}$'. 
Sometimes the window in which a player is trying to make a move may no longer exist. For example, in the position preceding the position of Figure 11, $\pp$ might have decided to make move $\alpha$ in Window $01$. However, before $\pp$ actually made this move, $\oo$ made a replicative move in the same window, which took us to the four-window position of Figure 11. $\pp$ may not notice this replicative move and complete its move $01.\alpha$ by the time when Window 01 no longer exists. This kind of a move is still considered legal, and its effect is making the move $\alpha$ in all (in our case both) of the descendants of the no-longer-existing Window $01$. The result will be 

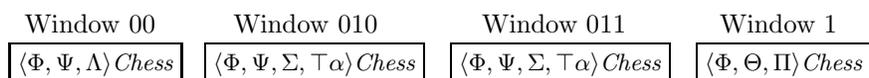
\begin{center}
\begin{picture}(327,36)
\put(269,24){Window $1$}
\put(260,10){\framebox{\small $\seq{\Phi,\Theta,\Pi}\chess$}}
\put(6,24){Window $00$}
\put(0,10){\framebox{\small $\seq{\Phi,\Psi,\Lambda}\chess$}}
\put(85,24){Window $010$}
\put(74,10){\framebox{\small $\seq{\Phi,\Psi,\Sigma,\pp\alpha}\chess$}}
\put(179,24){Window $011$}
\put(167,10){\framebox{\small $\seq{\Phi,\Psi,\Sigma,\pp\alpha}\chess$}}
\end{picture}
\end{center}

The initial position in the example that we have just discussed was one-window. This, generally, is not necessary. The operation $\st$ can be applied to any construction in the above style, such as, say,

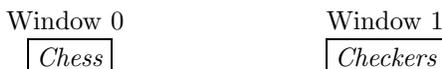
\begin{center}
\begin{picture}(246,36)
\put(156,24){Window $1$}
\put(156,10){\framebox{\checkers}}
\put(35,24){Window $0$}
\put(43,10){\framebox{\chess}}
\end{picture}
\end{center}

A play over this game, which our later-introduced notational conventions would denote by $\st(\chess\circ\checkers)$, will proceed in a way similar to what we saw, where more and more windows can be created, some (those whose names are 0-prefixed) displaying positions of chess, and some (with 1-prefixed names) displaying positions of checkers. In order to win, the machine will have to win all of the multiple parallel plays of chess and checkers that will be generated. 

As the dual of $\st$, $\cost$ can be characterized in exactly the same way as $\st$, only, in a $\cost$-game, it is $\pp$ who has the privilege of splitting windows, and for whom winning in just one of the multiple parallel plays is sufficient. 

To put together the above intuitions, let us agree that by a {\bf bitstring}\index{bitstring}\label{0bitstring} we mean a string of $0$s and $1$s, including infinite strings and the {\bf empty string}\index{empty string}\label{0empty string} $\epsilon$\index{$\epsilon$}\label{0epsilon}. 
We will be using the letters $w$, $u$, $v$ as metavariables for bitstrings. $\epsilon$\index{$\epsilon$}\label{0epsilon} will exclusively stand for the empty bitstring. The expression $uw$, meaningful only when $u$ is finite, will stand for the concatenation of strings $u$ and $w$. We write $u\preceq w$\index{$\preceq$}\label{0preceq} to mean that $u$ is an initial segment of $w$. 
And $u\prec w$ means that $u$ is a proper initial segment of $w$, i.e. that $u\preceq w$ and $u\not=w$. 

\begin{definition}\label{tree}\  

1. A {\bf bitstring tree}\index{bitstring tree (BT)}\label{0bitstring tree (BT)} ({\bf BT}) is a nonempty set $T$ of bitstrings,  called the {\bf branches}\index{branch of a BT}\label{0branch of a BT} of the tree (with finite branches also called 
{\bf nodes}),\index{node of a BT}\label{0node of a BT} such that, for all bitstrings $w,u$, the following three conditions\footnote{Due to a mechanical error, the third condition was lost in the published version of \cite{Jap03}.} are satisfied:

\hspace{30pt} a) If $w\in T$ and $u\prec w$, then  $u\in T$;

\hspace{30pt} b) $w0\in T$ iff $w1\in T$ (finite $w$).

\hspace{30pt} c) If $w$ is infinite and all $u$ with $u\prec w$ are in $T$, then so is $w$.

2. A {\bf complete branch}\index{complete branch of a BT}\label{0complete branch of a BT} of a BT \ $T$ is a branch $w$ of $T$ such that for no bitstring $u$ with $w\prec u$ do we have $u\in T$. A finite complete branch of $T$ is also said to be a {\bf leaf}\index{leaf}\label{0leaf} of $T$. Notice that $T$ (as a set of strings) is finite  iff all of its branches (as strings) are so. Hence, the terms ``complete branch" and ``leaf" are synonymic for finite BTs, as are ``branch'' and ``node''.

3. A {\bf decoration for}\index{decoration}\label{0decoration} a finite BT \ $T$ is a function $d$ that sends each leaf of $T$ to some game.

4. A  {\bf decorated bitstring tree} ({\bf DBT})\index{decorated bitstring tree (DBT)}\label{0decorated bitstring tree (DBT)} ${\cal T}$ is a pair $(T,d)$, where $T$ --- called the {\bf BT-structure}\index{BT-structure}\label{0treestructure of a DBT} of $\cal T$ --- is a finite BT, and $d$ ---
 called the {\bf decoration of $\cal T$} --- is a decoration for $T$. Such a $\cal T$  is said to be a {\bf singleton}\index{singleton DBT}\label{0singleton DBT} iff $T=\{\epsilon\}$. We identify each singleton DBT $(\{\epsilon\},d)$ with the game $d(\epsilon)$, and vice versa: think of each game $A$ as the singleton DBT $(\{\epsilon\},d)$ with $d(\epsilon)=A$. 
In some contexts, on the other hand, we may identify a DBT \ $\cal T$ with its BT-structure $T$, and say ``branch (leaf, etc.) of $\cal T$'' to mean ``branch (leaf, etc.) of $T$''.  
\end{definition}

 In Figure 11 we see an example of a DBT whose BT-structure is \ $\{\epsilon, 0,1,00,01,010,011\}$ \ and 
whose decoration is the function $d$ given by \ $d(00)=\seq{\Phi,\Psi,\Lambda}\chess$, \ $d(010)=d(011)=\seq{\Phi,\Psi,\Sigma}\chess$,  \  
$d(1)=\seq{\Phi,\Theta,\Pi}\chess$.

Drawing actual trees for DBTs is not very convenient, and an alternative way to represent a DBT ${\cal T}=(T,d)$ is the following:

\begin{itemize}
\item If $\cal T$ is a singleton with $d(\epsilon)=A$, then ${\cal T}$ is simply written as $A$.
\item Otherwise, ${\cal T}$ is written as $E_0\circ E_1$,\index{$\circ$}\label{0circ} where $E_0$ and $E_1$ represent the sub-DBTs of ${\cal T}$ rooted at $0$ and $1$, respectively. 
\end{itemize}

For example, the DBT of Figure 11 will be written as 
\[\mbox{\small $\Bigl(\bigl(\seq{\Phi,\Psi,\Lambda}\chess\bigr) \circ \bigl((\seq{\Phi,\Psi,\Sigma}\chess)\circ (\seq{\Phi,\Psi,\Sigma}\chess)\bigr)\Bigr)\circ \Bigl(\seq{\Phi,\Theta,\Pi}\chess\Bigr).$}\]

We are going to define $\st$ and $\cost$ as operations applicable not only to games, i.e. singleton DBTs, but 
to any DBTs as well.

\begin{definition}\label{prleg}
Let ${\cal T}=(T,d)$ be DBT. We define the notion of a {\bf prelegal position}\index{prelegal position}\label{0prelegal position} of $\st{\cal T}$ (resp. $\cost{\cal T}$), together with the function 
{\em $\tree{\hspace{1pt}\sti{\cal T}}{}$}\index{$\tree{}{}$}\label{0trtr} (resp. {\em $\tree{\hspace{1pt}\costi{\cal T}}{}$}) that associates a BT  
{\em $\tree{\hspace{1pt}\sti{\cal T}}{\seq{\Phi}}$} 
(resp. {\em $\tree{\hspace{1pt}\costi{\cal T}}{\seq{\Phi}}$}) with each such position $\Phi$, by the following induction:\vspace{5pt}

a) \  $\emptyrun$ is a prelegal position of $\st{\cal T}$ (resp. $\cost{\cal T}$), and 
\[\begin{array}{ll}
& \mbox{\em $\tree{\hspace{1pt}\sti{\cal T}}{\emptyrun}=T$}\\
\mbox{(resp.} & \mbox{{\em $\tree{\hspace{1pt}\costi{\cal T}}{\emptyrun}=T$})}.
\end{array}\]

b) \  $\seq{\Phi,\lambda}$ is a prelegal position of $\st{\cal T}$ (resp. $\cost{\cal T}$) iff $\Phi$ is so and one of the following two conditions is satisfied: 
\begin{description}

\item[1.] $\lambda=\oo \col{w}$ (resp. $\lambda=\pp\col{w}$) for some  leaf $w$ of {\em $\tree{\hspace{1pt}\sti{\cal T}}{\seq{\Phi}}$} (resp. {\em $\tree{\hspace{1pt}\costi{\cal T}}{\seq{\Phi}}$}). We call this sort of a labmove $\lambda$ or move $\col{w}$ {\bf replicative}.\index{replicative (lab)move}\label{0replicative (lab)move} In this case 
\[\begin{array}{ll}
 & \mbox{\em $\tree{\hspace{1pt}\sti{\cal T}}{\seq{\Phi,\oo\col{w}}}=\tree{\hspace{1pt}\sti{\cal T}}{\seq{\Phi}}\cup\{w0,w1\}$}\\
\mbox{(resp.} & 
\mbox{{\em $\tree{\hspace{1pt}\costi{\cal T}}{\seq{\Phi,\pp\col{w}}}=\tree{\hspace{1pt}\costi{\cal T}}{\seq{\Phi}}\cup\{w0,w1\}$}).}
\end{array}\]

\item[2.] $\lambda=\xx w.\alpha$ for some node $w$ of {\em $\tree{\hspace{1pt}\sti{\cal T}}{\seq{\Phi}}$} (resp. {\em $\tree{\costi{\hspace{1pt}\cal T}}{\seq{\Phi}}$}), player $\xx$ and move $\alpha$. We call this sort of a labmove $\lambda$ or move $w.\alpha$
{\bf nonreplicative}.\index{nonreplicative (lab)move}\label{0nonreplicative (lab)move} In this case 
\[\begin{array}{ll}
 & \mbox{\em $\tree{\hspace{1pt}\sti{\cal T}}{\seq{\Phi,\xx w.\alpha}}=\tree{\hspace{1pt}\sti{\cal T}}{\seq{\Phi}}$}\\
\mbox{(resp.} & \mbox{{\em $\tree{\hspace{1pt}\costi{\cal T}}{\seq{\Phi,\xx w.\alpha}}=\tree{\hspace{1pt}\costi{\cal T}}{\seq{\Phi}}$})}.
\end{array}\]
\end{description}
\end{definition}

As mentioned earlier, with a visualization in the style of Figure 11 in mind, the meaning of a replicative labmove $\xx \col{w}$ is that player $\xx$ splits leaf/window $w$ into two children windows $w0$ and $w1$; and the meaning of a nonreplicative labmove $\xx w.\alpha$ is that  $\xx$ makes  the move $\alpha$ in all windows whose names start with $w$.  Prelegality is a minimum 
condition that every legal run of a $\st$- or $\cost$-game should satisfy. In particular, prelegality means that new windows have only been created by the ``right player'' (i.e. $\oo$ in a $\st$-game, and $\pp$ in a $\cost$-game), and that no moves have been made in not-yet-created windows. 
As for $\tree{\hspace{1pt}\sti{\cal T}}{\seq{\Phi}}$ or $\tree{\hspace{1pt}\costi{\cal T}}{\seq{\Phi}}$, it shows to what we will be referring as the {\bf underlying BT-structure}\index{underlying BT-structure}\label{0underlying BT-structure} of the position to which $\Phi$ brings the game down. 
Note that, as can be seen from the definition, whether $\Phi$ is a prelegal position of $\st{\cal T}$ or $\cost{\cal T}$ and what the value of $\tree{\hspace{1pt}\sti{\cal T}}{\seq{\Phi}}$ or $\tree{\hspace{1pt}\costi{\cal T}}{\seq{\Phi}}$ is, only depends on the 
BT-structure of ${\cal T}$ and not on its decoration. 

The concept of a prelegal position of $\st{\cal T}$ can be generalized to all runs by stipulating that a {\bf prelegal run\index{prelegal run}\label{0prelegal run}  of $\st{\cal T}$} is a run whose every finite initial segment is a prelegal position of $\st{\cal T}$. Similarly, the function $\tree{\hspace{1pt}\sti{\cal T}}{}$ can be extended to all prelegal runs of $\st{\cal T}$ by stipulating that $\tree{\hspace{1pt}\sti{\cal T}}{\seq{\lambda_1,\lambda_2,\lambda_3,\ldots}}$ is the smallest BT containing all elements of the set 
\begin{equation}\label{apr15}\tree{\hspace{1pt}\sti{\cal T}}{\emptyrun}\cup\tree{\hspace{1pt}\sti{\cal T}}{\seq{\lambda_1}}\cup \tree{\hspace{1pt}\sti{\cal T}}{\seq{\lambda_1,\lambda_2}}\cup
\tree{\hspace{1pt}\sti{\cal T}}{\seq{\lambda_1,\lambda_2,\lambda_3}}\cup \ldots
\end{equation}
of bitstrings. In other words, $\tree{\hspace{1pt}\sti{\cal T}}{\seq{\lambda_1,\lambda_2,\lambda_3,\ldots}}$ is the result of adding to (\ref{apr15}) every infinite bitstring $w$ such that all finite initial segments of $w$ are in (\ref{apr15}). The concept of a prelegal position of $\cost{\cal T}$ and the function $\tree{\hspace{1pt}\costi{\cal T}}{}$ generalize to infinite runs in a similar way.

We now introduce an important notational convention that should be remembered. Let $u$ be a bitstring and $\Gamma$ any run. Then
\[\Gamma^{\preceq u}\index{\(\Gamma^{\preceq u}\)}\label{susu}\] 
will stand for the result of first removing from $\Gamma$ all labmoves except those that look like
$\xx w.\alpha$ for some  bitstring $w$ with $w\preceq u$, and then deleting this sort of prefixes `$w.$' 
in the remaining labmoves, i.e. replacing each remaining labmove $\xx w.\alpha$ (where $w$ is a bitstring) by $\xx\alpha$.
Example: If $u=101000\ldots$ and $\Gamma=\seq{\pp \epsilon.\alpha_1,\oo \col{\epsilon}, \oo 1.\alpha_2, \pp 0.\alpha_3,\oo \col{1},\pp 10.\alpha_4}$, then $\Gamma^{\preceq u}=\seq{\pp\alpha_1, \oo\alpha_2,\pp \alpha_4}$. 

Being a prelegal run of $\st{\cal T}$ is a necessary but not a sufficient condition for being a legal run of this game.
For simplicity, let us consider the case when ${\cal T}$ is singleton DBT $A$, where $A$ is a constant game. 
It was noted earlier that a legal run $\Gamma$ of $\st A$ can be thought of as consisting of multiple  legal runs of $A$. In particular, these runs will be the runs $\Gamma^{\preceq u}$, where $u$ is a complete branch of $\tree{\sti A}{\seq{\Gamma}}$. The labmoves of $\Gamma^{\preceq u}$ for such a $u$ are those $\xx\alpha$ that emerged as a result of making (nonreplicative) labmoves of the form $\xx w.\alpha$ with $w\preceq u$. For example, to branch $010$ in \mbox{Figure 11} corresponds run $\seq{\Phi,\Psi,\Sigma}$, where the labmoves of $\Phi$ originate from the nonreplicative labmoves of the form $\xx \epsilon.\alpha$ (i.e. $\xx .\alpha$) made before the first replicative move, the labmoves of $\Psi$ originate from the nonreplicative labmoves of the form $\xx w.\alpha$ with $w\preceq 0$ made between the first and the second replicative moves, and the labmoves of $\Sigma$ originate from the nonreplicative labmoves of the form $\xx w.\alpha$ with $w\preceq 01$ made between the second and the third replicative moves. Generally, for a prelegal run $\Gamma$ of $\st A$ to be a legal run, it is necessary and sufficient that all of the runs $\Gamma^{\preceq u}$, where $u$ is a complete branch of $\tree{\sti A}{\seq{\Gamma}}$, be legal runs of $A$. And for such a $\Gamma$ to be a $\pp$-won run, it is necessary and sufficient that all of those $\Gamma^{\preceq u}$ be $\pp$-won runs of $A$. 

When ${\cal T}$ is a non-singleton DBT, the situation is similar. For example, for $\Gamma$ to be a legal (resp. $\pp$-won) run of $\st(\chess\hspace{1pt}\circ \checkers)$, along with being a prelegal run, it is necessary that, for every complete branch $0u$ of $\tree{\hspace{1pt}\sti\mbox{\scriptsize ({\em Chess\hspace{1pt}$\circ$Checkers})}}{\seq{\Gamma}}$, $\Gamma^{\preceq 0u}$ be a legal (resp. $\pp$-won) run of $\chess$ and, for every complete branch $1u$ of the same tree,  $\Gamma^{\preceq 1u}$ be a legal (resp. $\pp$-won) run of $\checkers$.

Finally, the case with $\cost{\cal T}$, of course, is symmetric to that with $\st{\cal T}$.

All of the above intuitions are summarized in the following formal definitions of $\st$ and $\cost$, with Definition \ref{legst1} being for the simpler case when $\cal T$ is a singleton, and 
Definition \ref{legst} generalizing it to all DBTs. In concordance with the earlier remark that considering constant games is sufficient when 
defining propositional connectives, Definition \ref{legst1} assumes that $A$ is a constant game, and Definition \ref{legst} assumes that $\cal T$ is a {\bf constant DBT},\index{constant DBT}\label{0constant DBT} meaning a DBT whose decoration sends every leaf of its BT-structure to a constant game. 

\begin{definition}\label{legst1} Assume $A$ is a constant game.\vspace{5pt}

1. {\bf Branching recurrence $\st A$:}\index{branching recurrence}\label{0branching recurrence}\index{$\st$}\label{0st} 

\begin{itemize}
\item $\Gamma\in\legal{\sti A}{}$  iff $\Gamma$ is a prelegal run of $\st A$, and $\Gamma^{\preceq u}\in\legal{A}{}$ for every complete branch $u$ of {\em $\tree{\sti A}{\seq{\Gamma}}$}.
\item Whenever  $\Gamma\in\legal{\sti A}{}$, 
$\win{\sti A}{}\seq{\Gamma}=\pp$ iff $\win{A}{}\seq{\Gamma^{\preceq u}}=\pp$  for every complete branch $u$ of 
{\em $\tree{\sti A}{\seq{\Gamma}}$}.
\end{itemize}

2. {\bf Branching corecurrence $\cost A$:}\index{branching corecurrence}\label{0branching corecurrence}\index{$\cost$}\label{0cost} 

\begin{itemize}
\item $\Gamma\in\legal{\costi A}{}$  iff $\Gamma$ is a prelegal run of $\cost A$, and $\Gamma^{\preceq u}\in\legal{A}{}$ for every complete branch $u$ of {\em $\tree{\costi A}{\seq{\Gamma}}$}.
\item Whenever  $\Gamma\in\legal{\costi A}{}$, 
$\win{\costi A}{}\seq{\Gamma}=\oo$ iff $\win{A}{}\seq{\Gamma^{\preceq u}}=\oo$  for every complete branch $u$ of 
{\em $\tree{\costi A}{\seq{\Gamma}}$}.
\end{itemize}
\end{definition}

\begin{definition}\label{legst} \ Assume ${\cal T}=(T,d)$ is a constant DBT.\vspace{5pt}

1. {\bf Branching recurrence $\st {\cal T}$:} 

\begin{itemize}
\item $\Gamma\in\legal{\sti {\cal T}}{}$  iff $\Gamma$ is a prelegal run of $\st{\cal T}$, and $\Gamma^{\preceq wu}\in\legal{d(w)}{}$ for every complete branch $wu$ of {\em $\tree{\sti{\cal T}}{\seq{\Gamma}}$} where $w$ is a leaf of $T$.
\item Whenever  $\Gamma\in\legal{\sti {\cal T}}{}$, 
$\win{\sti {\cal T}}{}\seq{\Gamma}=\pp$ iff $\win{d(w)}{}\seq{\Gamma^{\preceq wu}}=\pp$  for every complete branch $wu$ of 
{\em $\tree{\sti{\cal T}}{\seq{\Gamma}}$} where $w$ is a leaf of $T$. 
\end{itemize}

2. {\bf Branching corecurrence $\cost {\cal T}$:} 

\begin{itemize}
\item $\Gamma\in\legal{\costi {\cal T}}{}$  iff $\Gamma$ is a prelegal run of $\cost{\cal T}$, and $\Gamma^{\preceq wu}\in\legal{d(w)}{}$ for every complete branch $wu$ of {\em $\tree{\costi{\cal T}}{\seq{\Gamma}}$} where $w$ is a leaf of $T$.
\item Whenever  $\Gamma\in\legal{\costi {\cal T}}{}$, 
$\win{\costi {\cal T}}{}\seq{\Gamma}=\oo$ iff $\win{d(w)}{}\seq{\Gamma^{\preceq wu}}=\oo$  for every complete branch $wu$ of 
{\em $\tree{\costi{\cal T}}{\seq{\Gamma}}$} where $w$ is a leaf of $T$. 
\end{itemize}
\end{definition}

Let us not forget to make our already routine observation that the definition of either operation $\st,\cost$ can be obtained from the definition of its dual by just interchanging $\pp$ with $\oo$.

Now it would be interesting to see how the moves of unilegal runs affect $\st$- and $\cost$-games. In fact, being able to describe the effect of such moves was our main motivation for defining $\st $ and $\cost$ in the general form as 
operations on DBTs. We need two preliminary definitions here.

\begin{definition}\label{replication}
Suppose ${\cal T}=(T,d)$ is a DBT and $w$ is a leaf of $T$. We define 
{\em \(\rep_w[{\cal T}]\)}\index{$\rep$}\label{0rep}
as the following DBT $(T',d')$:\vspace{5pt}

\noindent 1. $T'=T\cup\{w0,w1\}$.\vspace{5pt}

\noindent 2. $d'$ is the decoration for $T'$ such that:

(a) $d'(w0)=d'(w1)=d(w)$;

(b) for every other ($\not=w0,w1$) leaf $u$ of $T'$, $d'(u)=d(u)$.\vspace{4pt}
\end{definition}

Examples:\vspace{4pt} 

1. \(\rep_{0}[A\circ(B\circ C)]\ =\ (A\circ A)\circ(B\circ C)\);\vspace{2pt} 

2. \(\rep_{10}[A\circ(B\circ C)]\ =\ A\circ\bigl((B\circ B)\circ C\bigr);\)\vspace{2pt}

3. \(\rep_{11}[A\circ(B\circ C)]\ =\ A\circ\bigl(B\circ (C\circ C)\bigr).\)\vspace{2pt}

\begin{definition}\label{mov}
Suppose ${\cal T}=(T,d)$ is a DBT, $w$ is a node of $T$ and, for every leaf $u$ of $T$ with $w\preceq u$, 
 $\seq{\lambda}$ is a unilegal position of $d(u)$. We define {\em \(\mov_{w}^{\lambda}[{\cal T}]\)}\index{$\mov$}\label{0mov}
as the DBT $(T,d')$, where $d'$ is the decoration for $T$ such that:\vspace{5pt}

(a) For every leaf $u$ of $T$ with $w\preceq u$, \ $d'(u)=\seq{\lambda}d(u)$;

(b) for every other leaf $u$ of $T$, \ $d'(u)=d(u)$.
\end{definition}

Examples (assuming the appropriate unilegality conditions on $\seq{\lambda}$):\vspace{4pt} 

1. \(\mov_{10}^{\lambda}[A\circ(B\circ C)]\ =\ A\circ(\seq{\lambda}B \circ C)\);\vspace{2pt}  

2.  \(\mov_{1}^{\lambda}[A\circ(B\circ C)]\ =\ A\circ(\seq{\lambda}B \circ \seq{\lambda}C)\);\vspace{2pt} 

3. \(\mov_{\epsilon}^{\lambda}[A\circ(B\circ C)]\ =\ \seq{\lambda}A\circ(\seq{\lambda}B \circ \seq{\lambda}C)\).\vspace{8pt} 

The following theorem is a combination of Propositions 13.5 and 13.8 proven in \cite{Jap03}:

\begin{theorem}\label{pr80}
Suppose ${\cal T}=(T,d)$ is a DBT, and $\lambda$ any labmove.\vspace{5pt}  

\noindent 1. $\seq{\lambda}\in\Legal{\sti {\cal T}}$ iff one of the following conditions holds: 

(a) $\lambda$ is (the replicative labmove) $\oo \col{w}$, where $w$ is a leaf of $T$. In this case $\seq{\oo\col{w}}\st{\cal T}=\st  \mbox{Rep}_w[{\cal T}]$.

(b) $\lambda$ is (the nonreplicative labmove) $\xx w.\alpha$, where $\xx$ is either player, $w$ is a node of $T$ and, for every leaf $u$ of $T$ with $w\preceq u$, $\seq{\xx\alpha}$ is a unilegal position of $d(u)$. In this case $\seq{\xx w.\alpha}\st {\cal T}=\st \mbox{Nonrep}_{w}^{\xx\alpha}[{\cal T}]$.\vspace{5pt}

\noindent 2. $\seq{\lambda}\in\Legal{\costi {\cal T}}$ iff one of the following conditions holds: 

(a) $\lambda$ is (the replicative labmove) $\pp \col{w}$, where $w$ is a leaf of $T$. In this case $\seq{\pp\col{w}}\cost{\cal T}=\cost  \mbox{Rep}_w[{\cal T}]$.

(b) $\lambda$ is (the nonreplicative labmove) $\xx w.\alpha$, where $\xx$ is either player, $w$ is a node of $T$ and, for every leaf $u$ of $T$ with $w\preceq u$, $\seq{\xx\alpha}$ is a unilegal position of $d(u)$. In this case $\seq{\xx w.\alpha}\cost {\cal T}=\cost \mbox{Nonrep}_{w}^{\xx\alpha}[{\cal T}]$.
\end{theorem}

This theorem allows us to easily test whether a given run is a (uni)legal run of a given $\st$- or $\cost$-game, and if it is, to write out the corresponding sequence of games.

\begin{example}\label{ex10}
{\em Let $\Gamma=\seq{\oo \col{\epsilon},\oo 0.3,\pp 0.9,\oo \col{1},\oo 10.1,\pp 10.1}$, and $A_0=\st\ada x\ade y (y=x^2)$. In view of Theorem \ref{pr80}, $\Gamma$ is legal for $A_0$, and it brings the latter  
down to game $A_6$ as shown below:\vspace{5pt}

\noindent\(\begin{array}{ll}
\mbox{$A_0$:  } \st\Bigl(\ada x\ade y (y=x^2)\Bigr); & \\
\mbox{$A_1$:  } \st\Bigl(\ada x\ade y (y=x^2)\circ \ada x\ade y (y=x^2)\Bigr), & \mbox{i.e. $\seq{\oo\col{\epsilon}}A_0$}; \\
\mbox{$A_2$:  } \st\Bigl(\ade y (y=3^2)\circ \ada x\ade y (y=x^2)\Bigr), & \mbox{i.e. $\seq{\oo 0.3}A_1$};\\
\mbox{$A_3$:  } \st\Bigl((9=3^2)\circ \ada x\ade y (y=x^2)\Bigr), & \mbox{i.e. $\seq{\pp 0.9}A_2$};\\
\mbox{$A_4$:  } \st\Bigl((9=3^2)\circ\bigl( \ada x\ade y (y=x^2)\circ\ada x\ade y (y=x^2)\bigr)\Bigr), & \mbox{i.e. $\seq{\oo\col{1}}A_3$};\\
\mbox{$A_5$:  } \st\Bigl((9=3^2)\circ\bigl( \ade y (y=1^2)\circ\ada x\ade y (y=x^2)\bigr)\Bigr), & \mbox{i.e. $\seq{\oo 10.1}A_4$};\\
\mbox{$A_6$:  } \st\Bigl((9=3^2)\circ\bigl((1=1^2)\circ\ada x\ade y (y=x^2)\bigr)\Bigr), & \mbox{i.e. $\seq{\pp 10.1}A_5$}.\vspace{5pt}
\end{array}\)

The empty run $\emptyrun$ is a $\pp$-won run of each of the three $\circ$-components of $A_6$. It can be easily seen that then (and only then) $\emptyrun$ is a $\pp$-won run of $A_6$,
for $\st( \ldots\circ\ldots)$ essentially acts as parallel conjunction.  Hence $\Gamma$ is a $\pp$-won run of $A_0$. 
}\end{example}

The run that we see in the above example, though technically different, is still ``essentially the same'' as the one from Example \ref{apr18}. Indeed, as noted earlier, $\st$ and $\pst$ are 
equivalent when applied to traditional, low-interactivity problems such as $\ada x\ade y (y=x^2)$. What makes the resource $\st A$ stronger than $\pst A$ is 
$\oo$'s ability to try several different responses to a same move by $\pp$. In $\st \ada x\ade y (y=x^2)$, however, $\oo$ cannot take advantage of this flexibility because 
there are no legal runs of $\ada x\ade y (y=x^2)$ where $\oo$'s moves follow $\pp$'s moves. 

To get a feel of the substantial difference between $\st$ and $\pst$, let us consider, for simplicity, the {\bf bounded versions} $\st^b$, $\cost^b$, $\pst^b$, $\pcost^b$ of our recurrence operations. Here $b$ is a positive integer, 
setting the bound on the number of parallel plays of game $A$ that can be generated in a legal run of $\st A$ ($\cost A$, $\pst A$, $\pcost A$). That is, $\pst^b A$ and $\pcost^b A$ are nothing but the parallel conjunction and parallel disjunction of $b$ copies of $A$, respectively. And $\st^b A$ and $\cost^b A$ are defined as $\st A$ and $\cost A$, with the only difference that, in a legal run $\Gamma$,  a replicative move can be made at most $b-1$ times, so that $\tree{\sti^b A}{\seq{\Gamma}}$ or $\tree{\costi^b A}{\seq{\Gamma}}$ will have at most $b$ complete branches. 

We want to compare $\cost^2 D$ with $\pcost^2 D$, i.e. with $D\mld D$, where  
\[D\ = (\chess\add\gneg\chess)\adc(\checkers\add\gneg\checkers).\]

Winning $D\mld D$ is not easy for $\pp$ unless $\pp$ is a champion in either chess or checkers. Indeed, a smart environment may choose the left $\adc$-conjunct in the left occurrence of $D$ in $D\mld D$ while choose the right $\adc$-conjunct in the right occurrence. This will bring the game down to 
\[(\chess\add\gneg\chess)\mld(\checkers\add\gneg\checkers).\]
$\pp$ in trouble now. It can, say, make the moves `$1.1$' and `$2.2$, bringing the game down to $\chess\mld \gneg \checkers$. This will not help much though, as winning $\chess\mld \gneg \checkers$, unlike $\chess\mld\gneg\chess$,
is not any easier than winning either disjunct in isolation.

On the other hand, $\pp$ does have a nice winning strategy for $\cost^2 D$. At the beginning, $\pp$ waits till $\oo$ chooses one of the two $\adc$-conjuncts of $D$. This brings the game down to, say, $\cost^2(\chess\add\gneg\chess)$. 
Then and only then, $\pp$ 
makes a replicative move, thus creating two copies of $\chess\add\gneg\chess$. In one copy $\pp$ chooses the left $\add$-disjunct, and in the other copy chooses the right $\add$-disjunct. 
Now 
the game will have evolved to $\cost^1(\chess\circ\gneg\chess)$. With $\cost^1(A\circ\gneg A)$ essentially being nothing but $A\mld\gneg A$, mimicking in $\chess$ the moves made 
by $\oo$ in $\gneg\chess$ and vice versa guarantees a success for $\pp$. Among the runs consistent with this strategy is 
\[\seq{\oo .1, \pp \col{\epsilon},\pp 0.1,\pp 1.2,\oo 1.\alpha_1,\pp 0.\alpha_1,\oo 0.\alpha_2,\pp 1.\alpha_2,\oo 1.\alpha_3,\pp 0.\alpha_3,\ldots},\]
to which corresponds the following sequence of games:\vspace{5pt}

\noindent\(\begin{array}{l}
\cost^2 ((\chess\add\gneg\chess)\adc(\checkers\add\gneg\checkers));\\
\cost^2 (\chess\add\gneg\chess);\\
\cost^1 ((\chess\add\gneg\chess)\circ (\chess\add\gneg\chess));\\
\cost^1 (\chess\circ (\chess\add\gneg\chess));\\
\cost^1 (\chess\circ \gneg\chess);\\
\cost^1 (\chess\circ \seq{\oo\alpha_1}\gneg\chess);\\
\cost^1 (\seq{\pp\alpha_1}\chess\circ \seq{\oo\alpha_1}\gneg\chess);\\
\cost^1 (\seq{\pp\alpha_1,\oo\alpha_2}\chess\circ \seq{\oo\alpha_1}\gneg\chess);\\
\cost^1 (\seq{\pp\alpha_1,\oo\alpha_2}\chess\circ \seq{\oo\alpha_1,\pp\alpha_2}\gneg\chess);\\
\cost^1 (\seq{\pp\alpha_1,\oo\alpha_2}\chess\circ \seq{\oo\alpha_1,\pp\alpha_2,\oo\alpha_3}\gneg\chess);\\
\cost^1 (\seq{\pp\alpha_1,\oo\alpha_2,\pp\alpha_3}\chess\circ \seq{\oo\alpha_1,\pp\alpha_2,\oo\alpha_3}\gneg\chess);\\
\mbox{\Large $\cdots$}\vspace{5pt}
\end{array}\)

As we are going to see later, affine logic is sound with respect to the semantics of computability logic no matter whether the exponential operators $!,?$ of the former are understood as 
$\pst,\pcost$ or $\st,\cost$. Thus, affine logic cannot distinguish between the two groups of recurrence operations. But computability logic certainly sees a difference. As noted earlier, 
it validates \[(P\intimpl R)\adc (Q\intimpl R)\intimpl (P\add Q\intimpl R)\] while makes \[(P\pintimpl R)\adc (Q\pintimpl R)\pintimpl (P\add Q\pintimpl R)\] fail. 
 Here are two other examples of
principles that can be shown to be valid with one sort of recurrence while invalid with the other sort: 

\[\begin{array}{ll}
\st(P\add Q)\mli \st P\add\st Q & \mbox{\em  \ is valid};\\
\pst(P\add Q)\mli \pst P\add\pst Q & \mbox{\em  \ is not}.\vspace{5pt}\\
P\mlc \pst(P\mli Q\mlc P)\mli\pst Q & \mbox{\em  \ is valid};\\
P\mlc \st(P\mli Q\mlc P)\mli\st Q & \mbox{\em  \ is not}.
\end{array}\]    

As for how the strengths of $\st$ and $\pst$ relate, as we may guess, the situation is:

\[\begin{array}{ll}
\st P\mli \pst P & \mbox{\em  \ is valid};\\
\pst P\mli\st P & \mbox{\em  \ is not}.
\end{array}\]    

\section{Static games}\label{ss5}

Our games are obviously general enough to model anything that one would call a (two-agent, 
two-outcome) interactive problem.  However, they are a bit too general. There are games where the chances of a player to succeed essentially depend on the relative speed at which its adversary acts. A simple example would be the following game: 

  \begin{center}
\begin{picture}(136,100)

\put(69,92){\circle{16}}
\put(65,88){$\pp$}
\put(69,84){\line(-1,-1){34}}
\put(34,63){\scriptsize $\pp \alpha$}
\put(69,84){\line(1,-1){34}}
\put(92,63){\scriptsize $\oo \beta$}

\put(34,42){\circle{16}}
\put(30,38){$\pp$}

\put(103,42){\circle{16}}
\put(99,38){$\oo$}

\put(-6,10){{\bf Figure 12:} A non-static game}
\end{picture}
\end{center}
One cannot ask which player has a winning strategy here, for this game is a contest of speed rather than intellect: the winner will be whoever is fast enough to move first. 
CL does not want to consider this sort of games meaningful computational problems, and restricts its attention to the subclass of games that it calls {\em static}. Intuitively, static games
are ones where speed is irrelevant: in order to win, for either player only matters {\em what} to do (strategy) rather than {\em how fast} to do (speed). `These are games where, 
roughly speaking, it never hurts a player to postpone making moves'.\footnote{From the American Mathematical Society review of \cite{Jap03} by Andreas Blass.}

Static games are defined in terms of the auxiliary concept of {\em $\xx$-delay}. The notation $\Gamma^\pp$ used below means the result of deleting from $\Gamma$ all $\oo$-labeled (lab)moves. Symmetrically for $\Gamma^\oo$. 

\begin{definition}\label{delay} 
Let $\xx$ be either player, and $\Gamma,\Delta$ arbitrary runs. We say that $\Delta$ is a {\bf $\xx$-delay}\index{delay}\label{0delay} of $\Gamma$ iff the following two conditions are satisfied:

1. $\Delta^\pp=\Gamma^\pp$ and $\Delta^\oo=\Gamma^\oo$;

2. for any $k,n\geq 1$, if the {\em $k$th} $\pneg\xx$-labeled move is made earlier than (is to the left of) the {\em $n$th} $\xx$-labeled move in $\Gamma$, then so is it in $\Delta$. 

\end{definition}

Intuitively, ``$\Delta$ is a $\xx$-delay of $\Gamma$'' means that in $\Delta$ both players have played the same way as in $\Gamma$ (condition 1), only, in $\Delta$,  $\xx$ might have been acting with some delay, i.e. slower than in $\Gamma$ (condition 2). In more technical terms, $\Delta$ is the result of shifting in $\Gamma$ some (maybe all, maybe none) 
$\xx$-labeled moves to the right; in the process of shifting,  $\xx$-labeled moves can jump over $\pneg\xx$-labeled moves, but a $\xx$-labeled move can never jump over another $\xx$-labeled move. For example, the 
run \(\Gamma=\seq{\pp\alpha, \oo\beta,\pp\gamma,\oo\delta}\) has exactly the following five $\pp$-delays:
\[\begin{array}{l} 
\Delta_1=\seq{\pp\alpha, \oo\beta,\pp\gamma,\oo\delta} \ (=\Gamma);\\
\Delta_2=\seq{\oo\beta,\pp\alpha,\pp\gamma,\oo\delta};\\
\Delta_3=\seq{\pp\alpha, \oo\beta,\oo\delta, \pp\gamma};\\
\Delta_4=\seq{\oo\beta,\pp\alpha,\oo\delta, \pp\gamma};\\
\Delta_5=\seq{\oo\beta,\oo\delta, \pp\alpha,\pp\gamma}.
\end{array}\]

\begin{definition}\label{static}
A constant game $A$ is said to be {\bf static}\index{static game}\label{0static game} iff, for any player $\xx$ and any runs $\Gamma,\Delta$ such that $\Delta$ is a $\xx$-delay of $\Gamma$, we have:
\[\mbox{if $\win{A}{}\seq{\Gamma}=\xx$, then $\win{A}{}\seq{\Delta}=\xx$.}\]
This definition generalizes to all games by stipulating that a not-nece\-ssarily-constant game is {\bf static} iff every instance of it is so.  
\end{definition}

Looking at the game of Figure 12, $\seq{\oo \beta,\pp\alpha}$ is a $\pp$-delay of $\seq{\pp\alpha,\oo\beta}$. The latter is $\pp$-won while the former is not. So, that game is not static. On the other hand, all of the other examples of games we have seen or will see in this paper are static. This is no surprise. 
In view of the following theorem, the closure of the set of all strict games --- including all predicates --- under all of our game operations forms a natural 
family of static games:

\begin{theorem}\label{apr20} \ 

1. Every strict game (and hence every elementary game) is static.

2. Every game operation defined in this paper  
preserves the static property of games. 
\end{theorem} 

\begin{proof}  That all strict games are static has been proven in \cite{Jap03} (Proposition 4.8); and, of course, every elementary game is trivially strict. This takes care of clause 1. 
As for clause 2, it is a part of Theorem 14.1 of \cite{Jap03}. Even though the operations $\pst,\pcost,\mla,\mle$ were not officially introduced in \cite{Jap03}, 
they can be handled in exactly the same way as $\mlc,\mld$. \end{proof}
 
See Section 4 of \cite{Jap03} for arguments in favor of the belief that static games are adequate formal counterparts of our intuition of ``pure'', speed-independent interactive computational problems. Based on that belief, CL uses the terms ``static game'' and (interactive) ``{\bf computational problem}''\index{computational problem}\label{0computational problem3} as synonyms. We have been informally using the concept of validity, which in  
intuitive terms was characterized as being a scheme of ``always computable'' problems. As will be seen from the formal definition of validity given in Section \ref{ss7}, the exact meaning of 
a ``problem'' is a static --- rather than any --- game.   

All of the examples of winning strategies that we have seen so far shared one feature: for every position, the strategy had a strict prescription for a player regarding whether it 
should move or wait till the adversary makes a move. This might have given us the wrong illusion of being a common case, somehow meaning that static games, even when properly free, still can  
 always be ``adequately'' modeled as strict games.  Not really. Below is an example, borrowed from \cite{Jap03}, of an inherently free static game. The winning strategy for it substantially takes advantage of the flexibility offered by the free-game approach: the fact that it is not necessary for a player to precisely know  
whether in a given position it needs to move or wait. Any attempt to model such a game as a strict game would signify counterfeiting the true essence of the interactive computational problem that it represents.    

\begin{example}\label{exx}
{\em Let $A(x,z)$ be a decidable arithmetical predicate such that the predicate $\cla zA(x,z)$ is undecidable, and  let $B(x,y)$ be an undecidable arithmetical predicate. Consider the following computational problem:
\[\ada x \Bigl(\ada y\bigl(\cla zA(x,z)\mlc B(x,y)\bigr)\adc \ada z A(x,z)\  \mli\    
\cla zA(x,z)\mlc\ada y B(x,y) \Bigr).\]

After $\oo$ specifies a value $m$ for $x$, $\pp$ will seemingly have to decide what to do: to watch or to think. 
The `watch' choice is to wait till $\oo$ specifies a value $k$ for $y$ in the consequent, after which $\pp$ can select the $\adc$-conjunct $\ada y\bigl(\cla zA(m,z)\mlc B(m,y)\bigr)$ in the antecedent and specify $y$ as $k$ in it, thus bringing the play down to the always-won elementary game 
$\cla zA(m,z)\mlc B(m,k) \mli\cla zA(m,z)\mlc B(m,k)$.
 While being successful if $\cla zA(m,z)$ is true, the watch strategy is a bad choice when
$\cla zA(m,z)$ is false,
for  there is no guarantee that $\oo$ will indeed make a move in $\ada y B(m,y)$, and if not, the game will be lost.
When $\cla z A(m,z)$ is false, the following `think' strategy is successful: Start looking for a number $n$ for which $A(m,n)$ is false. This can be done by testing $A(m,z)$, in turn, for $z=1$, $z=2$, ... After you find $n$, select the $\adc$-conjunct $\ada zA(m,z)$ in the antecedent, specify $z$ as $n$ in it, and you are the winner. The trouble is that if $\cla z A(m,z)$ is true, such a number $n$ will never be found. Thus, which of the above two choices (watch or think) would be successful depends on whether $\cla z A(m,z)$ is true or false, and since $\cla z A(x,z)$ is undecidable, $\pp$ has no effective way to make the right choice. Fortunately, there is no need to choose. Rather, these two strategies can be pursued simultaneously: $\pp$  starts looking for a number $n$ which makes $A(m,n)$ false and, at the same time, periodically checks if $\oo$ has made a move in $\ada y B(m,y)$. If the number $n$ is found before $\oo$ makes such a move, $\pp$ continues as prescribed by the think strategy; if vice versa, $\pp$ continues as prescribed by the watch strategy; finally, if none of these two events ever occur, which, note, is only possible when $\cla z A(m,z)$ is true (for otherwise a number $n$ falsifying $A(m,n)$ would have been found), again $\pp$ will be the winner. This is so because, just as in the corresponding scenario of the watch strategy, $\pp$ will have won both of the conjuncts of the consequent. 
}\end{example}

\section{Winnability}\label{icp}

 Now that we know what computational problems are, it is time to explain what {\em computability}, i.e. {\em algorithmic solvability}, i.e. existence of an {\em algorithmic winning strategy} exactly means. The definitions given in this section are semiformal. The omitted technical details are rather standard or irrelevant, and can be easily restored by anyone familiar with Turing machines. 
If necessary, the corresponding detailed definitions can be found in Part II of \cite{Jap03}.

As we remember, the central point of our philosophy is to require that player $\pp$ (here identified with its {\em strategy}) be implementable as a computer program, with effective and fully determined behavior. On the other hand, the behavior of  $\oo$, including its speed, can be arbitrary. This intuition is captured by the model of interactive computation where $\pp$ is formalized as what we call 
{\bf HPM}.\index{HPM}\label{0HPM}\footnote{HPM stands for `Hard-Play Machine'. See \cite{Jap03} for a (little long) story about why ``hard". The name EPM for the other model defined shortly stands for ``Easy-Play Machine".} 

An HPM $\cal H$ is a Turing machine which, together with an ordinary read/write {\em work tape},\index{work tape}\label{0work tape} has two additional, read-only tapes: the {\em valuation tape}\index{valuation tape}\label{0valuation tape} and the {\em run tape}.\index{run tape}\label{0run tape}  The presence of these two tapes is related to the fact that 
the outcome of a play over a given game depends on two parameters: (1) the valuation that tells us which instance of the game is played, and (2) the run that is generated 
in the play. $\cal H$ should have full access to information about these two parameters, and this information is provided 
by the valuation and run tapes: the former spells a (the ``actual") valuation $e$ by listing constants in the lexicographic order of the corresponding variables, and the latter spells, at any given time, the current position, i.e. the sequence of the (labeled) moves made by the two players so far. 
Thus, both of these two tapes can be considered input tapes. The reason for our choice to keep them separate is the difference in the nature of the input that they provide. Valuation is a {\em static} input, known at the very beginning of a computation/play and remaining unchanged throughout the subsequent process. On the other hand, the input provided by the run tape is {\em dynamic}: 
every time one of the players makes a move, the move (with the corresponding label) is appended to the content of this  
tape, with such content being unknown and hence blank at the beginning of interaction. Technically the run tape is read-only: the machine has unlimited read access to this (as well as to the valuation) tape, but it cannot write directly on it. 
Rather, $\cal H$ makes a move $\alpha$ by constructing it at the beginning of its work tape, delimiting its end with a 
blank symbol, and entering one of the specially designated states called {\em move states}.\index{move state}\label{0move state} Once this happens, $\pp\alpha$ is automatically appended to  the current  position spelled on the run tape. 
While the frequency at which the machine can make moves is naturally limited by its clock cycle time (the time each computation step takes), there are no limitations to how often the environment can make a move, so, during one computation step of the machine, any finite number of any moves by the environment can be appended to 
the content of the run tape. This corresponds to the earlier-pointed-out intuition that not only the strategy, but also the relative speed of the  environment can be arbitrary. For technical clarity, we assume that the run tape remains stable during a clock cycle, and is updated only on a transition from one cycle to another. Specifically, where $\seq{\Phi}$ is the position spelled on the run tape during a given cycle and $\alpha_1,\ldots,\alpha_n$ (possibly $n=0$) is the sequence of the moves made by the environment during the cycle,
the content of the run tape throughout the next cycle will be either $\seq{\Phi,\oo\alpha_1,\ldots,\oo\alpha_n,\pp\beta}$ or 
$\seq{\Phi,\oo\alpha_1,\ldots,\oo\alpha_n}$, depending on whether the machine did or did not make a move $\beta$ during the previous cycle. Such a transition is thus nondeterministic, with nondeterminism being related to the different possibilities for the above sequence $\alpha_1,\ldots,\alpha_n$. 

A {\em configuration}\index{configuration}\label{0configuration} of an HPM \ $\cal H$ is defined in the standard way: this is a full description of the (``current") state of the machine, the locations of its three scanning heads
and the contents of its tapes, with the exception that, in order to make finite descriptions of configurations possible, we do not formally include a description of the unchanging 
(and possibly essentially infinite) content of the valuation tape as a part of configuration, but rather account for it in our definition of computation branch as this will be seen below. 
The {\em initial configuration}\index{initial configuration}\label{0initial configuration} is the configuration where $\cal H$ is in its start state and the work and run tapes are empty. A configuration $C'$ of $\cal H$ is said to be an {\bf $e$-successor}\index{successor configuration}\label{0successor configuration} of configuration $C$ if, when valuation $e$ is spelled on the valuation tape,  $C'$ can legally follow $C$ in the standard  sense, based on the transition function (which we assume to be deterministic) of the machine and accounting for the possibility of the above-described nondeterministic updates of the content of the run tape. An {\bf $e$-computation branch}\index{computation branch}\label{0computation branch} of $\cal H$ is a sequence of configurations of $\cal H$ where the first
configuration is the   initial configuration and each other configuration is an $e$-successor of the previous one.
Thus, the set of all $e$-computation branches captures all possible scenarios (on valuation $e$) corresponding to different behaviors by $\oo$.
Each $e$-computation branch $B$ of $\cal H$ incrementally spells --- in the sense that must be clear --- a run $\Gamma$ on the run tape, which we call the {\bf run spelled by $B$}.\index{run spelled by a computation branch}\label{0run spelled by a computation branch} 

\begin{definition}\label{feb2}
For games $A$ and $B$ we say that:

1. An HPM $\cal H$ {\bf wins 
$A$ on a valuation $e$}\index{win}\label{0win} iff,  whenever 
$\Gamma$ is the run spelled by some $e$-computation branch of $\cal H$, $\win{A}{e}\seq{\Gamma}=\pp$.  

2. An HPM $\cal H$ (simply) {\bf wins}
$A$ iff it wins $A$ on every valuation. 

3. $A$ is {\bf winnable}\index{winnable}\label{0winnable} 
iff there is an HPM that wins $A$. Such an HPM is said to be a {\bf solution}\index{solution}\label{0solution} for $A$. 

4. $A$ is \ {\bf reducible}\index{reducible}\label{0reducible}  to $B$ iff $B\mli A$ is winnable.  An HPM that wins $B\mli A$ is said to be a {\bf reduction}\index{reduction (as an HPM)}\label{0reduction (as an HPM)} of $A$ to $B$.

5. $A$ and $B$ are {\bf equivalent}\index{equivalent games}\label{0equivalent games} iff $A$ is reducible to $B$ and vice versa.  
\end{definition}

The HPM model of interactive computation seemingly strongly favors the environment in that the latter may 
be arbitrarily faster than the machine. What happens if we start limiting the speed of the environment? The answer is
 {\em nothing} as far as computational problems, i.e. static games, are concerned. The alternative model of computation called 
EPM takes the idea if limiting the speed of the environment to the extreme by always letting the machine to decide when
the environment can move and when it should wait; yet, as it turns out, the EPM model yields the same class of winnable static games as the HPM model does. 

An {\bf EPM}\index{EPM}\label{0EPM} is a machine defined in the same way as an HPM, with the only difference that now the environment can (but is not obligated to) make
a move only when the machine explicitly allows it to do so, the event called {\bf granting permission}.\index{granting permission}\label{0granting permission} Technically permission is granted by entering one of the specially designated states called {\bf permission states}.\index{permission state}\label{0permission state} The only requirement that the machine is expected to satisfy is that, as long as the adversary is playing legal, the machine should grant permission every once in a while; how long that ``while" lasts, however, is totally up to the machine. This amounts to having full control over the speed of the adversary. 

The above intuition is formalized as follows. After correspondingly redefining the `$e$-successor' relation --- in particular, 
accounting for the condition that now a (one single) $\oo$-labeled move may be appended to the contents of the run tape only when the machine is in a permission state --- the concepts of an 
{\em $e$-computation 
branch}\index{computation branch}\label{0computation branch2} of an EPM, the {\em run spelled} by such a branch, etc. are defined in exactly the same way as for HPMs. We say that 
a computation  branch $B$ of an EPM is {\bf fair}\index{fair computation branch}\label{0fair computation branch} iff permission is granted infinitely many times in $B$. 

\begin{definition}\label{feb2d}
For a game $A$ and an EPM $\cal E$, we say that:

1. $\cal E$ {\bf wins $A$ on a valuation $e$} iff, whenever $\Gamma$ is the run spelled by some $e$-computation branch $B$ of $\cal E$, unless $\Gamma$ is a
$\oo$-illegal run of $e[A]$,  $B$ is fair and $\win{A}{e}\seq{\Gamma}=\pp$.

2. $\cal E$ (simply) {\bf wins} $A$ iff it wins $A$ on every valuation.  
\end{definition}

We will be using the expressions \{HPMs\}\index{\{HPMs\}}\label{0hpms} and \{EPMs\}\index{\{EPMs\}}\label{0epms} for the sets of all HPMs and all EPMs, respectively.

The following fact, proven in \cite{Jap03} (Theorem 17.2), establishes equivalence between the two models of computation for static games:

\begin{theorem}\label{eq}
There is an effective function $f:$ \{EPMs\}$\longrightarrow$\{HPMs\} such that, for every EPM $\cal E$ and static game $A$ (and valuation $e$), whenever $\cal E$ wins $A$ (on $e$), so does $f({\cal E})$. 
And vice versa:  there is an effective function $f:$ \{HPMs\}$\longrightarrow$\{EPMs\} such that, for every HPM $\cal H$ and static game $A$ (and valuation $e$), whenever $\cal H$ wins $A$ (on $e$), so does $f({\cal H})$. 
 \end{theorem} 

The philosophical significance of this theorem is that it reinforces the belief according to which static games are games that allow us to make full abstraction from speed. Its technical importance is related to the fact that the EPM-model is much more convenient when it comes to describing interactive algorithms/strategies, as we will have a chance to see later.  
The two models also act as natural complements to each other: as shown in Section 20 of \cite{Jap03}, we can meaningfully talk about the (uniquely determined) play between a given HPM and a given EPM, while this is impossible if both players are EPMs or both are HPMs. This fact has been essentially exploited in the completeness theorems for
logic $\predell$ and its fragments proven in \cite{CL1,CL2,CL3,CL4}, where environment's strategies for the games represented by unprovable formulas were described as  
EPMs and then it was shown that no HPM can win against such EPMs. 

In view of Theorem \ref{eq}, winnability of a static game $A$ can be equivalently defined as existence of an EPM (rather than HPM) that wins $A$. Since we are only concerned with static games, from now on we will treat either definition as 
an equally official definition of winnability. And we extend the usage of the terms {\bf solution}\index{solution}\label{0solution2} and {\bf reduction}\index{reduction (as an EPM)}\label{0reduction (as an EPM)} (Definition \ref{feb2}) from HPMs to EPMs. 
For a static game $A$, valuation $e$ and HPM or EPM $\cal M$,  we write \[{\cal M}\models_e A, \ \ {\cal M}\models A\mbox{\ \ and\ \ } \models A
\index{$\models$}\label{0models}\] to mean that $\cal M$ wins $A$ on valuation $e$, that $\cal M$ (simply) wins $A$ and that $A$ is winnable, respectively. Also, we will be using the terms ``{\bf computable}''\index{computable}\label{0computable} or ``{\bf algorithmically solvable}''\index{algorithmically solvable}\label{0algorithmically solvable} as synonyms of ``winnable''.

One might guess that, just as the ordinary Turing machine model, our HPM and EPM models are highly rigid with respect to reasonable technical variations. For example, the models where only environment's moves are visible to the machine  
yield the same class of winnable static games. Similarly, there is no difference between whether we allow the scanning heads on the valuation and run tapes to move in either or only one (left to right) direction. Another variation is the one where an attempt by either 
player to make an illegal move has no effect: such moves are automatically rejected and/or filtered out by some interface hardware or software and thus illegal runs are never generated. Obviously in such a case a minimum requirement would be that 
the question of legality of moves be decidable (which is indeed ``very easily decidable'' for naturally emerging games, including all games from the closure of the set of predicates under all of our game operations). This again yields models equivalent to HPM and/or EPM. 

\section{Validity}\label{ss7}
While winnability is a property of games, validity is a property of logical formulas, meaning that the formula is a scheme of winnable static games. To define this concept formally, we need to agree on a formal language first. It is going to be an extension of the language of classical predicate calculus without identity or functional symbols. Our language is more expressive than the latter not only because it has non-classical operators in it such as $\adc,\ada,\st$ etc., but also due to the fact that we now have two sorts of atoms: {\em elementary}\index{elementary atom}\label{0elementary atom} and {\em general}.\index{general atom}\label{0general atom} Elementary atoms represent elementary games, i.e. predicates, while general atoms represent any computational problems, i.e. any (not-necessarily-elementary) static games. The point is that elementary problems are interesting and meaningful in their own right, and validate principles that may not be valid in the general case. We want to be able to analyze games at a reasonably fine level, which is among the main reasons for our choice to have the two sorts of atoms in the language.

More formally, for each integer $n\geq 0$, our language has infinitely many $n$-ary\index{arity of a letter}
\label{0arity of a letter}
 {\bf elementary letters}\index{elementary letter}\label{0elementary letter} and $n$-ary {\bf general letters}.\index{general letter}\label{0general letter} Elementary letters are what is called {\em predicate letters}\index{predicate letter}\label{0predicate letter} in ordinary logic.  We will consistently use the lowercase $p,q,r,s$ as metavariables for elementary letters, and the uppercase $P,Q,R,S$ as metavariables for general letters.  A {\bf nonlogical}\index{non-logical atom}\label{0non-logical atom} {\bf atom}\index{atom}\label{0atom} is $L(t_1,\ldots,t_n)$, where $L$ is an $n$-ary elementary or general letter, and each $t_i$ is a {\bf term},\index{term}\label{0term2} i.e. one of the {\bf variables}\index{variable}\label{0variable2} $v_1,v_2,v_3,\ldots$ or one of the {\bf constants}\index{constant}\label{0constant2} $1,2,3,\ldots$. Such an atom $L(t_1,\ldots,t_n)$ is said to be {\bf $L$-based}. When $L$ is $0$-ary,\index{arity of an atom}
\label{0arity of an atom}
 the only $L$-based atom will be usually written as $L$ rather than $L()$. An $L$-based atom is said to be elementary, general, $n$-ary, etc. if $L$ is so.  We also have two {\bf logical atoms}\index{logical atom}\label{0logical atom} $\twg$ and $\tlg$. {\bf Formulas} are constructed from atoms and variables in the standard way applying to them the unary connectives $\gneg,\pst,\pcost,\st,\cost$, the binary connectives $\mli,\pintimpl,\intimpl$, the variable- ($\geq 2$) arity connectives $\mlc,\mld,\adc,\add$, and
the quantifiers $\cla,\cle,\ada,\ade,\mla,\mle$. Throughout the rest of this paper, unless otherwise specified, ``formula'' will always mean a formula of this language, and letters $E,F,G,H$ will be used as metavariables for formulas.     We also continue using $x,y,z$ as metavariables for variables, $c$ for constants and $t$ for terms.

The definitions of a {\em bound occurrence} and a {\em free occurrence} of a variable are standard. They extend from variables to all terms by stipulating that an occurrence of a constant is always free.
When an occurrence of a variable $x$ is within the scope of ${\cal Q}x$ for several quantifiers $\cal Q$, then $x$ is considered bound by the quantifier ``nearest'' to it. For instance, the occurrence of $x$ within $Q(x,y)$ in $\cla x(P(x)\mld \ada x\mla yQ(x,y))$ is bound by $\ada x$ rather than $\cla x$, for the latter is overridden by the former. An occurrence of a variable that is bound by $\cla x$ or $\cle x$ is said to be {\bf blindly bound}.\index{blindly bound}\label{0blindly bound}

In concordance with a similar notational practice established earlier for games, sometimes we represent a  formula $F$ as 
$F(x_1,\ldots,x_n)$, where the $x_i$ are pairwise distinct variables. In the context set by such a representation, 
$F(t_1,\ldots,t_n)$ will mean the result of simultaneously replacing in $F$ all free occurrences of each variable $x_i$ 
($1\leq i\leq n$) by 
term $t_i$. In case each $t_i$ is a variable $y_i$, it may be not clear whether $F(x_1,\ldots,x_n)$ or $F(y_1,\ldots,y_n)$ was originally meant to represent $F$ in a given context.   Our disambiguating convention 
is that the context is set by the expression that was used earlier. That is, when we first mention $F(x_1,\ldots,x_n)$ and only after that 
use the expression $F(y_1,\ldots,y_n)$, the latter should be understood as the result of replacing variables in the former rather than vice versa. 
It should be noted that, when representing $F$ as $F(x_1,\ldots,x_n)$,
we do not necessarily mean that $x_1,\ldots,x_n$ are exactly the variables that have free occurrences in $F$.

An {\bf interpretation}\index{interpretation (as a function)}\label{0interpretation (as a function)} is a function $^*$ that sends each $n$-ary elementary (resp. general) letter $L$ to an elementary (resp. static) game with a fixed attached $n$-tuple $x_1,\ldots,x_n$ of variables. We denote such a game by $L^*(x_1,\ldots,x_n)$, and call the tuple $(x_1,\ldots,x_n)$ the {\bf canonical tuple of $L^*$}.\index{canonical tuple}\label{0canonical tuple} When we do not care about the canonical tuple, simply $L^*$ can be written instead of $L^*(x_1,\ldots,x_n)$.  
According to our earlier conventions, $x_1,\ldots,x_n$ have to be neither {\em all} nor the {\em only} variables on which the game $L^*=L^*(x_1,\ldots,x_n)$ depends; in fact, $L^*$ does not even have to be finitary here.  The canonical tuple is only used for setting a context, in which $L^*(t_1,\ldots,t_n)$ can be conveniently written later for $L^*(x_1/t_1,\ldots,x_n/t_n)$. This eliminates the need to have a special syntactic construct in the language for the operation of substitution of variables. 

Interpretations are meant to turn formulas into games. Not every interpretation is equally good for every formula though, and some precaution is necessary to avoid confusing collisions of variables, as well as to guarantee that $\cla x,\cle x$ are only applied to games for which they are defined, i.e. games unistructural in $x$. For this reason, we restrict interpretations 
to ``admissible'' ones. We say that an interpretation $^*$ is {\bf admissible for}\index{admissible interpretation}\label{0admissible interpretation} a formula $F$, or simply is {\bf $F$-admissible} iff, for every $n$-ary (general or elementary) letter $L$ occurring in $F$, the following two conditions are satisfied:

\begin{description}
\item[(i)] $L^*$ does not depend on any variables that are not among its canonical tuple but occur in $F$.
\item[(ii)] If the $i$th ($1\leq i\leq n$) term of an occurrence of an $L$-based atom in $F$ is blindly bound, then $L^*$ is unistructural in the $i$th variable of its canonical tuple. 
\end{description}

Every  interpretation $^*$ extends from letters to formulas for which $^*$ is admissible in the obvious way: 
\begin{itemize}
\item where $L$ is an $n$-ary letter with $L^*=L^*(x_1,\ldots,x_n)$ and $t_1,\ldots,t_n$ are any terms,
$\bigl(L(t_1,\ldots,t_n)\bigr)^*=L^*(t_1,\ldots,t_n)$;
\item $^*$ respects the meanings of logical operators (including logical atoms as $0$-ary operators) as the corresponding game operations; that is:  \ $\twg^*=\twg$; \ $(\gneg G)^*=\gneg (G^*)$; \ $(G\adc H)^*=(G^*)\adc (H^*)$; \ $(\cla xG)^*=\cla x(G^*)$; etc. 
\end{itemize}
When $F^*=A$, we say that $ ^*$ {\bf interprets}\index{interpret}\label{0interpret} $F$ as $A$, and that $F^*$ is {\bf an interpretation}\index{interpretation (as a game)}\label{0interpretation (as a game)} of $F$.

 Notice that condition (ii) of admissibility is automatically satisfied when $L$ is an elementary letter, because an elementary problem (i.e. $L^*$) is always unistructural and hence unistructural in all variables. 
In most typical cases we will be interested in interpretations $^*$ where $L^*$ is unistructural and does not depend on any variables other than those of its canonical tuple, so that both conditions (i) and (ii) will be automatically satisfied. With this remark in mind and in order to relax terminology, henceforth we may  sometimes omit ``$F$-admissible'' and simply say ``interpretation''; every time 
an expression $F^*$ is used in a context, it should be understood that the range of $^*$ is restricted to $F$-admissible interpretations.

\begin{definition}\label{valid}
We say that a formula $F$ is {\bf valid}\index{valid}\label{0valid} --- and write $\valid F$\index{$\valid$}\label{0validd} --- iff, for every $F$-admissible interpretation $^*$, the game $F^*$ is winnable. 
\end{definition}

The main technical goal of CL at this stage of its development is to find axiomatizations for the set of valid formulas or various nontrivial fragments of that set. A considerable progress has already been achieved in this direction; more is probably yet to come in the future.

\section{Uniform validity}\label{ss8}
If we disabbreviate ``$F^*$ is winnable'' as $\exists {\cal M} ({\cal M}\models F^*)$, validity in the sense of Definition \ref{valid} can be written as 
$\forall ^* \exists {\cal M}({\cal M}\models F^*)$,  where $\cal M$ ranges over HPMs 
or EPMs, and $^*$ over $F$-admissible interpretations. Reversing the order of quantification yields the following stronger property of uniform validity:

\begin{definition}\label{uvalid}
We say that a formula $F$ is {\bf uniformly valid}\index{uniformly valid}\label{0uniformly valid} --- and write $\uvalid F$\index{$\uvalid$}\label{0uvalid} --- iff there is an HPM or (equivalently) EPM $\cal M$ such that, for every $F$-admissible interpretation $^*$, ${\cal M} \models F^*$.

Such an HPM or EPM $\cal M$ is said to be a {\bf uniform solution}\index{uniform solution}\label{0uniform solution} for $F$, and ${\cal M}\uvalid F$ is written to express that $\cal M$ is a uniform solution for $F$.   
\end{definition}

Intuitively, a uniform solution $\cal M$ for a formula $F$ is an interpretation-independent winning strategy: since,  unlike valuation, the ``intended'' or ``actual'' interpretation $^*$ is not visible to the machine,  $\cal M$  has to play in some standard, uniform way that would be successful 
for any possible interpretation of $F$. 
 
The term ``uniform'' is borrowed from \cite{Abr94} as this understanding of validity in its spirit is close to that in Abramsky\index{Abramsky}\label{0Abramsky} and Jagadeesan's\index{Jagadeesan}\label{0Jagadeesan} tradition. The concepts of validity in Lorenzen's\index{Lorenzen}\label{0Lorenzen} \cite{Lor59} tradition, or in the sense of Japaridze \cite{Jap00,Jap02a}, also belong to this category. Common to those uniform-validity-style notions is that validity there 
is not defined as being ``always true" (true=winnable) as this is the case with the classical understanding 
of this concept; in those approaches the concept of truth is often simply absent, and validity is treated as a basic concept 
in its own rights. 
As for simply validity, it is closer to validities in the sense of Blass\index{Blass}\label{0Blass} \cite{Bla92} or Japaridze \cite{Jap97}, and presents a direct generalization of the corresponding classical concept in that it indeed means being ``true" in every particular setting.  

Which of our two versions of validity is more interesting depends on the motivational standpoint. 
It is validity rather than uniform validity that tells us what can be computed in principle. So, a 
com\-putability-theoretician would focus on validity. Mathematically, non-validity is generally by an order of magnitude more informative --- and correspondingly harder to prove --- than non-uniform-validity. Say, the non-validity of $p\add\gneg p$
 means existence of solvable-in-principle yet algorithmically unsolvable 
problems\footnote{Well, saying so is only accurate with the Strong Completeness clause of Theorem \ref{main5} (which, as conjectured in \cite{Jap03}, extends from $\predell$ to any other complete fragments of CL) in mind, according to which the non-validity of $p\add\gneg p$ implies the existence of a {\em finitary} predicate $A$ for which $A\add\gneg A$ has no algorithmic solution. As will be pointed out in a comment following Theorem \ref{main5}, without the finitarity restriction, a machine may fail to win $A\add\gneg A$ not (only) due to the fundamental limitations of algorithmic methods, but rather due to the fact that it can never finish 
reading all necessary information from the valuation tape to determine the truth status of $A$.} 
 --- the fact 
that became known to the mankind only as late as in the 20th century. 
As for 
the non-uniform-validity of $p\add\gneg p$, it is trivial: of course there is no way to choose one of the two disjuncts that would be true for all possible values of $p$ because, as the Stone Age intellectuals were probably aware, some $p$ are true and some are false.  

On the other hand, it is uniform validity rather than validity that is of interest in more applied areas of computer science such as knowledgebase systems  or systems for planning and action (see Section \ref{applc}).  
In this sort of applications we want a logic on which a universal problem-solving machine can be based. Such a machine would or should be able to solve problems represented by logical formulas  without any specific knowledge of the meanings of 
their atoms, i.e. without knowledge of the actual interpretation. Remembering what was said about the intuitive meaning of uniform validity, this concept is exactly what fits the bill.

Anyway, the good news is that the two concepts of validity appear to yield the same logic. This has been conjectured for 
the full language of CL in \cite{Jap03} (Conjecture 26.2), and by now, as will be seen from our Theorem \ref{valuval}, successfully verified for the rather expressive fragment of that language --- the language of logic $\predell$.

\section{Logic $\predell$}\label{ss9}

The language of logic $\predell$ is the fragment of the language of Section \ref{ss7}  obtained by forbidding the parallel group of quantifiers and the recurrence group of propositional connectives. This leaves us with the operators 
$\gneg$, $\mlc$, $\mld$, $\mli$, $\adc$, $\add$, $\cla$, $\cle$,  $\ada$, $\ade$, along with the logical atoms 
$\twg,\tlg$ and the two sorts (elementary and general) of nonlogical atoms. Furthermore, for safety and without loss of expressive power, we agree that a formula cannot contain both bound and free occurrences of the same variable. We refer to the formulas of this language as {\bf $\predell$-formulas}.\index{$\predell$-formula}\label{0predell-formula}

Our axiomatization of $\predell$ employs the following terminology.  Understanding $F\mli G$ as an abbreviation for $\gneg F\mld G$, a {\bf positive}\index{positive occurrence}\label{0positive occurrence} (resp. {\bf negative})\index{negative occurrence}\label{0negative occurrence} {\bf occurrence} of a subformula is
one that is in the scope of an even (resp. odd) number of occurrences of $\gneg$.
A {\bf surface occurrence}\index{surface occurrence}\label{0surface occurrence} of a subformula is an occurrence that is 
not in the scope of any choice operators. A $\predell$-formula not containing general atoms and choice operators --- i.e. a formula of the language of classical first-order logic --- is said to be {\bf elementary}.\index{elementary formula}\label{0elementary formula} The 
{\bf elementarization}\index{elementarization}\label{0elementarization} of a $\predell$-formula $F$ is the result of replacing
in $F$ all surface occurrences of each subformula of the form  $G_1\add\ldots\add G_n$ or $\ade xG$ by $\tlg$, all surface occurrences of each subformula  
of the form  $G_1\adc\ldots\adc G_n$ or $\ada xG$ by $\twg$, all positive surface occurrences of each general atom by 
$\tlg$, and all negative surface occurrences of each general atom by $\twg$. A $\predell$-formula is said to be {\bf stable}\index{stable formula}\label{0stable formula}
iff its elementarization is classically valid, i.e. provable in classical predicate calculus. Otherwise it is {\bf instable}.\index{instable formula}\label{0instable formula}

With ${\mathcal P}\mapsto C$\index{$\mapsto$}\label{0mapsto} meaning ``from premise(s) $\mathcal P$ conclude $C$'', logic $\predell$\index{$predell$}\label{0predell} is given by the following four rules where, as can be understood, both the premises and the conclusions range over $\predell$-formulas:

\begin{description}
\item[A\ \ ]   \label{RA}$\vec{H}\mapsto E$, where $E$ is stable and $\vec{H}$ is 
a set of formulas satisfying the following conditions:\vspace{-3pt}
\begin{description}
\item[(i)]  Whenever $E$ has a positive (resp. negative)  surface occurrence of a subformula $G_1\adc\ldots\adc G_n$ (resp. $G_1\add \ldots\add G_n$), for each 
$i\in\{1,\ldots,n\}$, $\vec{H}$ contains the result of replacing that occurrence in $E$ by $G_i$;
\item[(ii)] Whenever $E$ has a positive (resp. negative) surface occurrence of a subformula $\ada x G(x)$ (resp. $\ade xG(x)$), $\vec{H}$ contains the result of replacing that occurrence in $E$ by $G(y)$ for some variable $y$ not occurring in $E$.
\end{description}
\item[B1\ ]  \label{RB1}$H\mapsto E$, where $H$ is the result of replacing in $E$ a negative (resp. positive) surface occurrence of a subformula $G_1\adc\ldots\adc G_n$ (resp. $G_1\add\ldots\add G_n$) by $G_i$ for some $i\in\{1,\ldots, n\}$.
\item[B2\ ]  \label{RB2}$H\mapsto E$, where $H$ is the result of replacing in $E$ a negative (resp. positive) surface occurrence of a subformula $\ada xG(x)$ (resp. $\ade xG(x)$) by $G(t)$ for some term $t$ such that (if $t$ is a variable) 
neither the above occurrence of 
$\ada xG(x)$ (resp. $\ade xG(x)$) in $E$ nor any of the free occurrences of $x$ in $G(x)$ are in the scope of $\cla t$,
$\cle t$, $\ada t$ or $\ade t$. 
\item[C\ \ ]  \label{RC}$H\mapsto E$, where $H$ is the result of replacing in $E$ two --- one positive and one negative ---
surface occurrences of some $n$-ary general letter by an $n$-ary elementary letter that does not occur in $E$.
\end{description}

Axioms are not explicitly stated, but note that the set of premises of Rule {\bf A} sometimes can be empty, in which case the conclusion acts as an axiom. 
Looking at a few examples should help us get a syntactic feel of this most unusual deductive system. 

The following is a 
$\predell$-proof of $\ada x\ade y \bigl(P(x)\mli P(y)\bigr)$:\vspace{5pt}

$\begin{array}{ll}
1.\ \ p(z)\mli p(z) & \mbox{(from $\{\}$ by Rule {\bf A});}\\
2.\ \ P(z)\mli P(z) & \mbox{(from 1 by Rule {\bf C});}\\
3.\ \ \ade y\bigl(P(z)\mli P(y)\bigr) & \mbox{(from 2 by Rule {\bf B2});}\\
4.\ \ \ada x\ade y\bigl(P(x)\mli P(y)\bigr) & \mbox{(from \{3\} by Rule {\bf A}).}
\end{array}$\vspace{5pt}

On the other hand, $\predell\not\vdash \ade y\ada x\bigl(P(x)\mli P(y)\bigr)$. Indeed, obviously this instable formula cannot be the conclusion of any rule but {\bf B2}. If it is derived by this rule, the premise should be 
$\ada x\bigl(P(x)\mli P(t)\bigr)$ for some term $t$ different from $x$. $\ada x\bigl(P(x)\mli P(t)\bigr)$, in turn, could only be derived by Rule {\bf A} where, for some variable $z$ different from $t$, $P(z)\mli P(t)$ is a (the) premise. The latter 
is an instable formula and does not contain choice operators, so the only rule by which it can be derived is {\bf C}, 
where the premise is $p(z)\mli p(t)$ for some elementary letter $p$. Now we deal with a classically non-valid and hence instable elementary formula, and it cannot be derived by any of the four rules of $\predell$.  

Note that, in contrast, the ``blind version" $\cle y\cla x\bigl(P(x)\mli P(y)\bigr)$ of
$\ade y\ada x\bigl(P(x)\mli P(y)\bigr)$ is provable:\vspace{5pt}

$\begin{array}{ll}
1.\ \ \cle y\cla x \bigl(p(x)\mli p(y)\bigr) & \mbox{(from $\{\}$ by Rule {\bf A});}\\
2.\ \ \cle y\cla x \bigl(P(x)\mli P(y)\bigr) & \mbox{(from 1 by Rule {\bf C}).}\vspace{5pt}
\end{array}$

`{\em There is $y$ such that, for all $x$, $P(x)\mli P(y)$}' is true yet not in a constructive sense, thus 
belonging to the kind of principles that have been fueling controversies between the classically- and constructivistically-minded.  As noted in Section \ref{intr}, computability logic is offering a peaceful settlement, telling the arguing parties: ``There is no need to fight at all. It appears that you simply have two different concepts of `{\em there is}'/`{\em for all}\hspace{1pt}'. So, 
why not also use two different names: $\cle/\cla$ and $\ade/\ada$. 
Yes, $\cle y\cla x \bigl(P(x)\mli P(y)\bigr)$ is indeed right; and  yes, 
$\ade y\ada x\bigl(P(x)\mli P(y)\bigr)$ is indeed wrong."  Clauses 1 and 2 of Exercise \ref{Jan20} illustrate a 
similar solution for the law of the excluded middle, the most controversial principle of classical logic.

The above-said remains true with $p$ instead of $P$, for what is relevant there is the difference between the 
constructive and non-constructive versions of logical operators rather than how atoms are understood. Then how about the difference between the elementary and non-elementary versions of atoms? This 
distinction allows computability logic to again act in its noble role of a reconciliator/integrator, but this time between classical and linear logics, telling them: ``It appears that you have two different concepts of the objects that logic is meant to study. So, why not also use two different sorts of atoms to represent such objects: elementary atoms $p,q,\ldots$,  and general atoms $P,Q,\ldots$. Yes, $p\mli p\mlc p$ is indeed right; and yes, $P\mli P\mlc P$ (Exercise \ref{Jan20}(4)) is indeed wrong". However, as pointed out in Section \ref{intr}, the term ``linear logic" in this context should be understood in a very generous sense, referring not to the particular deductive system proposed by Girard but rather to the general philosophy and intuitions traditionally associated with it. The formula of clause 3 of the following exercise separates $\predell$ from linear logic. That formula is provable in affine logic though. Switching to affine logic, i.e. restoring the deleted (from classical logic) rule of weakening, does not however save the case: the $\predell$-provable formulas of clauses 10, 11 and 18 of the exercise are provable in neither linear nor affine logics.

\begin{exercise}\label{Jan20}
In clauses 14 and 15 below, ``\hspace{2pt}$\predell\vdash E\Leftrightarrow F$" stands for ``\hspace{2pt}$\predell\vdash E\mli F$ {\em and} $\predell\vdash F\mli E$". 
Verify that:\vspace{7pt}

\noindent 1. \ \hspace{2pt}$\predell\vdash P\mld\gneg P$.\vspace{3pt} 

\noindent 2. \ \hspace{2pt}$\predell\not\vdash P\add\gneg P$. Compare with 1.\vspace{3pt}

\noindent 3. \ \hspace{2pt}$\predell\vdash P\mlc P\mli P$.\vspace{3pt} 

\noindent 4. \ \hspace{2pt}$\predell\not\vdash P\mli P\mlc P$. Compare with 3,5.\vspace{3pt}

\noindent 5. \ \hspace{2pt}$\predell\vdash P\mli P\adc P$.\vspace{3pt} 

\noindent 6. \ \hspace{2pt}$\predell\vdash (P\add Q)\mlc(P\add R)\mli P\add(Q\mlc R)$.\vspace{3pt} 

\noindent 7. \ \hspace{2pt}$\predell\not\vdash P\add(Q\mlc R)\mli(P\add Q)\mlc(P\add R)$. Compare with 6,8.\vspace{3pt}

\noindent 8. \ \hspace{2pt}$\predell\vdash \hspace{2pt}p\hspace{1pt}\add(Q\mlc R)\mli(\hspace{1pt}p\hspace{1pt}\add Q)\mlc(\hspace{1pt}p\hspace{1pt}\add R)$.\vspace{3pt} 

\noindent 9. \ \hspace{2pt}$\predell\not\vdash \hspace{2pt}p\hspace{1pt}\adc(Q\mlc R)\mli(\hspace{1pt}p\hspace{1pt}\adc Q)\mlc(\hspace{1pt}p\hspace{1pt}\adc R)$. Compare with 8.\vspace{3pt} 

\noindent 10. $\predell\vdash (P\mlc P)\mld (P\mlc P)\mli (P\mld P)\mlc(P\mld P)$.\vspace{3pt}  

\noindent 11. $\predell\vdash (P\mlc(R\adc S))\adc(Q\mlc(R\adc S))\adc((P\adc Q)\mlc R)\adc ((P\adc Q)\mlc S) \mli$

\hspace{43pt}$(P\adc Q)\mlc(R\adc S)$.\vspace{3pt}

\noindent 12. $\predell\vdash \hspace{2pt}\cla\hspace{1pt} xP(x)\mli \ada xP(x)$.\vspace{3pt}

\noindent 13.  $\predell\not\vdash \ada xP(x)\mli \hspace{2pt}\cla\hspace{1pt} xP(x)$. Compare with 12.\vspace{3pt}
 
\noindent 14. $\predell\vdash \cle xP(x)\adc\cle x Q(x)\Leftrightarrow\cle x\bigl(P(x)\adc Q(x)\bigr)$. 

\hspace{43pt}Similarly for $\add$ instead of $\adc$,  and/or $\cla$ instead of $\cle$.\vspace{3pt} 

\noindent 15. $\predell\vdash \ada x\cle yP(x,y)\Leftrightarrow \cle y\ada xP(x,y)$.

\hspace{43pt}Similarly for $\ade$ instead of $\ada$, and/or $\cla$ instead of $\cle$.\vspace{3pt} 

\noindent 16. $\predell\vdash \hspace{1pt}\cla\hspace{2pt} x\bigl(P(x)\mlc Q(x)\bigr)\mli \hspace{2pt}\cla\hspace{1pt} xP(x)\mlc\hspace{1pt}\cla\hspace{1pt} xQ(x)$.\vspace{3pt}

\noindent 17. $\predell\not\vdash \ada x\bigl(P(x)\mlc Q(x)\bigr)\mli \ada xP(x)\mlc\ada xQ(x)$. Compare with 16.\vspace{3pt}

\noindent 18. $\predell\vdash \ada x\Bigl(\bigl(P(x)\mlc\ada xQ(x)\bigr)\adc\bigl(\ada xP(x)\mlc Q(x)\bigr)\Bigr)\mli$ 

\hspace{43pt}$\ada xP(x)\mlc \ada xQ(x)$.\vspace{3pt}

\noindent 19. $\predell\vdash$ formula (\ref{e1}) of Section \ref{ss4.4}.\vspace{3pt}

\noindent 20. $\predell\not\vdash$ formula (\ref{e2}) of Section \ref{ss4.4}. Compare with 19.\vspace{3pt}
\end{exercise}

Taking into account that classical validity and hence stability is recursively enumerable, from the way $\predell$ is axiomatized it is obvious that 
the set of theorems of $\predell$ is recursively enumerable.
Not so obvious, however, may be the following theorem proven in \cite{CL4}. As it turns out, the choice/constructive quantifiers 
$\ada,\ade$ are much better behaved than their blind/classical counterparts $\cla,\cle$, yielding a decidable first-order logic:

\begin{theorem}\label{dec}
 The $\cla,\cle$-free fragment of $($the set of theorems of$)$ $\predell$ is decidable.
\end{theorem}

Next, based on the straightforward 
observation that elementary formulas are derivable in  $\predell$ (in particular, from the empty set of premises by Rule {\bf A}) 
exactly when they are classically valid, we have: 

\begin{theorem}\label{dec18}
 $\predell$ is a conservative extension of classical predicate logic: the latter is nothing but the elementary fragment (i.e. the set of all elementary theorems) of the former. 
\end{theorem}

Remember that a predicate $A$ is said to be of arithmetical complexity $\Delta_2$ iff $A=\cle x\cla yB_1$ and $\gneg A=\cle x\cla y B_2$ for some decidable predicates $B_1$ and 
$B_2$. 

The following Theorem \ref{main5} is the strongest soundness and completeness result known so far in computability logic. Its proof has taken 
about half of the volume of \cite{CL3} and almost entire \cite{CL4}. A similar theorem for the propositional version {\bf CL2} of $\predell$ was proven in \cite{CL1}-\cite{CL2}.

\begin{theorem}\label{main5} 
$\predell\vdash F$ iff $F$ is valid (any $\predell$-formula $F$).
Furthermore:

{\bf Uniform-Constructive Soundness:}\index{uniform-constructive soundness}\label{0uniform-constructive soundness} There is an effective  procedure that takes a $\predell$-proof of an arbitrary $\predell$-formula $F$ and 
constructs a uniform solution for $F$.

{\bf Strong Completeness:}\index{strong completeness}\label{0strong completeness} If a $\predell$-formula $F$ is not provable in $\predell$,
 then $F^*$ is not computable  for some $F$-admissible \mbox{interpretation $^*$} that interprets all elementary atoms as finitary predicates of arithmetical complexity $\Delta_2$, and interprets all general atoms 
as \mbox{$\adc,\add$-combinations} of 
finitary predicates of arithmetical complexity $\Delta_2$.  
\end{theorem}

A non-finitary game generally depends on infinitely many variables, and appealing to this sort of games in a completeness proof could seriously weaken such a result: the reason for incomputability  of a non-finitary game could be just the fact that the machine can never finish reading all the relevant information from its valuation tape. Fortunately, in view 
of the Strong Completeness clause, it turns out that the question whether non-finitary games are allowed or not has no effect on the (soundness and) completeness of $\predell$; moreover, finitary games can be further restricted to the sort 
as simple as $\adc,\add$-combinations of finitary predicates. 

Similarly, the Uniform-Constructive Soundness clause dramatically strengthens the soundness result for $\predell$ and, as will be argued in the following section, opens application areas far beyond logic or the pure theory of computation. First of all, notice that it immediately implies a positive verification of the earlier-mentioned 
Conjecture 26.2 of \cite{Jap03} restricted to the language of $\predell$, according to which validity and uniform validity are extensionally equivalent. Indeed, if a $\predell$-formula $F$ is uniformly valid, then it is automatically also valid, as uniform validity is stronger than validity. Suppose now $F$ is valid. Then, by the completeness part of Theorem \ref{main5},  $\predell \vdash F$. But then, by  the Uniform-Constructive Soundness clause, $F$ is uniformly valid.  Thus, we have: 

\begin{theorem}\label{valuval}
A $\predell$-formula is valid if and only if it is uniformly valid.
\end{theorem}

But $\predell$ is sound in an even stronger sense. Knowing that a solution for a given problem exists might be of little practical importance without being able to actually find such a solution. No problem: according to the Uniform-Constructive Soundness clause, a uniform solution for a $\predell$-provable formula $F$ automatically comes with a $\predell$-proof of $F$. The earlier-mentioned soundness theorem for Heyting's intuitionistic calculus proven in \cite{int1} comes in the same uniform-constructive form, and so does the soundness theorem for affine logic (Theorem \ref{main}) proven later in this paper.

\section{Applied systems based on CL}\label{applc}

The original motivations behind CL were computability-theoretic: the approach provides a systematic  
answer to the question `what can be computed?', which is a fundamental question of computer science. Yet, the above discussion of the uniform-constructive nature of the known soundness theorems for various fragments of CL reveals that the CL paradigm is not only about {\em what} can be computed. It is equally about 
{\em how} problems can be computed/solved, suggesting that CL should have potential utility, with its application areas   
not limited to the theory of computation. In the present section we will briefly examine why 
and how CL is of interest in some other fields of study, specifically, knowledgebase systems
and constructive applied theories.

The reason for the failure of $p\add\neg p$ as a computability-theoretic principle is  
that the problem represented by this formula may have no effective solution --- that is, the predicate $p^*$ may be undecidable. The reason why this principle would fail in the context of knowledgebase systems, however, is much simpler. A knowledgebase system may fail to solve the problem
$\mbox{\em Female}\hspace{1pt}(\mbox{\em Dana})  \add \gneg\mbox{\em Female}\hspace{1pt}(\mbox{\em Dana})$ not because the latter has no effective solution (of course it has one), but because the system simply lacks sufficient knowledge to determine Dana's gender. On the other hand, any system would be able to ``solve" the problem
 $\mbox{\em Female}\hspace{1pt}(\mbox{\em Dana})  \mld   \gneg\mbox{\em Female}\hspace{1pt}(\mbox{\em Dana})$
as this is an automatically won elementary game so that there is nothing to solve at all. 
Similarly, while $\cla y\cle x\mbox{\em Father}\hspace{1pt}(x,y)$ is an automatically solved elementary problem expressing the almost tautological knowledge that every person has a father, ability to solve the problem  
$\ada y\ade x\mbox{\em Father}\hspace{1pt}(x,y)$ implies the nontrivial knowledge of everyone's actual father. 
Obviously the knowledge expressed by $A\add B$ or $\ade xA(x)$ is generally stronger than the knowledge expressed by $A\mld B$ or $\cle xA(x)$, yet the formalism of classical logic fails to capture this difference --- the difference whose relevance hardly requires any explanation. The traditional approaches to knowledgebase systems 
(\cite{Kon89,Lev00,Moo85} etc.) 
try to mend this gap by augmenting the language of classical logic with special epistemic constructs, such as the 
 modal ``know that" operator $\kn$\hspace{1pt}, after which probably $\kn A
 \mld \kn B$ would be suggested as a translation for $A \add B$  and $\cla y\cle x \kn A(x,y)$ for $\ada y\ade xA(x,y)$. Leaving it for the philosophers to argue whether,
say, $\cla y\cle x \kn A(x,y)$ really expresses the constructive meaning 
of $\ada y\ade xA(x,y)$, and forgetting that epistemic constructs often yield unnecessary and very unpleasant  complications such as messiness and non-semidecidability of the resulting logics, some of the major issues still do not seem to be taken care of. Most of the actual knowledgebase and information systems are interactive, and what we really need is a logic of {\em interaction} rather than just a logic 
of knowledge. Furthermore, a knowledgebase logic needs to be {\em resource-conscious}. The informational resource
expressed by $\ada x(\mbox{\em Female}\hspace{1pt}(x)  \add   \gneg\mbox{\em Female}\hspace{1pt}(x))$ is not as strong as
the one expressed by  $\ada x(\mbox{\em Female}\hspace{1pt}(x)  \add   \gneg\mbox{\em Female}\hspace{1pt}(x))\mlc\ada x(\mbox{\em Female}\hspace{1pt}(x)  \add   \gneg\mbox{\em Female}\hspace{1pt}(x))$: the former implies the resource provider's commitment to tell only one (even though an arbitrary one) person's gender, while the latter is about telling any two people's genders. A reader having difficulty in understanding why this difference is relevant, may try to replace {\em Female}$\hspace{1pt}(x)$ with {\em Acid}$\hspace{1pt}(x)$, and then think of a (single) 
piece of litmus paper. Neither classical logic nor its standard epistemic extensions have the ability to account for such differences. But CL promises to be adequate. 
It {\em is} a logic of interaction, it {\em is} resource-conscious, and it {\em does} 
capture the relevant differences between truth and actual ability to find/compute/know truth.

When CL is used as a logic of knowledgebases, its formulas represent interactive queries. A formula whose main operator is $\add$ or $\ade$ can be understood as a question asked by the user, and a formula whose main operator is $\adc$ or $\ada$ as a question asked by the system.
Consider the problem $\ada x \ade y \mbox{\em Has}\hspace{1pt}(x, y)$, where $\mbox{\em Has}\hspace{1pt}(x,y)$ means ``patient $x$ has disease  $y$" (with {\em Healthy} counting as one of the possible ``diseases"). This formula is the following question asked by the system: ``Who do you want me to diagnose?"  The user's response can be ``Dana". This move brings the game down to 
$\ade y \mbox{\em Has}\hspace{1pt}(\mbox{\em Dana}, y)$. This is now a question asked by the user: ``What does Dana have?". The system's response can be ``flu", taking us to the terminal position \(\mbox{\em Has}\hspace{1pt}(\mbox{\em Dana}, \mbox{\em Flu}).\) The system has been successful iff Dana really has a flu. 

Successfully solving the above problem $\ada x \ade y \mbox{\em Has}\hspace{1pt}(x, y)$ requires having all relevant medical information for each possible patient, which in a real diagnostic system would hardly be the case. Most likely, such a system, after
receiving a request to diagnose $x$, would make counterqueries regarding $x$'s symptoms, blood pressure, test results, age, gender, etc., so that the query that the system will be solving would have a higher degree of interactivity than the two-step query $\ada x \ade y \mbox{\em Has}\hspace{1pt}(x, y)$ does, with questions and counterquestions interspersed in some complex fashion. Here is when other computability-logic operations come into play. 
$\gneg$ turns queries into counterqueries; parallel operations generate combined queries, with $\mli$ acting as a query reduction operation; $\st,\pst$ allow repeated queries, etc. Here we are expanding our example. 
Let {\em Sympt}$\hspace{1pt}(x,s)$ mean ``patient $x$ has (set of) symptoms $s$", and $\mbox{\em Pos}\hspace{1pt}(x,t)$ mean ``patient $x$ tests 
positive for test $t$". Imagine a diagnostic system that can diagnose any particular patient $x$, but needs some additional information. Specifically, it needs to know $x$'s symptoms; plus, the system may require to have $x$ taken a test $t$ that it selects dynamically in the course of a dialogue with the user depending on what responses it received. The interactive task/query that such a system is performing/solving can then be expressed by the formula\vspace{-3pt} 
\begin{equation}\label{diagn}\ada x\Bigl(\ade s\mbox{\em Sympt}\hspace{1pt}(x,s)\mlc \ada t\bigl(\mbox{\em Pos}\hspace{1pt}(x,t)\add\gneg \mbox{\em Pos}\hspace{1pt}(x,t)\bigr)\mli \ade y\mbox{\em Has}\hspace{1pt}(x,y)\Bigr).\vspace{-3pt}\end{equation}
A possible scenario of playing the above game is the following. At the beginning, the system waits until the user specifies a patient $x$ to be diagnosed. We can think of this stage as systems's requesting the user to select a particular (value of) $x$, remembering that the presence of $\ada x$ automatically implies such a request. 
After a patient $x$ --- say $x=X$ --- is selected,
the system requests to specify $X$'s symptoms. Notice that our game rules 
make the system successful if the user fails to provide this information, i.e. specify a (the true) value for $s$ in 
$\ade s\mbox{\em Sympt}\hspace{1pt}(X,s)$. 
Once a response --- say, $s=S$ --- is received, the system selects a test $t=T$ and asks the user to perform it on $X$, i.e. to choose the true disjunct of $\mbox{\em Pos}\hspace{1pt}(X,T)\add\gneg \mbox{\em Pos}\hspace{1pt}(X,T)$. Finally, provided that the user gave correct answers to all counterqueries 
(and if not, the user has lost), the system
makes a diagnostic decision, i.e. specifies a value $Y$ for $y$ in   $\ade y\mbox{\em Has}(X,y)$ for which 
$\mbox{\em Has}(X,Y)$ is true. 

The presence of a single ``copy" of $\ada t\bigl(\mbox{\em Pos}\hspace{1pt}(x,t)\add\gneg \mbox{\em Pos}\hspace{1pt}(x,t)\bigr)$
in the antecedent of (\ref{diagn})  means that the system may request testing a given patient only once. If $n$ tests were potentially needed instead, 
this would be expressed by taking the $\mlc$-conjunction of $n$ identical conjuncts $\ada t\bigl(\mbox{\em Pos}\hspace{1pt}(x,t)$ $\add\gneg \mbox{\em Pos}\hspace{1pt}(x,t)\bigr)$. 
And if the system potentially needed 
an unbounded number of tests, then we would write $\pst\ada t\bigl(\mbox{\em Pos}\hspace{1pt}(x,t)\add\gneg \mbox{\em Pos}\hspace{1pt}(x,t)\bigr)$, thus further weakening (\ref{diagn}): a system that performs this weakened task is not as good as the one performing (\ref{diagn}) as it requires stronger external (user-provided) informational resources. Replacing the main 
quantifier $\ada x$ by $\cla x$, on the other hand, would strengthen (\ref{diagn}), signifying the system's ability 
to diagnose a patent purely on the basis of his/her symptoms and test result without knowing who the patient really is. However, if in its diagnostic decisions the system uses some additional information on patients such 
their medical histories stored in its knowledgebase and hence needs to know the patient's identity,
$\ada x$ cannot be upgraded to $\cla x$. Replacing $\ada x $ by $\mla x$ would be a yet another way to strengthen (\ref{diagn}), 
signifying the system's ability to diagnose 
all patients rather than any particular one; obviously effects of at least the same strength would be achieved by 
just prefixing (\ref{diagn}) with $\pst$ or $\st$. 

As we just mentioned system's {\bf knowledgebase},\index{knowledgebase}\label{0knowledgebase} let us make clear what it means. Formally, this is a finite 
$\mlc$-conjunction {\em KB}
of formulas, which can also be thought of as the (multi)set of its conjuncts. We call the elements of this set the {\bf internal informational resources}\index{internal informational resource}\label{0internal informational resource} of the system. Intuitively, {\em KB} represents all of the 
nonlogical knowledge available to the system, so that (with a fixed built-in logic in mind) the strength of the former 
determines the query-solving power of the latter. Conceptually, however, we do not think of {\em KB} as a 
part of the system properly. The latter is just ``pure", logic-based problem-solving software of universal utility that 
initially comes to the user without any nonlogical knowledge whatsoever. Indeed, built-in nonlogical knowledge 
would make it no longer universally applicable: Dana can be a female in the world of one potential user while a male in the 
world of another user, and $\forall x\forall y(x\times y=y\times x)$ can be false to a user who understands $\times$ as Cartesian rather than number-theoretic product. It is the user who selects and maintains {\em KB} for the system, putting into it all informational resources that (s)he believes are relevant, correct and maintainable. Think of the formalism of CL as a highly declarative programming language, and the process of creating {\em KB} as programming in it.

The knowledgebase {\em KB} of the system may include atomic elementary formulas expressing factual knowledge, such as $\mbox{\em Female}\hspace{1pt}(\mbox{\em Dana})$, or non-atomic elementary formulas expressing general knowledge, such as $\cla x
\bigl(\cle y\mbox{\em Father}\hspace{1pt}(x,y)\mli\mbox{\em Male}\hspace{1pt}(x)\bigr)$ or $\cla x\cla 
y\bigl(x\times(y+1)=(x\times y)+x\bigr)$; it can also include nonclassical formulas such as 
$\st\ada x\bigl(\mbox{\em Female}(x)\add\mbox{\em
 Male}(x)\bigr)$, expressing potential knowledge of everyone's gender, or $\st\ada x\ade y(x^2=y)$, 
expressing ability to repeatedly compute the square function, or something more complex and more interactive, such as formula (\ref{diagn}). With each resource $R\in${\em KB} is associated (if not physically, at least conceptually) its 
{\bf provider}\index{provider}\label{0provider} --- an agent that solves the query $R$ for the system, i.e. plays the game $R$ against the system. 
Physically the provider could be a computer program allocated to the system, or a network server having the system 
as a client, or another knowledgebase system to which the system has querying access, or even human personnel servicing the system. E.g., the provider for $\st\ada x\ade y\mbox{\em Bloodpressure}\hspace{1pt}(x,y)$ would probably be a team of nurses repeatedly performing the task of measuring the blood pressure of a patient specified by the system and reporting the outcome back to the system. Again, we do not think of providers as a part of the system itself. The latter only sees {\em what} resources are available to it, without knowing or caring about {\em how} the corresponding providers do their job; furthermore,
the system does not even care {\em whether} the providers really do their job right.   
The system's responsibility is only to correctly solve queries for the user {\em as long as} none of the providers 
fail to do their job.  Indeed, if the system misdiagnoses a patient because a nurse-provider gave it wrong information 
about that patient's blood pressure, the hospital (ultimate user) is unlikely to  fire the system and demand refund from its vendor; more likely, it would fire the nurse. 
Of course, when $R$ is elementary, the 
provider has nothing to do, and its successfully playing $R$ against the system simply means that $R$ is true. 
Note that in the picture that we have just presented, the system plays each game $R\in${\em KB} in the role of $\oo$, so that, from the system's perspective, the game that it plays against the provider of $R$ is $\gneg R$ rather than $R$.

The most typical internal informational resources, such as factual knowledge or queries solved by computer programs, can be reused an arbitrary number of times and with unlimited branching capabilities, i.e. in the strong sense captured by $\st$, and thus they would be prefixed with a $\st$ as we did with $\ada x\bigl(\mbox{\em Female}(x)\add\mbox{\em
 Male}(x)\bigr)$ and $\ada x\ade y(x^2=y)$. There was 
no point in $\st$-prefixing $\mbox{\em Female}\hspace{1pt}(\mbox{\em Dana})$, $\cla x
\bigl(\cle y\mbox{\em Father}\hspace{1pt}(x,y)\mli\mbox{\em Male}\hspace{1pt}(x)\bigr)$  or $\cla x\cla y\bigl(x\times(y+1)=(x\times 
y)+x\bigr)$ because every elementary game $A$ is equivalent to $\st A$ and hence remains ``recyclable" even without recurrence operators. As noted in Section \ref{ss4.6}, there is no difference between $\st$ and $\pst$ as long as ``simple" 
resources such as $\ada x\ade y(x^2=y)$ are concerned. However, 
in some cases --- say, when a resource with a high degree of interactivity is supported by an unlimited number of independent providers each of which however allows to run only one single ``session"  --- the weaker operator $\pst$ will have to be used instead of $\st$. 
Yet, some of the internal informational resources could be essentially non-reusable. A single provider possessing 
a single item of disposable pregnancy test device would apparently be able to support the resource  $\ada x(\mbox{\em Pregnant}(x)\add\gneg \mbox{\em Pregnant}\hspace{1pt}(x)\bigr)$ but not  $\st \ada x(\mbox{\em Pregnant}\hspace{1pt}(x)\add\gneg \mbox{\em Pregnant}\hspace{1pt}(x)\bigr)$
and not even $\ada x(\mbox{\em Pregnant}\hspace{1pt}(x)\add\gneg \mbox{\em Pregnant}\hspace{1pt}(x)\bigr)\mlc \ada x(\mbox{\em Pregnant}\hspace{1pt}(x)\add\gneg \mbox{\em Pregnant}\hspace{1pt}(x)\bigr)$. Most users, however, would try to refrain from including this sort of a resource into 
{\em KB}, and rather make it a part (antecedent) of possible queries. Indeed, knowledgebases with non-recyclable resources would tend to weaken from query to query and require more careful maintainance and updates. 
Whether recyclable or not, all of the resources of {\em KB} can be used independently and in parallel. This is exactly what allows us to identify {\em KB} with the $\mlc$-conjunction of its elements. 

Assume {\em KB}\hspace{2pt}$=R_1\mlc\ldots\mlc R_n$, and let us now try to visualize a system solving a query $F$ for the user. The designer would probably select an interface where the user only sees the moves made by the system in 
$F$, and hence gets the illusion that the system is just playing $F$. But in fact the game that the system is really playing 
is {\em KB}\hspace{2pt}$\mli F$, i.e. $\gneg R_1\mld\ldots\mld\gneg R_n\mld F$. Indeed, the system is not only interacting with the user 
in $F$, but --- in parallel --- also with its resource providers against whom, as we already know, it plays $\gneg R_1,\ldots,\gneg R_n$. As long as those providers do not fail to do their job, the system loses each of  
the games $\gneg R_1,\ldots,\gneg R_n$. Then our semantics for $\mld$ implies that the system wins its play over  
the ``big game" $\gneg R_1\mld\ldots\mld\gneg R_n\mld F$ if and only if it wins it in the $F$ component, i.e. 
successfully solves the query $F$. 

Thus, the system's ability to solve a query $F$ reduces to its ability to generate a solution for {\em KB}\hspace{2pt}$\mli F$, i.e. a reduction of $F$ to {\em KB}. What would give the system such an ability is built-in knowledge of CL --- in particular, a {\bf uniform-constructively sound 
axiomatization}\index{uniform-constructively sound axiomatization}\label{0uniform-constructively sound axiomatization}
of it, by which we mean a deductive system $S$ (with effective proofs of its theorems) that satisfies the Uniform-Constructive Soundness clause of Theorem \ref{main5} with ``$S$" in the role of $\predell$. According to the uniform-constructive soundness property, it would be sufficient for the system to find a proof of {\em KB}$\mli F$, which would allow it to (effectively) construct a machine $\cal M$ and then run it on {\em KB}$\mli F$ with a guaranteed success. 
  
Notice that it is uniform-constructive soundness rather than simple soundness of the 
the built-in (axiomatization of the) logic that allows the knowledgebase system to function. Simple soundness just means that every provable formula is valid. This is not sufficient for two reasons. 

One reason is that validity of a formula $E$ only implies that, for every interpretation $^*$, a solution for the problem $E^*$ exists. It may be the case, however, that 
different interpretations require different solutions, so that choosing the right solution requires knowledge of the 
actual interpretation, i.e. the {\em meaning}, of the atoms of $E$. Our assumption is that the system has no 
nonlogical knowledge, which, in more precise terms, means nothing but   that it has no knowledge of the interpretation $^*$. Thus, a solution that the system 
generates for $E^*$ should be successful for any possible interpretation $^*$. In other words, it should be a uniform solution for $E$. This is where uniform-constructive soundness of the underlying logic becomes  
critical, by which every provable formula is not only valid, but also uniformly valid. Going back to the example with which this section started, the reason why 
$p\add\gneg p$ fails in the context of computability theory is that it is not valid. On the other hand, the reason for the failure 
of this principle in the context of knowledgebase systems is that it is not uniformly valid: a solution for it, 
even if such existed for each interpretation $^*$, would depend on whether $p^*$ is true or false, and the system would  be unable to 
figure out the truth status of $p^*$ unless this information was explicitly or implicitly contained in {\em KB}. Thus, for knowledgebase systems the primary semantical concept of interest is uniform validity rather than validity.

The other reason why simple soundness of the built-in logic would not be sufficient for a knowledgebase system to 
function --- even if every provable formula was known to be uniformly valid --- is the following. With simple soundness, after finding a proof of $E$, even though the system would know that a solution for $E^*$ exists, it might have no way to actually find such a solution. On the other hand, uniform-constructive soundness guarantees that a (uniform) solution for every provable formula not only exists, but can be effectively extracted from a proof.

As for completeness of the built-in logic,  unlike uniform-constructive soundness, it is a desirable but not necessary condition. So far a complete axiomatization has been 
found only for the fragment of CL limited to the language of $\predell$. We hope that the future will bring completeness results for more expressive fragments as well. But even if not, we can still certainly 
succeed in finding ever stronger axiomatizations that are uniform-constructively sound even if not necessarily complete. 
Extending $\predell$ with some straightforward rules such as the ones that allow to replace $\st F$ by $F\mlc \st F$ and $\pst F$ by $F\mlc\pst F$, the rules $F\mapsto \st F$, $F\mapsto \pst F$, etc. 
would already immensely strengthen the logic. Our soundness proof for the incomplete affine logic given later is another result in 
a similar direction. 
It should be remembered that, when it 
comes to practical applications in the proper sense, the logic that will be used is likely to be far from complete anyway. For example, the popular classical-logic-based systems and programming languages are incomplete, and the reason is not that 
a complete axiomatization for classical logic is not known, but rather the unfortunate fact of life that often 
efficiency only comes at the expense of completeness. 

But even $\predell$, despite the absence of recurrence operators in it, is already very powerful. Why don't we see a simple example to get  
the taste of it as a query-solving logic. Let {\em Acid}$(x)$ mean 
``solution $x$ contains acid", and {\em Red}$(x)$ mean ``litmus paper turns red in solution $x$". 
Assume that the knowledgebase {\em KB} of a $\predell$-based system contains $
\cla x\bigl(\mbox{\em Red}(x)\leftrightarrow\mbox{\em Acid}(x)\bigr)$ and
$\ada x\bigl(\mbox{\em Red}(x)\add\gneg \mbox{\em Red}(x)\bigr)$, accounting for knowledge of the fact that a solution contains acid iff the litmus paper turns red in it, and for  
availability of a provider who possesses a piece of litmus paper that it can dip into any solution and report the paper's color to the system. 
Then the system 
can solve the acidity query $\ada x\bigr(\mbox{\em Acid}(x)\add\gneg\mbox{\em Acid}(x)\bigr)$. 
This follows from the fact, left as an exercise for the reader to verify, that $\predell\vdash \mbox{\em KB}\mli \ada x\bigr(\mbox{\em Acid}(x)\add\gneg\mbox{\em Acid}(x)\bigr)$.
 
Section 26 of \cite{Jap03} outlines how the context of knowledgebase systems can be further extended to 
systems for planning and action. Roughly, the formal semantics in such applications remains the same, and what changes is only the underlying philosophical assumption that the truth values of predicates and propositions are fixed or  predetermined. 
Rather, those values in CL-based planning systems are viewed as something that interacting agents may be able to manage. That is, predicates or propositions 
there stand for {\em tasks} rather than {\em facts}. E.g. {\em Pregnant}$(${\em Dana}$)$ --- or, perhaps,
{\em Impregnate}$(${\em Dana}$)$ instead --- can be seen as having no predetermined truth value, with Dana or her mate being in control of whether to make it true or not. And the nonelementary formula $\ada x\mbox{\em Hit}(x)$ describes the task of hitting any one target $x$ selected by the environment/commander/user. Note how naturally resource-consciousness arises here: while 
$\ada x\mbox{\em Hit}(x)$ is a task accomplishable with one ballistic missile, the stronger task  $\ada x\mbox{\em Hit}(x)\mlc\ada x\mbox{\em Hit}(x)$ would require two missiles instead. All of the other operators of CL, too, have natural interpretations as operations on physical (as opposed to just informational) tasks, with $\mli$ acting as a task reduction operation. To get a feel of this, 
let us look at the task

\begin{center}
$\begin{array}{l}
\mbox{\em Give me a wooden stake $\adc$ Give me a silver bullet}\\  
\mli \ \ \mbox{\em Destroy the vampire\ $\adc$ Kill the werewolf}.
\end{array}$
\end{center}
This is a task accomplishable by an 
agent who has  a mallet and a gun as well as sufficient time, energy and bravery  to chase and eliminate any one (but not both) of the two monsters, and 
only needs  a wooden stake and/or a silver bullet to complete his noble mission. Then the story told by the legal run $\seq{\oo 2.2,\pp 1.2}$
of the above game is that the environment asked the agent to kill the werewolf, to which the agent replied by the counterrequest to give him a silver bullet. The task will be considered eventually accomplished by the agent iff 
he indeed killed the werewolf as long as a silver bullet was indeed given to him. 

The fact that CL is a conservative extension of classical logic also makes the former a reasonable and appealing alternative to the latter in its most traditional and unchallenged application areas. In particular, it makes perfect sense to base applied theories --- such as, say, Peano arithmetic (axiomatic number theory) --- on CL instead of classical logic. Due to conservativity, no old information 
would be lost or weakened this way. On the other hand, we would get by an order of magnitude more expressive, constructive and computationally meaningful theories than their classical-logic-based versions. Let us see a little more precisely 
what we mean by a CL-based applied theory. 
For simplicity, here we restrict our considerations to the cases when the set {\em AX} of nonlogical {\bf axioms} of the
applied theory is finite. As we did with {\em KB}, we identify {\em AX} with the $\mlc$-conjunction of its elements. 
From (the problem represented by) {\em AX} --- or, equivalently, each conjunct of it --- we require to be computable in our sense, i.e. come with an HPM or EPM that solves it. So, notice, all axioms of the old, classical-logic-based version of the theory could be automatically included into the new set {\em AX} because they represent true and hence computable elementary problems. Many of those old axioms can be constructivized by, say, replacing blind or parallel operators with their choice equivalents. E.g., we would want to rewrite the axiom $\cla x\cle y(y=x+1)$ of arithmetic as the more informative $\ada x\ade y(y=x+1)$. And, of course, to the old axioms or their constructivized versions could be added some essentially new axioms expressing basic computability principles specific to (the particular interpretation underlying) the theory. Provability (theoremhood) of a formula $F$ in such a  
theory we understand as provability of the formula {\em AX}$\mli F$ in the underlying axiomatization of CL which, 
as in the case of knowledgebase systems, is assumed to be uniform-constructively sound. The rule of modus ponens 
has been shown in \cite{Jap03} (Proposition 21.3)\footnote{In the official formulation of Proposition 21.3 in \cite{Jap03}, the first argument of $h$ was an HPM. In view of Theorem \ref{eq}, however, replacing ``HPM'' with ``EPM'' is perfectly legitimate.} to preserve computability in the following constructive sense:

\begin{theorem}\label{june27}
 There is an effective  function  \ $h$: \{EPMs\}$\times$\{EPMs\} $\longrightarrow$ \{EPMs\} such that, for any EPMs $\cal E$,$\cal C$, static games 
$A$,$B$ and valuation $e$, if ${\cal E}\models_e A$ and ${\cal C}\models_e A\mli B$, then $h({\cal E},{\cal C})\models_e B$.  
\end{theorem}

\noindent This theorem, together with our assumptions that {\em AX} is computable and that the underlying logic is 
uniform-constructively sound, immediately implies that the problem represented by any theorem $F$ of the applied theory is computable and that, furthermore, a solution for such a problem can be effectively constructed from a 
proof of $F$.
So, for example, once a formula $\ada x\ade y\hspace{1pt} p(x,y)$ has been proven, we would know that, for every 
$x$, a $y$ with $p(x,y)$ not only exists, but can be algorithmically found; furthermore, we would be able to actually construct such an algorithm. Similarly, a reduction --- in the sense of Definition \ref{feb2}(4) --- of the 
acceptance problem to the halting problem would automatically come with a proof of 
$\ada x\ada y\bigl(H(x,y)\add\gneg H(x,y)\bigr)\mli
\ada x\ada y\bigl(A(x,y)\add\gneg A(x,y)\bigr)$ in such a theory. 
Is not this exactly what the constructivists have been calling for?..

\section{Affine logic}\label{ss11}
 
Linear logic and its variations such as affine logic have only one group $!,?$ of exponential operators. The semantics of CL induces at least two equally natural ``counterparts'' of $!,?$: the parallel group $\pst,\pcost$ 
and the branching group $\st,\cost$ of recurrence operators. Hence, when rewritten in terms of computability logic, 
each $(!,?)$-involving rule of linear logic produces two identical versions: one with $(\pst,\pcost)$ and one with $(\st,\cost)$.

Precisely, the language of what we here call {\bf affine logic}\index{affine logic}\label{0affine logic2} $\al$ is obtained from the more expressive language of Section \ref{ss7} by forbidding nonlogical elementary atoms (but not the logical elementary atoms $\twg$ and $\tlg$), and restricting the operators of the language to 
$\gneg$, $\mlc$, $\mld$, $\adc$, $\add$, $\pst$, $\pcost$, $\st$, $\cost$, 
$\ada$, $\ade$. 
For simplicity, this list does not officially include $\mli$ or other definable operators such as $\pintimpl$ and 
$\intimpl$. If we write $F\mli G$, it should be understood as an abbreviation of $\gneg F\mld G$. 
Furthermore, without loss of expressive power, we allow
$\gneg$ to be applied only to nonlogical atoms, in all other cases understanding $\gneg F$ as an abbreviation defined by: \ 
$\gneg\twg = \tlg$; \ 
$\gneg \tlg=\twg$; \ 
$\gneg\gneg F=F$; \ 
$\gneg(F_1\mlc\ldots\mlc F_n)=\gneg F_1\mld\ldots\mld\gneg F_n$; \ 
$\gneg(F_1\mld\ldots\mld F_n)=\gneg F_1\mlc\ldots\mlc\gneg F_n$; \ 
$\gneg(F_1\adc\ldots\adc F_n)=\gneg F_1\add\ldots\add\gneg F_n$; \ 
$\gneg(F_1\add\ldots\add F_n)=\gneg F_1\adc\ldots\adc\gneg F_n$; \
$\gneg\pst F=\pcost\gneg F$; \ 
$\gneg\pcost F=\pst\gneg F$; \ 
$\gneg\st F=\cost\gneg F$; \ 
$\gneg\cost F=\st\gneg F$; \  
$\gneg\ada x F=\ade x\gneg F$; \ 
$\gneg\ade x F=\ada x\gneg F$.
The formulas of this language will be referred to as {\bf $\al$-formulas}.\index{$\al$-formula}\label{0al-formula}

Let $x$ be a variable, $t$ a term and $F(x)$ a formula. We say that $t$ is {\bf free for $x$ in $F(x)$} iff 
none of the free occurrences of $x$ in $F(x)$ is in the scope of ${\cal Q} t$ for some quantifier $\cal Q$. Of course, when $t$ is a constant, this condition is always satisfied.

A {\bf sequent}\index{sequent}\label{0sequent} is a nonempty finite sequence of $\al$-formulas. We think of each sequent $F_1,\ldots,F_n$ as the formula 
$F_1\mld\ldots\mld F_n$.
This allows us to automatically extend the concepts of validity, uniform validity, free occurrence, etc. from formulas to sequents. A formula $F$ is considered {\bf provable} in $\al$ iff $F$, understood as a one-element sequent, is provable. 

Deductively logic $\al$\index{$\al$}\label{0al} is given by the following 16 rules, where:
$\s{G},\s{H}$ are arbitrary (possibly empty) sequences of $\al$-formulas; $\s{\pcost G}$ is an arbitrary (possibly empty) sequence of $\pcost$-prefixed $\al$-formulas; $\s{\cost G}$ is an arbitrary (possibly empty) sequence of $\cost$-prefixed $\al$-formulas; $n\geq 2$; $1\leq i\leq n$; $x$ is any variable; $E$, $F$, $E_1$, \ldots, $E_n$, $E(x)$ are any $\al$-formulas; $y$ is any variable not occurring (whether free or within $\ada y$ or $\ade y$) in the conclusion
of the rule; and $t$ is any term free for $x$ in $E(x)$.\vspace{10pt}

\begin{center}
\begin{picture}(250,32)
\put(0,20){\bf Identity Axiom:}
\put(120,22){\line(1,0){30}}
\put(120,8){$\gneg E,E$}
\end{picture}
\end{center}

\begin{center}
\begin{picture}(250,32)
\put(0,20){\bf $\twg$-Axiom:}
\put(120,22){\line(1,0){30}}
\put(131,8){$\twg$}
\end{picture}
\end{center}

\begin{center}
\begin{picture}(250,40)
\put(120,30){$\s{G},E,F,\s{H}$}
\put(0,20){\bf Exchange:}
\put(120,22){\line(1,0){50}}
\put(120,8){$\s{G},F,E,\s{H}$}
\end{picture}
\end{center}

\begin{center}
\begin{picture}(250,40)
\put(126,30){$\s{G}$}
\put(0,20){\bf Weakening:}
\put(120,22){\line(1,0){22}}
\put(120,8){$\s{G},E$}
\end{picture}
\end{center}

\begin{center}
\begin{picture}(250,40)
\put(120,30){$\s{G},\pcost E,\pcost E$}
\put(0,20){\bf $\pcost$-Contraction:}
\put(120,22){\line(1,0){46}}
\put(129,8){$\s{G},\pcost E$}
\end{picture}
\end{center}

\begin{center}
\begin{picture}(250,40)
\put(120,30){$\s{G},\cost E,\cost E$}
\put(0,20){\bf $\cost$-Contraction:}
\put(120,22){\line(1,0){48}}
\put(130,8){$\s{G},\cost E$}
\end{picture}
\end{center}

\begin{center}
\begin{picture}(250,40)
\put(144,30){$\s{G},E_i$}
\put(0,20){\bf $\add$-Introduction:}
\put(120,22){\line(1,0){74}}
\put(120,8){$\s{G},E_1\add\ldots\add E_n$}
\end{picture}
\end{center}

\begin{center}
\begin{picture}(250,40)
\put(120,30){$\s{G},E_1$\hspace{20pt}{\Huge$\ldots$}\hspace{20pt}$\s{G},E_n$}
\put(0,20){\bf $\adc$-Introduction:}
\put(120,22){\line(1,0){120}}
\put(144,8){$\s{G},E_1\adc\ldots\adc E_n$}
\end{picture}
\end{center}

\begin{center}
\begin{picture}(250,40)
\put(125,30){$\s{G},E_1,\ldots,E_n$}
\put(0,20){\bf $\mld$-Introduction:}
\put(120,22){\line(1,0){75}}
\put(120,8){$\s{G},E_1\mld\ldots\mld E_n$}
\end{picture}
\end{center}

\begin{center}
\begin{picture}(250,40)
\put(120,30){$\s{G_1},E_1$\hspace{20pt}{\Huge$\ldots$}\hspace{20pt}$\s{G_n},E_n$}
\put(0,20){\bf $\mlc$-Introduction:}
\put(120,22){\line(1,0){130}}
\put(126,8){$\s{G_1},\ldots,\s{G_n},E_1\mlc\ldots\mlc E_n$}
\end{picture}
\end{center}

\begin{center}
\begin{picture}(250,40)
\put(123,30){$\s{G}, E$}
\put(0,20){\bf $\pcost$-Introduction:}
\put(120,22){\line(1,0){28}}
\put(120,8){$\s{G},\pcost E$}
\end{picture}
\end{center}

\begin{center}
\begin{picture}(250,40)
\put(123,30){$\s{G}, E$}
\put(0,20){\bf $\cost$-Introduction:}
\put(120,22){\line(1,0){28}}
\put(120,8){$\s{G},\cost E$}
\end{picture}
\end{center}

\begin{center}
\begin{picture}(250,40)
\put(123,30){$\s{\pcost G}, E$}
\put(0,20){\bf $\pst$-Introduction:}
\put(120,22){\line(1,0){33}}
\put(120,8){$\s{\pcost G},\pst E$}
\end{picture}
\end{center}

\begin{center}
\begin{picture}(250,40)
\put(123,30){$\s{\cost G}, E$}
\put(0,20){\bf $\st$-Introduction:}
\put(120,22){\line(1,0){33}}
\put(120,8){$\s{\cost G},\st E$}
\end{picture}
\end{center}

\begin{center}
\begin{picture}(250,40)
\put(130,30){$\s{G}, E(t)$}
\put(0,20){\bf $\ade$-Introduction:}
\put(120,22){\line(1,0){52}}
\put(120,8){$\s{G},\ade x E(x)$}
\end{picture}
\end{center}

\begin{center}
\begin{picture}(250,40)
\put(130,30){$\s{G}, E(y)$}
\put(0,20){\bf $\ada$-Introduction:}
\put(120,22){\line(1,0){52}}
\put(120,8){$\s{G},\ada x E(x)$}
\end{picture}
\end{center}

Unlike any other results that we have surveyed so far, the soundness and completeness of affine logic, while claimed already in \cite{Jap03}, has never been officially proven. For this reason, the following theorem comes with a full proof, to which most of the remaining part of this paper is devoted. 
 
\begin{theorem}\label{main}
If $\al\vdash S$, then $\valid S$ $($any sequent $S$$)$. Furthermore:

 {\bf Uniform-Constructive Soundness:}  There is an effective procedure that takes any $\al$-proof of any sequent $S$ and constructs a uniform solution for $S$. 
\end{theorem}

As mentioned earlier, a similar (uniform-constructive) soundness theorem for Heyting's intuitionistic calculus
has been proven in \cite{int1}, with 
intuitionistic implication understood as $\intimpl$, and intuitionistic conjunction, disjunction and quantifiers as $\adc,\add,\ada,\ade$.

\section{Soundness proof for affine logic}\label{snew}
This technical section is devoted to a proof of Theorem \ref{main}. It also contains a number of useful lemmas that could be employed in other proofs.
 
\subsection{$\predell$-derived validity lemmas}\label{ss12}
 
In our proof of Theorem \ref{main} we will need a number of lemmas concerning uniform validity of certain formulas. Some of such validity proofs will be given directly in Sections \ref{mvl} and \ref{prf}. But some proofs come ``for free", based on the already known soundness of $\predell$. In fact, here we will only exploit the propositional fragment $\clt$\index{$\clt$}\label{0clt} of $\predell$. 
The former is obtained from the latter by mechanically restricting its language to $0$-ary letters, and disallowing the 
(now meaningless) usage of quantifiers. Let us call the formulas of such a language {\bf $\clt$-formulas}.\index{$\clt$-formula}\label{0clt-formula} Restricting the language to $\clt$-formulas simplifies the formulation of $\predell$: Rule {\bf B2} disappears, and so does clause (ii) of Rule {\bf A}. $\predell$ is a conservative extension of $\clt$, so, for a $\clt$-formula $F$, it does not matter whether we say $\predell\vdash F$ or $\clt\vdash F$. 

In Section \ref{ss11} we started using the notation $\s{G}$ for sequences of formulas. We also agreed to identify sequences 
of formulas with $\mld$-disjunctions of those formulas. So, from now on, an underlined expression such as $\s{G}$ will mean 
an arbitrary formula $G_1\mld \ldots\mld G_n$ for some $n\geq 0$. The expression $\s{G}\mld E$ should be read as 
$G_1\mld \ldots\mld G_n\mld E$ rather than as $(G_1\mld \ldots\mld G_n)\mld E$. The number of disjuncts in $\s{G}$ may be empty. When this is a possibility, $\s{G}$ will  
 usually occur as a disjunct within a bigger expression such as $\s{G}\mld E$ or $\s{G}\mli E$, both of which simply mean $E$.  

As we agreed that $p,q,r,s$ (with no tuples of terms attached) stand for nonlogical elementary atoms and $P,Q,R,S$ for general atoms, 
$\s{p},\s{q},\s{r},\s{s}$ will correspondingly stand for $\mld$-disjunctions of elementary atoms, and $\s{P},\s{Q},\s{R},\s{S}$ for 
$\mld$-disjunctions of general atoms. 

We will also be underlining complex expressions such as $G\mli H$,  $\ade xG(x)$ or $\cost G$. $\s{G\mli H}$ should be understood as $(G_1\mli H_1)\mld\ldots\mld (G_n\mli H_n)$, \ $\s{\ade xG(x)}$ as $\ade x G_1(x)\mld\ldots\mld\ade xG_n(x)$ (note that only the $G_i$ vary but not $x$), \ $\s{\cost G}$ as $\cost G_1\mld\ldots\mld \cost G_n$, 
$\cost\vspace{1pt}\s{\cost G}$ as $\cost(\cost G_1\mld\ldots\mld\cost G_n)$, etc.

A $\clt$-formula $E$ is said to be {\bf general-base}\index{general-base formula}\label{0general-base formula} iff it does not contain any elementary atoms. 
A {\bf substitution}\index{substitution}\label{0substitution} is a function $\sigma$ that sends every general atom $P$ of the language of $\clt$ 
to a (not necessarily $\clt$-) formula $\sigma(P)$. This mapping extends to all general-base $\clt$-formulas by stipulating that $\sigma$ commutes with each operator: $\sigma(\gneg E)=\gneg \sigma(E)$,  
$\sigma(E_1\adc\ldots\adc E_k)=\sigma(E_1)\adc\ldots\adc \sigma(E_k)$, etc.
We say that a formula $F$ is a {\bf substitutional instance}\index{substitutional instance}\label{0substitutional instance} of a general-base $\clt$-formula $E$ iff $F=\sigma(E)$ for some substitution $\sigma$. 
Thus, ``$F$ is a substitutional instance of $E$'' means that $F$ has the same form as $E$. 

In the following lemma, we assume $n\geq 2$, and $1\leq i\leq n$. Note that the expressions given in clauses (d)-(k) are schemata of formulas rather than formulas, for the lengths of their underlined expressions, as well as $i$ and $n$,  may vary. Strictly speaking, the expressions of clauses (a)-(c) are so as well, because $P,Q,R$ are metavariables for general atoms rather than particular general atoms.

\begin{lemma}\label{l8}
All substitutional instances of all $\clt$-formulas given by the following schemata are uniformly valid. Furthermore, there is an effective procedure that takes any particular formula matching a given scheme and constructs 
an EPM that is a uniform solution for all substitutional instances of that formula.\vspace{7pt} 

a) $\gneg P\mld P$;\vspace{4pt}

b) $P\mld Q \mli Q\mld P$;\vspace{4pt}

c) $(P\mli Q)\mlc (Q\mli R)\mli (P\mli R)$;\vspace{4pt}

d) $(\s{Q_1}\mld P_1)\mlc\ldots\mlc (\s{Q_n}\mld P_n)\mli\s{Q_1}\mld\ldots\mld\s{Q_n}\mld (P_1\mlc\ldots\mlc P_n)$;\vspace{4pt}

e) $\s{(P\mli Q)}\mli (\s{R}\mld \s{P}\mld\s{S}\mli \s{R}\mld\s{Q} \mld\s{S})$;\vspace{4pt}

f) $\s{Q}\mld\s{R}\mld\s{S}\mli\s{Q}\mld(\s{R})\mld\s{S}$;\vspace{4pt}

g) $\s{Q}\mld(\s{R})\mld\s{S}\mli\s{Q}\mld\s{R}\mld\s{S}$;\vspace{4pt}

h) $\bigl(P_1\mlc P_2\mlc \ldots\mlc P_n \mli Q\bigr)\mli \bigl(P_1\mli(P_2\mli\ldots(P_n\mli Q)\ldots)\bigr)$;\vspace{4pt}

i) $\s{Q}\mli\s{Q}\mld P$;\vspace{4pt}

j) $P_i\mli P_1\add\ldots\add P_n$;\vspace{4pt}

k) $(\s{Q}\mld P_1)\mlc\ldots\mlc(\s{Q}\mld P_n)\mli \s{Q}\mld (P_1\adc\ldots\adc P_n)$.

\end{lemma}

\begin{proof} 
In order to prove this lemma, it would be sufficient to show that all formulas given by the above schemata are provable in $\predell$ (in fact, $\clt$). Indeed, if we succeed in doing so, then an effective procedure whose existence is claimed in the present lemma could be designed to work as follows. First, the procedure finds a $\predell$-proof of a given formula $E$. Then, based on that proof and using the procedure whose existence is stated in Theorem \ref{main5}, it finds a uniform solution $\cal E$ for that formula. It is not hard to see that the same $\cal E$ will automatically be a uniform solution for every substitutional instance of $E$ as well. 
So, now it remains to do the simple syntactic exercise of checking $\predell$-provabilities for each clause of the lemma. 

Notice that every formula $E$ given by one of the clauses (a)-(h) has --- more precisely, we may assume that it has --- exactly two, one negative and one positive, occurrences of each (general) atom, with all occurrences being surface ones. For such an $E$, let $E'$ be the result of rewriting each general atom $P$ of $E$ into a nonlogical elementary atom $p$ in such a way that different general atoms are rewritten as different elementary atoms. Then $E$ follows from $E'$ in $\predell$ by a series of applications of Rule {\bf C}, specifically, as many applications as the number of different atoms of $E$. In turn, observe that for each of the clauses (a)-(h), the formula $E'$ would be a classical tautology. Hence $E'$ follows from the empty set of premises by Rule {\bf A}. Thus, $\predell\vdash E$.

For clause (i), let $\s{q}$ be the result of replacing in $\s{Q}$ all atoms by pairwise distinct nonlogical elementary atoms. The formula $\s{q}\mli\s{q}\mld P$ is stable and choice-operator-free, so it follows from $\{\}$ by Rule {\bf A}. From the latter, applying Rule {\bf C} as many times as the number of disjuncts in $\s{Q}$, we obtain 
the desired $\s{Q}\mli\s{Q}\mld P$.

For clause (j), the following is a $\predell$-proof of the corresponding formula(s):

\hspace{20pt}1. $p_i\mli p_i$ (from $\{\}$ by Rule {\bf A});

\hspace{20pt}2. $P_i\mli P_i$ (from 1 by Rule {\bf C});

\hspace{20pt}3.  $P_i\mli P_1\add\ldots\add P_n$ (from 2 by Rule {\bf B1}). 

For clause (k), note that $(\s{Q}\mld P_1)\mlc\ldots\mlc(\s{Q}\mld P_n)\mli \s{Q}\mld (P_1\adc\ldots\adc P_n)$ is stable. Hence it follows by Rule {\bf A} from $n$ premises, where each premise is 
$(\s{Q}\mld P_1)\mlc\ldots\mlc(\s{Q}\mld P_n)\mli \s{Q}\mld P_i$ for one of the $i\in\{1,\ldots,n\}$. Each such formula, in turn, can be obtained by a series of applications of Rule {\bf C} from
\[(\s{Q}\mld P_1)\mlc\ldots\mlc(\s{Q}\mld P_{i-1})\mlc(\s{q}\mld p_i)\mlc (\s{Q}\mld P_{i+1})\mlc\ldots\mlc(\s{Q}\mld P_n)\mli \s{q}\mld p_i,\]
where $p_i$ is an elementary nonlogical atom and $\s{q}$ is obtained from $\s{Q}$ by replacing its general atoms by pairwise distinct (and distinct from $p_i$) elementary nonlogical atoms. In turn, the above formula can be seen to be stable and hence, as it does not contain choice operators, derivable from the empty set of premises by Rule {\bf A}. 
\end{proof}

\subsection{Closure lemmas}\label{ss13}

In this section we let $n$ range over positive integers, $x$ over any variables, $E,F,G$ (possibly with subscripts)  over any $\al$-formulas, and $\cal E$, $\cal C$, $\cal D$ (possibly with subscripts) over any EPMs. Unless otherwise specified, in each context these 
metavariables are assumed to be universally quantified. 

First two of the following three lemmas have been proven in Section 21 of \cite{Jap03}. 
Here we provide a proof only for the third, never officially proven, one. 

\begin{lemma}\label{stclos}  
For any static game $A$, if $\models A$, then $\models \st A$.

Moreover, there is an effective function $h: \ \{\mbox{EPMs}\}\longrightarrow\{\mbox{EPMs}\}$  such that, for any EPM 
$\cal E$, static game $A$ and valuation $e$, if ${\cal E}\models_e A$, then $h({\cal E})\models_e \st A$.
\end{lemma} 

\begin{lemma}\label{adaclos}  
For any static game $A$, if $\models A$, then $\models \ada x A$.

Moreover, there is an effective function $h: \ \{\mbox{EPMs}\}\longrightarrow\{\mbox{EPMs}\}$  such that, for any EPM 
$\cal E$ and static game $A$, if ${\cal E}\models A$, then $h({\cal E})\models \ada x A$.
\end{lemma}

\begin{lemma}\label{pstclos}  
For any static game $A$, if $\models A$, then $\models \pst A$.

Moreover, there is an effective function $h: \ \{\mbox{EPMs}\}\longrightarrow\{\mbox{EPMs}\}$  such that, for any EPM 
$\cal E$, static game $A$ and valuation $e$, if ${\cal E}\models_e A$, then $h({\cal E})\models_e \pst A$.
\end{lemma}

\begin{proof} 
Intuitively the idea here is simple: if we (machine $\cal E$) know how to win $A$, then, applying the same strategy to each conjunct separately, we (machine $h({\cal E})$) can win the infinite conjunction $\pst A=A\mlc A\mlc A\mlc\ldots$ as well. 

To give a detailed description of the machine $h({\cal E})$ that materializes this idea, we need some preliminaries. 
Remember the $e$-successor relation between HPM configurations from Section \ref{icp}. In the context of a fixed HPM $\cal H$, valuation $e$ and configuration $C$, the transition from $C$ to a successor ($e$-successor) configuration $C'$ is nondeterministic because it depends on the sequence $\Psi$ of the moves (labeled with $\oo$) made by the environment while the machine was in configuration $C$. Once such a $\Psi$ is known, however, the value of $C'$ becomes determined and can be calculated from $C$, (the relevant finite part of) $e$ and (the transition function of) $\cal H$. We call the $e$-successor of $C$ uniquely determined by such $\Psi$ the {\bf $(e,\Psi)$-successor} of $C$ (in $\cal H$).  

On the way of constructing the EPM $h({\cal E})$, we first turn $\cal E$ into an HPM $\cal H$ such that, for every static game $A$ and valuation $e$, \mbox{${\cal H}\models_e A$} whenever ${\cal E}\models_e A$.  According to Theorem \ref{eq}, such an $\cal H$ can be constructed effectively. Now, using $\cal H$, we define $h({\cal E})$ to be the EPM which, with a valuation $e$ spelled on its valuation tape, acts as follows. Its work consists in iterating the following procedure 
ITERATION$(k)$ infinitely many times, starting from $k=1$ and incrementing $k$ by one at every new step. During each ITERATION$(k)$ step, $h({\cal E})$ maintains $k-1$ records $C_1,\ldots,C_{k-1}$ and creates one new record $C_k$, with 
each such $C_i$ holding a certain configuration of $\cal H$. Here is how ITERATION$(k)$ proceeds:\vspace{10pt}

\noindent {\bf Procedure} ITERATION$(k)$:\vspace{5pt} 

\noindent 1. Grant permission. Let $\Psi=\seq{\oo\alpha}$ if the adversary responds by a move $\alpha$, and $\Psi=\emptyrun$ if there is no response.\vspace{5pt}

\noindent 2. For $i=1$ to $i=k-1$, do the following:

a) If $\cal H$ makes a move $\beta$ in configuration $C_i$, make the move $i.\beta$;

b) Update $C_i$ to the $(e,\Psi^{i.})$-successor\footnote{For $\Psi^{i.}$, remember the notation $\Gamma^\alpha$ from page \pageref{apr2}.} of $C_i$.\vspace{5pt} 

\noindent 3. Let $C$ be the initial configuration of $\cal H$, and $\Phi$ the position currently spelled on the run tape.

a) If $\cal H$ makes a move $\beta$ in configuration $C$, make the move $k.\beta$;

b) Create the record $C_k$ and initialize it to the $(e,\Phi^{k.})$-successor of $C$.\vspace{10pt}

Obviously (the description of) $h({\cal E})$ can be effectively obtained from $\cal H$ and hence from $\cal E$, so that, as promised, 
$h$ is indeed an effective function. What remains to verify is that, whenever $\cal E$ wins a static game $A$ on  
a valuation $e$, we have $h({\cal E})\models_e \pst A$. Consider any such $A$ and $e$, and suppose $h({\cal E})\not\models_e \pst A$. We want to show that then ${\cal E}\not\models_e A$.
Let $B$ be an arbitrary $e$-computation branch of $h({\cal E})$, and let $\Gamma$ be the run spelled by $B$. Permission is granted at the beginning of each of the infinitely many routines ITERATION$(k)$, so $B$ is fair. Therefore, $h({\cal E})\not\models_e A$ simply means that $\win{\psti A}{e}\seq{\Gamma}=\oo$. The latter, in turn, implies that for some $n\in\{1,2,3,\ldots\}$, $\win{A}{e}\seq{\Gamma^{n.}}=\oo$. This can be easily seen from the fact that every move that $h({\cal E})$ makes starts with an `$n.$' for some $n$. But an analysis of the procedure followed by $h({\cal E})$ can convince us that $\Gamma^{n.}$ is the run spelled by some $e$-computation branch of $\cal H$. This means that ${\cal H}\not\models_e A$. Remembering that 
 ${\cal H}\models_e A$ whenever ${\cal E}\models_e A$, we find that ${\cal E}\not\models_e A$. 
\end{proof}

\begin{lemma}\label{pl10}
If $\uvalid E$, then $\uvalid \pst E$.

Moreover, there is an effective function $h: \ \{\mbox{EPMs}\}\longrightarrow\{\mbox{EPMs}\}$ 
such that, for any ${\cal E}$ and $E$, 
if ${\cal E}\uvalid E$, then $h({\cal E})\uvalid \pst E$. 
\end{lemma}

\begin{proof} As Lemma \ref{pstclos} asserts (or rather implies),  
 there is an effective function $h: \ \{\mbox{EPMs}\}\longrightarrow\{\mbox{EPMs}\}$  such that, for any EPM 
$\cal E$ and any static game $A$, if ${\cal E}\models A$, then $h({\cal E})\models \pst A$. We now claim for that very function $h$ that,  if ${\cal E}\uvalid E$, then $h({\cal E})\uvalid \pst E$. Indeed, assume ${\cal E}\uvalid E$. Consider any $\pst E$-admissible interpretation $^*$. Of course, the same interpretation is also $E$-admissible. Hence, ${\cal E}\uvalid E$ implies 
${\cal E}\models E^*$. But then, by the known behavior of $h$, we have $h({\cal E})\models \pst E^*$. Since $^*$ was arbitrary, we conclude that $h({\cal E})\uvalid \pst E$.
\end{proof}

\begin{lemma}\label{l10}
If $\uvalid E$, then $\uvalid \st E$.

Moreover, there is an effective function $h: \ \{\mbox{EPMs}\}\longrightarrow\{\mbox{EPMs}\}$ 
such that, for any ${\cal E}$ and $E$, 
if ${\cal E}\uvalid E$, then $h({\cal E})\uvalid \st E$. 
\end{lemma}

\begin{proof} Similar to Lemma \ref{pl10}, only use Lemma \ref{stclos} instead of Lemma \ref{pstclos}.
\end{proof}

\begin{lemma}\label{l10a}
If $\uvalid E$, then $\uvalid \ada x E$.

Moreover, there is an effective function $h: \ \{\mbox{EPMs}\}\longrightarrow\{\mbox{EPMs}\}$ 
such that, for any ${\cal E}$, $x$ and $E$, 
if ${\cal E}\uvalid E$, then $h({\cal E})\uvalid \ada x E$. 
\end{lemma}

\begin{proof} Similar to Lemma \ref{pl10}, only use Lemma \ref{adaclos} instead of Lemma \ref{pstclos}. 
\end{proof}

\begin{lemma}\label{l1a} {\bf (Modus ponens)} 
If $\uvalid F$ and $\uvalid F\mli E$, then $\uvalid E$. 

Moreover, there is an effective function $h: \ \{\mbox{EPMs}\}\times\{\mbox{EPMs}\}\longrightarrow\{\mbox{EPMs}\}$ 
such that, for any ${\cal E}$, $\cal C$, $F$ and $E$, 
if ${\cal E}\uvalid F$  and 
${\cal C}\uvalid F\mli E$, then $h({\cal E},{\cal C})\uvalid E$. 
\end{lemma}

\begin{proof} According to Theorem \ref{june27}, 
there is an effective  function  \ $h$: \{EPMs\}$\times$\{EPMs\} $\rightarrow$ \{EPMs\} such that, 
for any static games $A$,$B$, valuation $e$ and EPMs ${\cal E}$ and ${\cal C}$, 
\begin{equation}\label{oct9}\mbox{\em if ${\cal E}\models_e A$ and ${\cal C}\models_e A\mli B$, then $h({\cal E},{\cal C})\models_e B$.}
\end{equation}
We claim that such a function $h$ also behaves as our lemma states. To see this, 
assume ${\cal E}\uvalid F$  and 
${\cal C}\uvalid F\mli E$, and consider an arbitrary valuation $e$ and an arbitrary $E$-admissible interpretation $^*$.
Our goals is to show that $h({\cal E},{\cal C})\models_e E^*$, which obviously means the same as 
\begin{equation}\label{oct9a}
h({\cal E},{\cal C})\models_e e[E^*].
\end{equation}
We define the new interpretation $^\dagger$ by stipulating that, for every $k$-ary letter $L$ with $L^*= L^*(x_1,\ldots,x_k)$,  $L^\dagger$ is the game $L^\dagger (x_1,\ldots,x_k)$ such that, for any tuple $c_1,\ldots,c_k$ of constants, \[L^\dagger(c_1,\ldots,c_k)=e[L^*(c_1,\ldots,c_k)].\] Unlike $L^*(x_1,\ldots,x_k)$ that may depend on some ``hidden" variables (those that are not among $x_1,\ldots,x_k$),  obviously 
$L^\dagger (x_1,\ldots,x_k)$ does not depend on any variables other that $x_1,\ldots,x_k$. This makes $^\dagger$ admissible
for any $\al$-formula, including $F$ and $F\mli E$. Then our assumptions ${\cal E}\uvalid F$  and 
${\cal C}\uvalid F\mli E$ imply  ${\cal E}\models_e F^\dagger$  and 
${\cal C}\models_e F^\dagger\mli E^\dagger$. Consequently, by  
(\ref{oct9}), $h({\cal E},{\cal C})\models_e E^\dagger$,
i.e. $h({\cal E},{\cal C})\models_e e[E^\dagger]$. Now, with some thought, we can see that $e[E^\dagger]=e[E^*]$. Hence (\ref{oct9a}) is true. 
\end{proof}

\begin{lemma}\label{l1} {\bf (Generalized modus ponens)}
If $\uvalid F_1$, \ldots, $\uvalid F_n$ and $\uvalid F_1\mlc\ldots\mlc F_n\mli E$, then $\uvalid E$.

Moreover, 
there is an effective function 
$h: \ \{\mbox{EPMs}\}^{n+1}\longrightarrow\{\mbox{EPMs}\}$ 
such that, for any ${\cal E}_1$,\ldots,${\cal E}_n$,$\cal C$,$F_1$,\ldots,$F_n$,$E$, 
if ${\cal E}_1\uvalid F_1$, \ldots, ${\cal E}_n\uvalid F_n$  and 
${\cal C}\uvalid F_1\mlc\ldots\mlc F_n\mli E$, then $h({\cal E}_1,\ldots,{\cal E}_n, {\cal C})\uvalid E$. Such a function, in turn, can be effectively constructed for each particular $n$. 
\end{lemma}

\begin{proof}  
In case $n=1$, $h$ is the function whose existence is stated in Lemma \ref{l1a}. Below we will construct $h$ for the   
case $n=2$. It should be clear how to generalize that construction to 
any greater $n$. 

Assume ${\cal E}_1\uvalid F_1$, ${\cal E}_2\uvalid F_2$  and 
${\cal C}\uvalid F_1\mlc F_2\mli E$. 
By Lemma \ref{l8}(h), the formula $(F_1\mlc F_2\mli E)\mli (F_1\mli(F_2\mli E))$ has a uniform solution. 
Lemma \ref{l1a} allows us to combine that solution with ${\cal C}$ and find a uniform solution ${\cal D}_1$ for $F_1\mli(F_2\mli E)$. Now applying 
Lemma \ref{l1a} to ${\cal E}_1$ and ${\cal D}_1$, we find a uniform solution ${\cal D}_2$ for $F_2\mli E$. Finally,
 applying the same lemma to ${\cal E}_2$ and ${\cal D}_2$, we find a 
uniform solution $\cal D$ for $E$. Note that $\cal D$ does not depend on $F_1,F_2,E$, and that we constructed $\cal D$ in an effective way from the arbitrary ${\cal E}_1$, ${\cal E}_2$ and ${\cal C}$. Formalizing this construction yields the
function $h$ whose existence is claimed by our present lemma.
\end{proof}

\begin{lemma}\label{l1c} \ {\bf (Transitivity)} \  
If \ $\uvalid F\mli E$ \ and \ $\uvalid E\mli G$, \ then \ $\uvalid F\mli G$. 

Moreover, there is an effective function $h: \ \{\mbox{EPMs}\}\times\{\mbox{EPMs}\}\longrightarrow\{\mbox{EPMs}\}$ 
such that, for any ${\cal E}_1$, ${\cal E}_2$, $F$, $E$ and $G$, 
if ${\cal E}_1\uvalid F\mli E$  and 
${\cal E}_2\uvalid E\mli G$, then $h({\cal E}_1,{\cal E}_2)\uvalid F\mli G$. 
\end{lemma}

\begin{proof} Assume ${\cal E}_1\uvalid F\mli E$ and ${\cal E}_2\uvalid E\mli G$. By Lemma \ref{l8}(c), we also have 
${\cal C}\uvalid (F\mli E)\mlc (E\mli G)\mli (F\mli G)$ for some  ${\cal C}$.  Lemma \ref{l1} allows us to combine the three uniform solutions and construct a uniform solution $\cal D$ for $F\mli G$. Formalizing this construction yields the
function $h$ whose existence is claimed by our lemma.
\end{proof}

\subsection{More validity lemmas}\label{mvl}
In this section we will prove a number of winnability facts by describing winning strategies in terms of EPMs. When trying to show that a given EPM wins a given game, it is always safe to assume that the runs that the machine generates are never $\oo$-illegal, i.e. that the environment never makes a (first) illegal move, for if it does, the machine automatically wins. This assumption, that we call the {\bf clean environment assumption},\index{clean environment assumption}\label{0clean environment assumption} will always be explicitly or implicitly present in our winnability proofs.

We will often employ a uniform solution 
for $P\mli P$ called the {\bf copy-cat strategy}\index{copy-cat strategy}\label{0copy-cat strategy} (${\cal CCS}$).\index{$\cal CCS$}\label{0ccs} This strategy, sketched for $\gneg\mbox{\em Chess}\mld \mbox{\em Chess}$ in Section \ref{ss4.3}, consists in mimicking, in the antecedent, the moves made by the 
environment in the consequent, and vice versa. More formally, the algorithm that ${\cal CCS}$ follows is an infinite loop, on every iteration of which ${\cal CCS}$ keeps granting permission until the environment 
makes a move $1.\alpha$ (resp. $2.\alpha$), to which 
the machine responds by the move $2.\alpha$ (resp. $1.\alpha$). As shown in the proof of Proposition 22.1 of \cite{Jap03},
this strategy guarantees success in every game of the form $A\mld \gneg A$ and hence 
$A\mli A$. A perhaps important detail is that $\cal CCS$ never looks at the past history of the game, 
i.e. the movement of its scanning head on the run tape is exclusively left-to-right. This guarantees that, even if the original game was something else and it only evolved to $A\mli A$ later as a result of making a series of moves, switching to $\cal CCS$ after the game has been brought down to $A\mli A$ ensures success no matter what happened in the past. 
    
Throughout this section, $E$ and $F$ (possibly with indices and attached tuples of variables) range over $\al$-formulas, $x$ and $y$ over variables, $t$ over terms, $n,k,i,j$ over nonnegative integers, $w$ over bitstrings, and $\alpha,\gamma$ over moves. These metavariables are assumed to be universally quantified in each context unless otherwise specified. In accordance with our earlier convention, $\epsilon$ stands for the empty bitstring. 

Next, $^*$ always means an arbitrary but fixed interpretation admissible for the formula whose uniform validity we are trying to prove. For readability, we will sometimes omit this parameter and write, say, $E$ instead of $E^*$. From the context it will be usually clear whether ``$E$'' stands for the formula $E$ or the game $E^*$. Similarly, in our winnability proofs $e$ will always stand for  an arbitrary but fixed valuation --- specifically, the valuation spelled on the valuation tape of the machine under question. Again, for readability, we will typically omit $e$ when it is irrelevant, and write $E^*$ (or just $E$) instead of $e[E^*]$.

\begin{lemma}\label{pl6a}
$\uvalid E\mli \pcost E$. 

Moreover, there is an EPM $\cal E$ such that, for any $E$, \ ${\cal E}\uvalid E\mli \pcost E$.
\end{lemma}

\begin{proof} The idea of a uniform solution $\cal E$ for $E\mli \pcost E$ is simple: seeing the consequent as the infinite disjunction $E\mld E\mld E\mld\ldots...$, ignore all of its disjuncts but the first one, and 
play $E\mli \pcost E$ as if it was just $E\mli E$. 

In more precise terms, $\cal E$ follows the following procedure LOOP which, notice, only differs from $\cal CCS$ in that the move prefix `$2.$' is replaced by `$2.1.$':\vspace{5pt}

{\bf Procedure} LOOP: Keep granting permission until the environment makes a move `$1.\alpha$' or `$2.1.\alpha$'; 
in the former case respond by the move `$2.1.\alpha$', and in the latter case respond by the move `$1.\alpha$'; then  repeat LOOP.\vspace{5pt}  

Consider an arbitrary $e$-computation branch $B$ of $\cal E$, and let $\Theta$ be the run spelled by $B$. Obviously permission is granted  infinitely many times in $B$, so $B$ is fair. Hence, in order to show that $\cal E$ wins $E^*\mli\pcost E^*$ (on the irrelevant valuation $e$ which we, according to our conventions, are omitting and pretend that $E^*$ is a constant game so that $e[E^*]=E$), it would suffice to show 
that $\win{E^*\mli\pcosti E^*}{}\seq{\Theta}=\pp$. 

Let $\Theta_i$ denote the initial segment of $\Theta$ consisting of the (lab)moves made by the players by the beginning of the $i$th iteration of LOOP in $B$ (if such an iteration exists). By induction on $i$, based on the clean environment 
assumption and applying a routine analysis of the behavior of LOOP, 
one can easily find that

\begin{equation}\label{oct3bc}
\begin{array}{l}
\mbox{a) } \Theta_i\in\legal{E^*\mli \pcosti E^*}{};\\
\mbox{b) } \rneg\Theta_{i}^{1.}=\Theta_{i}^{2.1.}.  
\end{array}
\end{equation} 

If LOOP is iterated infinitely many times, then the above obviously extends from $\Theta_i$ to $\Theta$, because 
every initial segment of $\Theta$ is an initial segment of some $\Theta_i$, and similarly for $\Theta^{1.}$ and $\Theta^{2.1.}$. Suppose now LOOP is iterated only a finite number 
$m$ of times. Then $\Theta=\seq{\Theta_m,\Gamma}$, where $\Gamma$ entirely consists of $\oo$-labeled moves none of which has the prefix `$1.$' or `$2.1$'. This is so because the environment cannot  make a move $1.\alpha$ or $2.1.\alpha$ during the $m$th iteration (otherwise there would be a next iteration) and, since $\cal E$'s moves are only triggered by the above two sorts of moves, $\cal E$ does not move during the $m$th iteration of LOOP. But then, in view of the clean environment assumption,  $\Theta$ inherits condition (a) of (\ref{oct3bc}) from $\Theta_m$, because there are no $\pp$-labeled moves in $\Gamma$; and the same is the case with condition (b), because 
$\seq{\Theta_m,\Gamma}^{1.}=\Theta_{m}^{1.}$ and 
$\seq{\Theta_m,\Gamma}^{2.1.}=\Theta_{m}^{2.1.}$. 
Thus, no matter whether LOOP is iterated a finite or infinite number of times, we have:

\begin{equation}\label{oct3bb}
\begin{array}{l}
\mbox{a) } \Theta\in\legal{E^*\mli \pcosti E^*}{};\\
\mbox{b) } \rneg\Theta^{1.}=\Theta^{2.1.}.
\end{array} 
\end{equation} 

Since $\Theta\in\legal{E^*\mli \pcosti E^*}{}$, in order to show that 
$\win{E^*\mli \pcosti E^*}{}\seq{\Theta}=\pp$, i.e. show that $\win{\gneg E^*\mld \pcosti E^*}{}\seq{\Theta}=\pp$, by the definition of $\mld$, it would suffice to verify that either 
$\win{\gneg E^*}{}\seq{\Theta^{1.}}=\pp$ or $\win{\pcosti E^*}{}\seq{\Theta^{2.}}=\pp$.
 So, assume $\win{\gneg E^*}{}\seq{\Theta^{1.}}\not=\pp$, i.e. $\win{\gneg E^*}{}\seq{\Theta^{1.}}=\oo$, i.e. 
$\win{E^*}{}\seq{\rneg\Theta^{1.}}=\pp$.
 Then, by clause (b) of (\ref{oct3bb}),
$\win{E^*}{}\seq{\Theta^{2.1.}}=\pp$. But then, by the definition of $\pcost$, $\win{\pcosti E^*}{}\seq{\Theta^{2.}}=\pp$.

Thus, ${\cal E}\models E^*\mli \pcost E^*$. Since $^*$  was arbitrary and the work 
of $\cal E$ did not depend on it,  
 we conclude that ${\cal E}\uvalid E\mli \pcost E$.
\end{proof}

\begin{lemma}\label{l6a}
$\uvalid E\mli \cost E$. 

Moreover, there is an EPM $\cal E$ such that, for any $E$, \ ${\cal E}\uvalid E\mli \cost E$.
\end{lemma}

\begin{proof} Again, the idea of a uniform solution $\cal E$ for $E\mli \cost E$ is simple: just act as $\cal CCS$, never making any replicative moves in the consequent and pretending that the latter is $E$ rather than (the easier-to-win) 
$\cost E$.
The following formal description of the interactive algorithm that $\cal E$ follows is virtually the same as that of 
$\cal CCS$, with the only difference that the move prefix `$2.$' is replaced by `$2.\epsilon.$'.\vspace{5pt}

{\bf Procedure} LOOP: Keep granting permission until the environment makes a move `$1.\alpha$' or `$2.\epsilon.\alpha$'; 
in the former case respond by the move `$2.\epsilon.\alpha$', and in the latter case respond by the move `$1.\alpha$'; then  repeat LOOP.\vspace{5pt} 

Consider an arbitrary $e$-computation branch $B$ of $\cal E$. Let $\Theta$ be the run spelled by $B$. As in the proof of the previous lemma, clearly permission will be granted  infinitely many times in $B$, so this branch is fair.  Hence, in order to show that $\cal E$ wins the game, it would suffice to show 
that $\win{E^*\mli\costi E^*}{}\seq{\Theta}=\pp$. 

Let $\Theta_i$ denote the initial segment of $\Theta$ consisting of the (lab)moves made by the players by the beginning of the $i$th iteration of LOOP in $B$. By induction on $i$, based on the clean environment 
assumption and applying a routine analysis of the behavior of LOOP and the definitions of the relevant game operations,
one can easily find that

\[\begin{array}{l}\label{oct3}
\mbox{a) } \Theta_i\in\legal{E^*\mli \costi E^*}{};\\
\mbox{b) } \rneg\Theta_{i}^{1.}=\Theta_{i}^{2.\epsilon.}\ ;\\
\mbox{c) } \mbox{\em All of the moves in $\Theta_{i}^{2.}$ have the prefix `$\epsilon.$'.} 
\end{array}\] 

If LOOP is iterated infinitely many times, then the above obviously extends from $\Theta_i$ to $\Theta$, because 
every initial segment of $\Theta$ is an initial segment of some $\Theta_i$, and similarly for $\Theta^{1.}$ and $\Theta^{2.\epsilon.}$. And if LOOP is iterated only a finite number $m$ of times, then $\Theta=\Theta_m$. This is so because the environment cannot  make a move $1.\alpha$ or $2.\epsilon.\alpha$ during the $m$th iteration (otherwise there would be a next iteration), and any other move would violate the clean environment assumption; and, as long as the environment does not move during a given iteration, neither does the machine.
Thus, no matter whether LOOP is iterated a finite or infinite number of times, we have:

\begin{equation}\label{oct3a}
\begin{array}{l}
\mbox{a) } \Theta\in\legal{E^*\mli \costi E^*}{};\\
\mbox{b) } \rneg\Theta^{1.}=\Theta^{2.\epsilon.}\ ;\\
\mbox{c) } \mbox{\em All of the moves in $\Theta^{2.}$ have the prefix `$\epsilon.$'.} 
\end{array} 
\end{equation} 

Since $\Theta\in\legal{E^*\mli \costi E^*}{}$, in order to show that 
$\win{E^*\mli \costi E^*}{}\seq{\Theta}=\pp$, it would suffice to verify that either 
$\win{\gneg E^*}{}\seq{\Theta^{1.}}=\pp$ or $\win{\costi E^*}{}\seq{\Theta^{2.}}=\pp$.
 So, assume $\win{\gneg E^*}{}\seq{\Theta^{1.}}\not=\pp$, i.e. $\win{\gneg E^*}{}\seq{\Theta^{1.}}=\oo$, i.e. 
$\win{E^*}{}\seq{\rneg\Theta^{1.}}=\pp$.
 Then, by clause (b) of (\ref{oct3a}),
$\win{E^*}{}\seq{\Theta^{2.\epsilon.}}=\pp$.
 Pick any complete branch $w$ of $\tree{\costi E^*}{\seq{\Theta^{2.}}}$. In view of clause (c) of (\ref{oct3a}), 
we obviously have 
$\Theta^{2.\epsilon.}=(\Theta^{2.})^{\preceq w}$ (in fact, $w=\epsilon$). Hence
$\win{E^*}{}\seq{(\Theta^{2.})^{\preceq w}}=\pp$. Then, by the definition of $\cost$,
$\win{\costi E^*}{}\seq{\Theta^{2.}}=\pp$. 

Thus, ${\cal E}\models E^*\mli \cost E^*$ and, as $^*$ was arbitrary and the work of $\cal E$ did 
not depend on it, we conclude that ${\cal E}\uvalid E\mli \cost E$.
\end{proof}

Having already seen two examples, in the remaining uniform validity proofs we will typically limit ourselves to just constructing interactive algorithms, 
leaving routine verification of  their correctness to the reader. An exception will be the proof of Lemma 
\ref{l5} given separately in Section \ref{prf} where, due to the special complexity of the case, correctness verification will be done even more rigorously than 
we did this in the proofs of Lemmas \ref{pl6a} and \ref{l6a}. 

\begin{lemma}\label{pl4}
$\uvalid \pst(E\mli F)\mli(\pst E\mli \pst F)$. 

Moreover, there is an EPM $\cal E$ such that, for every $E$ and $F$, \ \[{\cal E}\uvalid \pst(E\mli F)\mli(\pst E\mli \pst F).\]
\end{lemma}

\begin{proof} Here is a strategy for $\cal E$ to follow:\vspace{5pt}

{\bf Procedure} LOOP: Keep granting permission until the adversary makes a move $\gamma$. Then:  

If $\gamma=1.i.1.\alpha$ (resp. $\gamma=2.1.i.\alpha$), then make the move $2.1.i.\alpha$ (resp. $1.i.1.\alpha$), and repeat LOOP;

If $\gamma=1.i.2.\alpha$ (resp. $\gamma=2.2.i.\alpha$), then make the move $2.2.i.\alpha$ (resp. $1.i.2.\alpha$), and repeat LOOP.
\end{proof}

\begin{lemma}\label{l4}
$\uvalid \st(E\mli F)\mli(\st E\mli \st F)$. 

Moreover, there is an EPM $\cal E$ such that, for every $E$ and $F$, \ \[{\cal E}\uvalid \st(E\mli F)\mli(\st E\mli \st F).\]
\end{lemma}

\begin{proof} 
A relaxed description of a uniform solution $\cal E$ 
for $\st(E\mli F)\mli(\st E\mli \st F)$
is as follows.
In $\st(E^*\mli F^*)$ and $\st E^*$ the machine is making exactly the same replicative moves (moves of the form $\col{w}$) as the environment is making in $\st F^*$. This ensures that the underlying BT-structures of the three $\st$-components of the game stay identical, and now all the machine needs for a success is to win the game 
$(E^*\mli F^*)\mli(E^*\mli F^*)$ within each branch of those trees. This can be easily achieved by applying copy-cat methods to the two occurrences of $E$ and the two occurrences of $F$. 

In precise terms, the strategy that $\cal E$ follows is described by the following interactive algorithm.\vspace{5pt}

{\bf Procedure} LOOP: Keep granting permission until the adversary makes a move $\gamma$. Then:  

If $\gamma=2.2.\col{w}$, then make the moves $1.\col{w}$ and $2.1.\col{w}$, and repeat LOOP;

If $\gamma=2.2.w.\alpha$ (resp. $\gamma=1.w.2.\alpha$), then  make the move $1.w.2.\alpha$ (resp. $2.2.w.\alpha$), and repeat LOOP; 

If $\gamma=2.1.w.\alpha$ (resp. $\gamma=1.w.1.\alpha$), then make the move $1.w.1.\alpha$ (resp. $2.1.w.\alpha$), and repeat LOOP.
\end{proof}

\begin{lemma}\label{pl4a} 
$\uvalid \pcost (E_1\mld\ldots\mld E_n)\mli \pcost E_1\mld\ldots\mld \pcost E_n$. 

Moreover, there is an effective procedure that takes any particular value of $n$ and constructs an EPM $\cal E$ such that, for any $E_1,\ldots,E_n$, \ ${\cal E}\uvalid \pcost (E_1\mld\ldots\mld E_n)\mli \pcost E_1\mld\ldots\mld \pcost E_n$.

\end{lemma}
\begin{proof} We let $\cal E$ act as the following strategy prescribes, with $i$ ranging over $\{1,2,3,\ldots\}$ and $j$ over $\{1,\ldots,n\}$:\vspace{5pt}

{\bf Procedure} LOOP: Keep granting permission until the adversary makes a move $1.i.j.\alpha$ (resp. $2.j.i.\alpha$); then make the move $2.j.i.\alpha$ (resp. $1.i.j.\alpha$), and repeat LOOP.
\end{proof}

\begin{lemma}\label{l4a} 
$\uvalid \cost (E_1\mld\ldots\mld E_n)\mli \cost E_1\mld\ldots\mld \cost E_n$. 

Moreover, there is an effective procedure that takes any particular value of $n$ and constructs an EPM $\cal E$ such that, for any $E_1,\ldots,E_n$, \ ${\cal E}\uvalid \cost (E_1\mld\ldots\mld E_n)\mli \cost E_1\mld\ldots\mld \cost E_n$.

\end{lemma}

\begin{proof} Here is the algorithm for $\cal E$:\vspace{5pt}

{\bf Procedure} LOOP: Keep granting permission until the adversary makes a move $\gamma$. Then:

If $\gamma=1.\col{w}$, then make the $n$ moves $2.1.\col{w},\ldots,2.n.\col{w}$, and repeat LOOP;

If $\gamma=1.w.j.\alpha$ (resp. $\gamma=2.j.w.\alpha$) where $1\leq j\leq n$, then make the move $2.j.w.\alpha$ 
(resp. $1.w.j.\alpha$), and repeat LOOP.
\end{proof}

\begin{lemma}\label{pl6c}
$\uvalid \pcost E\mld\pcost E\mli\pcost E$. 

Moreover, there is an EPM $\cal E$ such that, for any $E$, \ ${\cal E}\uvalid 
\pcost E\mld\pcost E\mli\pcost E$.
\end{lemma}

\begin{proof} We let $\cal E$ work as the following strategy prescribes:

{\bf Procedure} LOOP: Keep granting permission until the adversary makes a move $\gamma$. Then act depending on which of the following cases applies, and after that repeat LOOP:  

If $\gamma=1.1.i.\alpha$, then make the move $2.j.\alpha$ where $j=2i-1$;

If $\gamma=1.2.i.\alpha$, then make the move $2.j.\alpha$ where $j=2i$;

If $\gamma=2.j.\alpha$ where $j=2i-1$, then make the move $1.1.i.\alpha$;

If $\gamma=2.j.\alpha$ where $j=2i$, then make the move $1.2.i.\alpha$.
\end{proof}

\begin{lemma}\label{l6c}
$\uvalid \cost E\mld\cost E\mli\cost E$. 

Moreover, there is an EPM $\cal E$ such that, for any $E$,  ${\cal E}\uvalid 
\cost E\mld\cost E\mli\cost E$.

\end{lemma}

\begin{proof} The idea of a strategy for $\cal E$  is to first replicate the consequent turning it into 
$\cost(E^*\circ E^*)$, which is essentially the same as $\cost E^*\mld \cost E^*$, and then switch to a strategy that is essentially the same as the ordinary copy-cat strategy. 
Precisely, here is how $\cal E$ works: it makes the move $2.\col{\epsilon}$ (replicating the consequent), after which it follows the following algorithm:\vspace{7pt}

{\bf Procedure} LOOP: Keep granting permission until the adversary makes a move $\gamma$. Then:

If $\gamma=2.0\alpha$ (resp. $\gamma=1.1.\alpha$), then make the move $1.1.\alpha$ (resp. $2.0\alpha$), and repeat LOOP;

If $\gamma=2.1\alpha$ (resp. $\gamma=1.2.\alpha$), then make the move $1.2.\alpha$ (resp. $2.1\alpha$), and repeat LOOP;

If $\gamma=2.\epsilon.\alpha$, then make the  moves $1.1.\epsilon.\alpha$ and $1.2.\epsilon.\alpha$, and repeat LOOP.
\end{proof}

\begin{lemma}\label{pl5}
$\uvalid \pcost\pcost E\mli \pcost E$. 

Moreover, there is an EPM $\cal E$ such that, for any $E$, \ ${\cal E}\uvalid \pcost \pcost E \mli \pcost E$.
\end{lemma}

\begin{proof} We select any effective one-to-one function $f$ from the set of all pairs of nonnegative integers onto the set of all nonnegative integers. Below is the  interactive algorithm that $\cal E$ follows:

{\bf Procedure} LOOP: Keep granting permission until the adversary makes a move $\gamma$. Then:

If $\gamma=1.i.j.\alpha$, then make the move $2.k.\alpha$ where $k=f(i,j)$,  and repeat LOOP;

If $\gamma=2.k.\alpha$, then make the move $2.i.j.\alpha$ where $k=f(i,j)$,  and repeat LOOP.
\end{proof}

\begin{lemma}\label{l5}
$\uvalid \cost\cost E\mli \cost E$. 

Moreover, there is an EPM $\cal E$ such that, for any $E$, \ ${\cal E}\uvalid \cost \cost E \mli \cost E$.
\end{lemma}

\begin{proof} Our proof of this lemma, unlike that of the ``similar'' Lemma \ref{pl5}, is fairly long. For this reason, it is given separately in Section \ref{prf}. 
\end{proof}

In what follows, we will be using the expressions $E^*(x)$, $E^*(t)$, etc. to mean the same as the more clumsy $\bigl(E(x)\bigr)^*$,   $\bigl(E(t)\bigr)^*$, etc. Also,   
remember from Section \ref{ss3} that, when $t$ is a constant, $e(t)=t$.

\begin{lemma}\label{oct5a}
$\uvalid \ada x\bigl(E(x)\mli F(x)\bigr)\mli \bigl(\ada xE(x)\mli\ada xF(x)\bigr)$. 

Moreover, there is an EPM $\cal E$ such that, 
for any  $E(x)$ and $F(x)$,     
\[{\cal E}\uvalid \ada x\bigl(E(x)\mli F(x)\bigr)\mli \bigl(\ada xE(x)\mli\ada xF(x)\bigr).\]
\end{lemma}

\begin{proof} 
Strategy: Wait till the environment makes the move `$2.2.c$' for some constant $c$. This brings the 
$\ada xF^*(x)$ component down to $F^*(c)$ and hence the entire game to \[\ada x\bigl(E^{*}(x)\mli F^*(x)\bigr)\mli \bigl(\ada xE^{*}(x)\mli F^*(c)\bigr).\]
Then make the same move $c$ in the antecedent and in $\ada xE^{*}(x)$, i.e. make the two moves 
`$1.c$' and `$2.1.c$'. The game will be brought down to $\bigl(E^{*}(c)\mli F^*(c)\bigr)\mli \bigl(E^{*}(c)\mli F^*(c)\bigr)$. Finally, switch to $\cal CCS$. 
\end{proof}

\begin{lemma}\label{oct5b} 
Assume $t$ is free for $x$ in $E(x)$. Then $\uvalid E(t)\mli \ade xE(x)$. 

Moreover, there is an effective function that takes any $t$ and constructs an EPM $\cal E$ such that, for any $E(x)$, whenever $t$ is free for $x$ in $E(x)$, \ 
${\cal E}\uvalid E(t)\mli \ade xE(x)$.
\end{lemma}

\begin{proof} Strategy: Let $c=e(t)$. Read $c$ from the valuation tape if necessary (i.e. if $t$ is a variable). Then make the move `$2.c$', bringing the game down to $E^*(c)\mli E^*(c)$. Then switch to $\cal CCS$.  \end{proof}

\begin{lemma}\label{oct5c} 
Assume $E$ does not contain $x$. Then
$\uvalid E\mli \ada xE$. 

Moreover, there is an EPM $\cal E$ such that, for any 
$E$ and $x$, as long as $E$ does not contain $x$, \ ${\cal E}\uvalid E\mli \ada xE$. 
\end{lemma}

\begin{proof} In this case we prefer to explicitly write the usually suppressed parameter $e$. Consider an arbitrary $E$ not containing $x$, and an arbitrary interpretation $^*$ admissible for $E\mli \ada xE$. The formula $E\mli \ada xE$ contains $x$ yet $E$ does not. Therefore, from the definition of admissibility and with a little thought we can see that  $E^*$ does not depend on $x$. In turn, this means --- as can be seen with some additional thought --- that the move $c$ by the environment 
(whatever constant $c$) in $e[\ada x E^*]$ brings this game down to 
$e[E^*]$. With this observation in mind, the following strategy can be seen to be successful:
Wait till the environment makes the move `$2.c$' for some constant $c$. Then read 
the sequence `$1.\alpha_1$', \ldots, `$1.\alpha_n$' of (legal) moves possibly made by the environment before it made the above move `$2.c$', and make the $n$ moves `$2.\alpha_1$', \ldots, `$2.\alpha_n$'. It can be seen that now the original 
game $e[E^*]\mli e[\ada xE^*]$ will have been brought down to $\seq{\Phi}e[E^*]\mli\seq{\Phi}e[E^*]$, where 
$\Phi=\seq{\pp \alpha_1,\ldots,\pp\alpha_n}$. So, switching to $\cal CCS$ at this point guarantees success.
\end{proof}

\begin{lemma}\label{oct99} 
Assume $E(x)$ does not contain $y$. Then $\uvalid \ada yE(y)\mli\ada xE(x)$. 
In fact, ${\cal CCS}\uvalid \ada yE(y)\mli\ada xE(x)$.
\end{lemma}

\begin{proof} Assuming that $E(x)$ does not contain $y$ and analyzing the relevant definitions, it is not hard to see that, for any interpretation $^*$ admissible for $\ada yE(y)\mli\ada xE(x)$,  we simply have $\bigl(\ada yE(y)\bigr)^*=\bigl(\ada x E(x)\bigr)^*$. So, we deal with a game of the form $A\mli A$, for which the ordinary copy-cat strategy is successful.
\end{proof}

\subsection{Iteration principle for branching recurrence}\label{prf}
 
The computability principles expressed by the formulas $\pcost\pcost E\mli\pcost E$ and $\cost\cost E\mli \cost E$, that can be equivalently rewritten as $\pst E\mli\pst\pst E$ and $\st E\mli \st\st E$, we call {\bf iteration principles}\index{iteration principle}\label{0iteration principle} 
(for $\pst$ and $\st$, respectively). This section is entirely devoted to a proof of Lemma \ref{l5}, i.e. the iteration principle for $\st$. We start with some auxiliary definitions.

A {\bf colored bit}\index{colored bit}\label{0colored bit} $b$ is a pair $(c,d)$, where $c$, called the {\bf content}\index{content (of a colored bit)}\label{0content (of a colored bit)} of $b$, is in $\{0,1\}$, and $d$, called the {\bf color}\index{color (of a bit)}\label{0color (of a bit)} of $b$, is in \{{\em blue,yellow}\}. We will be using the notation $\blu{c}$ (``blue $c$") for the colored bit whose content is $c$ and color is {\em blue}, and $\yel{c}$ (``yellow $c$") for the bit whose content is $c$ and color is {\em yellow}. The four colored bits 
will be treated as symbols, from which, just as from ordinary bits, we can form strings.

A {\bf colored bitstring}\index{colored bitstring}\label{0colored bitstring} is a finite or infinite string of colored bits. 
Consider a colored bitstring $v$. The {\bf content}\index{content (of a colored bitstring)}\label{0content (of a colored bitstring)} of $v$ is the result of ``ignoring the colors" in $v$, i.e. 
replacing every bit of $v$ by the content of that bit. The {\bf blue content}\index{blue content}\label{0blue content} of $v$ is the content of the string that results from deleting in $v$ all but blue bits. {\bf Yellow content}\index{yellow content}\label{0yellow content} is defined similarly. 
We use $\cont{v}$,\index{$\cont{v}$}\label{0contv} $\blu{v}$\index{$\blu{v}$}\label{0bluv} and $\yel{v}$\index{$\yel{v}$}\label{0yelv} to denote the content, blue content and yellow content of $v$, respectively. 
Example: if $v=\blu{1}\yel{0}\yel{0}\blu{0}\yel{1}$, we have $\cont{v}=10001$, $\blu{v}=10$ and $\yel{v}=001$. 
As in the case of ordinary bitstrings, $\epsilon$ stands for the empty colored bitstring, and $u\preceq w$ means that $u$ is a (not necessarily proper) initial segment of $w$.

\begin{definition}\label{coltree}
A {\bf colored bitstring tree}\index{colored bitstring tree (CBT)}\label{0colored bitstring tree (CBT)} ({\bf CBT}) is a set $T$ of colored bitstrings, called its {\bf branches}, such that the following conditions are satisfied:

a) The set $\{\cont{v}\ |\ v\in T\}$, which we denote by $\cont{T}$,\index{$\cont{T}$}\label{0ttt}  is a BT in the sense of Definition \ref{tree}. 

b) For any $w,u\in T$, if $\cont{w}=\cont{u}$, then $w=u$.

c) For no (finite) $v\in T$ do we have $\{v\blu{0},v\yel{1}\}\subseteq T$ or $\{v\yel{0},v\blu{1}\}\subseteq T$.

\noindent A colored bitstring $v$ is said to be a {\bf leaf} 
of $T$ iff $\cont{v}$ is a leaf 
of $\cont{T}$.
\end{definition}

 When represented in the style of (the underlying tree of) Figure 11 of Section \ref{ss4.6}, a CBT will look like an ordinary BT, with the only difference that now every edge will have one of the colors {\em blue} or {\em yellow}. Also, by condition (c), both of the outgoing 
edges (``sibling" edges) of any non-leaf node will have the same color. 
  
\begin{lemma}\label{sep19}
Assume $T$ is a CBT, and $w,u$ are branches of $T$ with $\blu{w}\preceq \blu{u}$ and $\yel{w}\preceq \yel{u}$. Then 
$w\preceq u$.
\end{lemma}

\begin{proof} Assume $T$ is a CBT, $w,u\in T$, and $w\not\preceq u$. We want to show that then $\blu{w}\not\preceq \blu{u}$ or $\yel{w}\not\preceq\yel{u}$. Let $v$ be the longest common initial segment of $w$ and $u$, so we have $w=vw'$ and $u=vu'$ for some 
$w',u'$ such that $w'$ is nonempty and $w'$ and $u'$ do not have a nonempty common initial segment. 
Assume the first bit of $w'$ is $\blu{0}$ (the cases when it is $\blu{1}$, $\yel{0}$ or $\yel{1}$, of course, will be similar). If $u'$ is empty, then $w$ obviously contains more blue bits than $u$ does, and we are done. Assume now $u'$ is nonempty, in particular, $b$ is the first bit of $u'$. Since $w'$ and $u'$ do not have a nonempty common initial segment, $b$ should be different from $\blu{0}$. By condition (b) of Definition \ref{coltree}, the content of $b$ cannot be $0$ (for otherwise we would have $v\blu{0}=vb$ and hence $b=\blu{0}$). Consequently, $b$ is either $\blu{1}$ or $\yel{1}$. The case $b=\yel{1}$ is ruled out by condition (c) of Definition \ref{coltree}. Thus, $b=\blu{1}$. But the blue content of $v\blu{0}$ is $\blu{v}0$ while the blue content of $v\blu{1}$ is $\blu{v}1$. Taking into account the obvious fact that 
the former is an initial segment of $\blu{w}$ and the latter is an initial segment of $\blu{u}$, we find $\blu{w}\not\preceq \blu{u}$.
\end{proof}
 
The uniform solution $\cal E$ for $\cost\cost E\mli\cost E$ that we are going to construct essentially uses a copy-cat strategy between the antecedent and the consequent. Of course, however, this strategy cannot be applied directly in the form of $\cal CCS$. The problem is that while a position of $\cost E^*$ is a decorated tree $\cal T$
in the style of Figure 11 of Section \ref{ss4.6}, in the case of $\cost\cost E^*$ it is a tree ${\cal T}'$ of trees such as, say, the one shown below:

\begin{center}
\begin{picture}(317,160)
\put(110,45){\framebox{\small $\seq{\Phi_3}E^*$}}
\put(150,45){\framebox{\small $\seq{\Phi_4}E^*$}}
\put(200,45){\framebox{\small $\seq{\Phi_5}E^*$}}

\put(105,33){\line(1,0){137}}
\put(105,90){\line(1,0){137}}
\put(105,33){\line(0,1){57}}
\put(242,33){\line(0,1){57}}
\put(128,55){\line(3,2){45}}
\put(168,55){\line(-3,2){20}}
\put(218,55){\line(-3,2){45}}
\put(130,60){\small 0}
\put(162,60){\small 1}
\put(152,75){\small 0}
\put(189,75){\small 1}

\put(0,45){\framebox{\small $\seq{\Phi_1}E^*$}}
\put(50,45){\framebox{\small $\seq{\Phi_2}E^*$}}
\put(-5,33){\line(1,0){97}}
\put(-5,90){\line(1,0){97}}
\put(-5,33){\line(0,1){57}}
\put(92,33){\line(0,1){57}}
\put(18,55){\line(3,2){25}}
\put(68,55){\line(-3,2){25}}
\put(21,62){\small 0}
\put(61,62){\small 1}

\put(280,45){\framebox{\small $\seq{\Phi_6}E^*$}}
\put(275,33){\line(1,0){47}}
\put(275,90){\line(1,0){47}}
\put(275,33){\line(0,1){57}}
\put(322,33){\line(0,1){57}}

\put(43,90){\line(5,2){129}}
\put(173,90){\line(-5,2){65}}
\put(300,90){\line(-5,2){129}}

\put(72,105){\small 0}
\put(138,106){\small 1}

\put(134,130){\small 0}
\put(207,130){\small 1}

\put(76,10){{\bf Figure 13:} A position ${\cal T}'$ of $\cost\cost E^*$}
\end{picture}
\end{center}

 The trick that $\cal E$ uses is that it sees the $BT$-structure of $\cal T$ as a colored bitstring tree, letting 
such a $\cal T$ ``simulate'' ${\cal T}'$.   
Specifically, $\cal E$ tries to maintain a one-to-one correspondence between the leaves (windows) of  $\cal T$ and the leaves of the leaves (small windows) of ${\cal T}'$, with the positions of $E^*$ in each pair of corresponding windows being identical. Figure 14 shows a possible $\cal T$, whose six windows, as we see, have the same contents as the six small windows of ${\cal T}'$, even if some permutation in the order of windows has taken place. 

\begin{center}
\begin{picture}(317,140)

\put(0,45){\framebox{\small $\seq{\Phi_1}E^*$}}
\put(50,45){\framebox{\small $\seq{\Phi_3}E^*$}}
\put(90,45){\framebox{\small $\seq{\Phi_4}E^*$}}
\put(150,45){\framebox{\small $\seq{\Phi_2}E^*$}}
\put(200,45){\framebox{\small $\seq{\Phi_5}E^*$}}
\put(280,45){\framebox{\small $\seq{\Phi_6}E^*$}}

\put(18,55){\line(2,1){141}}
\put(299,55){\line(-2,1){141}}

\put(68,55){\line(2,1){20}}
\put(108,55){\line(-2,1){45}}
\put(168,55){\line(2,1){25}}
\put(219,55){\line(-2,1){100}}

\put(68,60){\small $\yel{0}$}
\put(104,60){\small $\yel{1}$}
\put(176,62){\small $\blu{0}$}
\put(208,62){\small $\blu{1}$}
\put(42,70){\small $\blu{0}$}
\put(80,70){\small $\blu{1}$}

\put(76,90){\small $\yel{0}$}
\put(155,90){\small $\yel{1}$}

\put(130,115){\small $\blu{0}$}
\put(182,115){\small $\blu{1}$}

\put(24,10){{\bf Figure 14:} A possible corresponding position $\cal T$ of $\cost E^*$}
\end{picture}
\end{center}

The way this mapping works is that to a leaf $y$ of a leaf $x$ of ${\cal T}'$ corresponds a (the) leaf of $\cal T$ 
whose yellow content is $y$ and blue content is $x$, and vice versa. E.g., look at the small window containing 
$\seq{\Phi_3}E^*$ in Figure 13. It is leaf $00$ of leaf $01$ of ${\cal T}'$; and the window containing the same 
$\seq{\Phi_3}E^*$ in Figure 14 is leaf $\blu{0}\yel{0}\blu{1}\yel{0}$ of $\cal T$, whose yellow content is indeed 
$00$ and blue content is indeed $01$.  $\cal E$ has a way to maintain such a correspondence. Let us see how it 
acts when the antecedent and the consequent of $\cost\cost E^*\mli \cost E^*$ have evolved to the positions shown in
 Figures 13 and 14, respectively. If, say, the environment makes a non-replicative move $\alpha$ in the $\seq{\Phi_3}E^*$ component of 
${\cal T}'$ (resp. $\cal T$), $\cal E$ responds by the same move $\alpha$ in the $\seq{\Phi_3}E^*$ component of ${\cal T}$ 
(resp. ${\cal T}'$). This ensures that the positions of $E^*$ in the two components remain identical. Of course, the 
environment can make a move $\alpha$ in several components at once. For example, the move  
$1.00.\epsilon.\alpha$ by the environment would amount to making move $\alpha$ in the two components $\seq{\Phi_1}E^*$ and 
$\seq{\Phi_2}E^*$ of ${\cal T}'$. No problem, $\cal E$ can respond by the two consecutive moves $2.000.\alpha$ and 
$2.010.\alpha$. Now let us see how $\cal E$ handles replicative moves. If the environment replicates a small window of ${\cal T}'$, $\cal E$ replicates the corresponding window of $\cal T$, and colors the two newly emerged edges 
into yellow. In fact, the environment can replicate several small windows at once, as, say, would be the case if it makes the move $0.\col{1}$ in the antecedent, amounting to replicating the two windows containing $\seq{\Phi_2}E^*$ and $\seq{\Phi_5}E^*$. But, again, this is not a problem, for $\cal E$ can respond by two (or whatever number of small 
windows of the antecedent have been replicated) consecutive replicative moves in the consequent.
Suppose now the environment replicates a large window of ${\cal T}'$, such as, say, window  $00$. Then $\cal E$ responds by replicative moves in all leaves of ${\cal T}$ whose blue content is $00$, specifically, leaves 
$000$ and $010$, and colors the newly emerged edges into blue. With some thought one can see that, with this strategy, 
it is guaranteed that to any leaf $y$ of any leaf $x$ of the updated ${\cal T}'$ again corresponds the leaf of the updated $\cal T$ whose yellow content is $y$ and blue content is $x$, and vice versa. 

In an attempt to understand what replicative moves (ignoring all non-replicative moves inbetween) could have resulted in the tree of Figure 13 and the corresponding tree of Figure 14, we find the following. First, the environment made the replicative move $\col{\epsilon}$ in the antecedent of $\cost\cost E^*\mli \cost E^*$. To this $\cal E$ responded by the replicative move $\col{\epsilon}$ in the consequent. Then the environment made the (``deep'') replicative move $0.\col{\epsilon}$ in the antecedent. To this $\cal E$ responded by $\col{0}$ in the consequent. Next the environment made the replicative move $\col{0}$ in the antecedent. $\cal E$ responded by $\col{00}$ and $\col{01}$ in the consequent. Finally, the environment made (the ``deep'') replicative move $01.\col{0}$ in the antecedent, and $\cal E$ responded by $\col{001}$ in the consequent.

Keeping, according to the above scenario, all runs of $E^*$ in the branches of branches of the antecedent identical with runs of $E^*$ in the corresponding branches of the consequent 
can be eventually seen to guarantee a win for $\cal E$.

Now we describe $\cal E$ in precise terms. At the beginning, this EPM creates a record $T$ of the type `finite 
CBT', and initializes it to $\{\epsilon\}$. After that, $\cal E$ follows the following procedure:\vspace{7pt}

{\bf Procedure} LOOP: Keep granting permission until the adversary makes a move $\gamma$. If $\gamma$ satisfies the conditions of one of the following four cases, act as the corresponding case prescribes. Otherwise go to an infinite loop in a permission state. 

{\em Case (i):}   $\gamma=1.\col{w}$ for some bitstring $w$. Let $v_1,\ldots,v_k$ be\footnote{In each of the four cases 
we assume that the list $v_1,\ldots,v_k$ is arranged lexicographically.} all of the leaves $v$ of $T$ with $w=\blu{v}$. Then make the moves 
$2.\col{\cont{v}_1},\ldots,2.\col{\cont{v}_k}$, update $T$ to $T\cup\{v_1\blu{0},v_1\blu{1},\ldots,v_k\blu{0},v_k\blu{1}\}$, and repeat LOOP.

{\em Case (ii):} $\gamma=1.w.\col{u}$ for some bitstrings $w,u$. Let $v_1,\ldots,v_k$ be all of the leaves $v$ of $T$ 
such that $w\preceq \blu{v}$ 
and $u=\yel{v}$. Then make the moves 
$2.\col{\cont{v}_1},\ldots,2.\col{\cont{v}_k}$, update $T$ to $T\cup\{v_1\yel{0},v_1\yel{1},\ldots,v_k\yel{0},v_k\yel{1} \}$, and repeat LOOP.

{\em Case (iii):} $\gamma=1.w.u.\alpha$ for some bitstrings $w,u$ and move $\alpha$. Let $v_1,\ldots,v_k$ be all of the leaves $v$ of $T$ 
such that 
$w\preceq \blu{v}$ and $u\preceq\yel{v}$. Then make the moves $2.\cont{v}_1.\alpha,\ldots,2.\cont{v}_k.\alpha$, and repeat LOOP.

{\em Case (iv):} $\gamma=2.w.\alpha$ for some bitstring $w$. 
Let $v_1,\ldots,v_k$ be all of the leaves $v$ of $T$ 
with $w\preceq \cont{v}$.  
Then make the moves 
$1.\blu{v}_1.\yel{v}_1.\alpha,\ldots,1.\blu{v}_k.\yel{v}_k.\alpha$, and repeat LOOP.

\

Pick an arbitrary interpretation $^*$ admissible for $\cost\cost E\mli \cost E$, an arbitrary valuation $e$ and an arbitrary $e$-computation branch $B$ of $\cal E$. Let $\Theta$ be the run spelled by $B$. The work of $\cal E$  
does not depend on $e$. And, as $e$ is going to be fixed, we can safely omit this parameter (as we usually did in the previous section) and just write $E^*$ instead of $e[E^*]$. 
Of course, $\cal E$ is interpretation-blind, so, as long as it wins $\cost\cost E^*\mli \cost E^*$, it is a uniform solution for $\cost\cost E\mli \cost E$.

 From the description of LOOP it is immediately clear that $B$ is a fair. Hence, in order to show that $\cal E$ wins, 
it would be sufficient to verify that $\win{\costi\costi E^*\mli \costi E^*}{}\seq{\Theta}=\pp$.  
 
Let $N=\{1,\ldots,m\}$ if LOOP is iterated the finite number $m$ of times in $B$, and $N=\{1,2,3,\ldots\}$ otherwise.  For $i\in N$, we 
let $T_i$ denote the value of record $T$ at the beginning of the $i$th iteration of LOOP. Next, $\Theta_i$ will mean the 
initial segment of $\Theta$ consisting of the (lab)moves made by the beginning of the $i$th iteration of LOOP.
 Finally, $\Psi_i$ will stand for $\rneg\Theta_{i}^{1.}$ and $\Phi_i$ for $\Theta_{i}^{2.}$.

From the description of LOOP it is obvious that, 
for each $i\in N$, $T_i$ is a finite colored tree, and that $T_1\subseteq T_2\subseteq \ldots\subseteq T_i$. In our subsequent reasoning we will implicitly rely on this fact.

\begin{lemma}\label{sep21b}
For every $i$ with $i\in N$, we have:\vspace{3pt}

a) $\Phi_i$ is a prelegal position of $\cost E^*$, and {\em $\tree{\costi E^*}{\seq{\Phi_i}}=\cont{T_i}$}.

b) $\Psi_i$ is a prelegal position of $\cost\cost E^*$.

c) For every leaf $x$ of {\em $\tree{\costi\costi E^*}{\seq{\Psi_i}}$}, $\Psi_{i}^{\preceq x}$ is a prelegal position of $\cost E^*$.

d) For every leaf $z$ of $T_i$, $\blu{z}$ is a leaf of {\em $\tree{\costi \costi E^*}{\seq{\Psi_{i}}}$} and $\yel{z}$ is a leaf of 
{\em $\tree{\costi E^*}{\seq{\Psi_{i}^{\preceq \blu{z}}}}$}.

e) For every leaf $x$ of {\em $\tree{\costi \costi E^*}{\seq{\Psi_{i}}}$} and every leaf $y$ of {\em $\tree{\costi E^*}{\seq{\Psi_{i}^{\preceq x}}}$}, 
there is a leaf $z$ of $T_i$ such that $x=\blu{z}$ and $y=\yel{z}$.  By Lemma \ref{sep19}, such a $z$ is unique.

f) For every leaf $z$ of $T_i$, $\Phi_{i}^{\preceq \cont{z}}=(\Psi_{i}^{\preceq \blu{z}})^{\preceq \yel{z}}$.

g) $\Theta_i$ is a legal position of $\cost\cost E^*\mli \cost E^*$; hence, $\Phi_i\in\legal{\costi E^*}{}$ and 
$\Psi_i\in\legal{\costi\costi E^*}{}$.
\end{lemma}

\begin{proof} We proceed by induction on $i$. The basis case with $i=1$ is rather straightforward for each clause of the lemma and we do not discuss it. For the inductive step, assume $i+1\in N$, and the seven clauses of the lemma are true for $i$. 

{\em Clause (a):} By the induction hypothesis, $\Phi_i$ is a prelegal position of $\cost E^*$ and $\tree{\costi E^*}{\seq{\Phi_i}}=\cont{T}_i$. 
Assume the $i$th iteration of LOOP deals with Case (i), so that $\Phi_{i+1}=\seq{\Phi_i,\pp\col{\cont{v}_1},\ldots,\pp\col{\cont{v}_k}}$.\footnote{With $v_1,\ldots,v_k$ here and in later cases being as in the corresponding clause of the description of LOOP.} Each of $\cont{v}_1,\ldots,\cont{v}_k$ is a leaf of
$\cont{T}_i$, i.e. a leaf of $\tree{\costi E^*}{\seq{\Phi_i}}$. This guarantees that $\Phi_{i+1}$ is a prelegal position of $\cost E^*$.  Also,  
by the definition of function {\em Tree}, we have   
$\tree{\costi E^*}{\seq{\Phi_{i+1}}}=\tree{\costi E^*}{\seq{\Phi_i}}\cup\{\cont{v}_10,\cont{v}_11,\ldots,\cont{v}_k0,\cont{v}_k1\}$. But the latter is nothing but $\cont{T}_{i+1}$ as can be seen from the description of how Case (i) updates $T_i$ to $T_{i+1}$. 
A similar argument applies when the $i$th iteration of LOOP deals with Case (ii). Assume now the $i$th iteration of LOOP 
deals with Case (iii). Note that the moves made in the consequent of $\cost\cost E^*\mli\cost E^*$ 
(the moves that bring $\Phi_i$ to $\Phi_{i+1}$) are nonreplicative --- specifically, look like 
$\cont{v}.\alpha$ where $\cont{v}\in \cont{T}_i=\tree{\costi E^*}{\seq{\Phi_i}}$. Such moves do not destroy prelegality nor do they change the value of {\em Tree}, so  $\tree{\costi E^*}{\seq{\Phi_i}}=\tree{\costi E^*}{\seq{\Phi_{i+1}}}$. It remains to note that 
$T$ is not updated in this subcase, so that we also have $\cont{T}_{i+1}=\cont{T}_{i}$. Hence $\tree{\costi E^*}{\seq{\Phi_{i+1}}}=\cont{T}_{i+1}$. Finally, suppose the $i$th iteration of LOOP 
deals with Case (iv). It is the environment who moves in the 
consequent of $\cost\cost E^*\mli\cost E^*$, and does so before the machine makes any moves (in the antecedent). Then the clean environment assumption, in conjunction with the induction hypothesis for clause (g), implies that such a move by the environment cannot bring $\Phi_i$ to an illegal and hence non-prelegal position of $\cost E^*$. So, $\Phi_{i+1}$ remains a prelegal position of $\cost E^*$. As for 
$\tree{\costi E^*}{\seq{\Phi_{i+1}}}=\cont{T}_{i+1}$, it holds for the same reason as in the previous case. 

{\em Clause (b):} If the $i$th iteration of LOOP deals with Case (i), (ii) or (iii), it is the environment who moves in the antecedent of $\cost \cost E^*\mli\cost E^*$, and does so before the machine makes any moves. Therefore the clean environment assumption, with the induction hypothesis for clause (g) in mind, guarantees that $\Psi_{i+1}$ is a legal and hence prelegal position of $\cost\cost E^*$. Assume now that the $i$th iteration of LOOP deals with Case (iv), so that  
$\Psi_{i+1}=\seq{\Psi_i,\oo\blu{v}_1.\yel{v}_1.\alpha,\ldots,\oo\blu{v}_k.\yel{v}_k.\alpha}$.
By the induction hypothesis, $\Psi_i$ is a prelegal position of $\cost\cost E^*$. And, by the induction hypothesis for clause (d), 
each $\blu{v}_j$ ($1\leq j\leq k$) is a leaf of $\tree{\costi\costi E^*}{\seq{\Psi_i}}$, so adding the (lab)moves $\oo\blu{v}_1.\yel{v}_1.\alpha,\ldots,\oo\blu{v}_k.\yel{v}_k$ does not 
bring $\Psi_i$ to a non-prelegal position. $\Psi_{i+1}$ thus remains a prelegal position of $\cost\cost E^*$. As an aside, note also that those moves, being nonreplicative, do not modify $\tree{\costi\costi E^*}{\seq{\Psi_i}}$.

{\em Clause (c):} Just as in the previous clause, when the $i$th iteration of LOOP deals with Case (i), (ii) or (iii), the 
desired conclusion follows from the clean environment assumption in conjunction with the induction hypothesis for clause (g).  Assume now that the $i$th iteration of LOOP deals with Case (iv). Consider any leaf $x$ of $\tree{\costi\costi E^*}{\seq{\Psi_{i+1}}}$. As noted at the end of our proof of Clause (b), we have $\tree{\costi\costi E^*}{\seq{\Psi_i}}=\tree{\costi\costi E^*}{\seq{\Psi_{i+1}}}$, so $x$ is also a leaf of 
$\tree{\costi\costi E^*}{\seq{\Psi_i}}$. Therefore, 
if $\Psi_{i+1}^{\preceq x}=\Psi_{i}^{\preceq x}$, the conclusion that $\Psi_{i+1}^{\preceq x}$ is a prelegal 
position of $\cost E^*$ 
follows from the induction hypothesis. Suppose now $\Psi_{i+1}^{\preceq x}\not=\Psi_{i}^{\preceq x}$. Note that then,
in view of the induction hypothesis for clause (d),  $\Psi_{i+1}^{\preceq x}$ looks like $\seq{\Psi_{i}^{\preceq x},\oo y_1.\alpha,\ldots,\oo y_m.\alpha}$, where for each $y_j$ ($1\leq j\leq m$)  
we have $\blu{z}=x$ and $\yel{z}=y_j$ for some leaf $z$ of $T_i$, with  
$y_j$ being a leaf of $\tree{\costi E^*}{\seq{\Psi_{i}^{\preceq x}}}$. By the induction hypothesis for the present clause, 
$\Psi_{i}^{\preceq x}$ is a prelegal position of $\cost E^*$. Adding to such a position the nonreplicative labmoves 
$\oo y_1.\alpha,\ldots,\oo y_m.\alpha$ --- where the $y_j$ are leaves of $\tree{\costi E^*}{\seq{\Psi_{i}^{\preceq x}}}$ --- cannot bring it to a non-prelegal position. Thus, $\Psi_{i+1}^{\preceq x}$ remains a prelegal position of $\cost E^*$. 

{\em Clauses (d) and (e):} If the $i$th iteration of LOOP deals with Cases (iii) or (iv), $T_i$ is not modified, and no moves of the form $\col{x}$ or $x.\col{y}$ (where $x,y$ are bitstrings) are made in the antecedent of 
$\cost\cost E^*\mli\cost E^*$, so $\tree{\costi \costi E^*}{\seq{\Psi_i}}$ and $\tree{ \costi E^*}{\seq{\Psi_{i}^{\preceq x}}}$ (any leaf $x$ of
$\tree{\costi\costi E^*}{\Psi_i}$) are not affected, either. Hence Clauses (d) and (e) for $i+1$ are automatically inherited from the induction hypothesis for these clauses. 
This inheritance also takes place --- even if no longer ``automatically" --- when the $i$th iteration of LOOP deals with Case (i) or (ii). This can be verified by a routine analysis of how Cases (i) and (ii) modify $T_i$ and 
the other relevant parameters. Details are left to the reader.  

{\em Clause (f):} Consider any leaf $z$ of $T_{i+1}$. When the $i$th iteration of LOOP deals with Case (i) or (ii), no moves of the form $x.\alpha$ 
 are made in the consequent of $\cost \cost E^*\mli\cost E^*$, and no moves of the form $x.y.\alpha$ are made in the antecedent (any bitstrings $x,y$). Based on this, it is easy to see that for all
bitstrings $x,y$ 
we have $\Phi_{i+1}^{\preceq x}=\Phi_{i}^{\preceq x}$ and $(\Psi_{i+1}^{\preceq x})^{\preceq y}=
(\Psi_{i}^{\preceq x})^{\preceq y}$. Hence clause (f) for $i+1$ is inherited from the same clause for $i$. Now suppose  
the $i$th iteration of LOOP deals with Case (iii). Then $T_{i+1}=T_i$ and hence $z$ is also a leaf of $T_i$. From the description of Case (iii) one can easily see that if $w\not\preceq\blu{z}$ or $u\not\preceq\yel{z}$, we have
$\Phi_{i+1}^{\preceq \cont{z}}=\Phi_{i}^{\preceq \cont{z}}$ and 
$(\Psi_{i+1}^{\preceq\blu{z}})^{\preceq \yel{z}}=(\Psi_{i}^{\preceq\blu{z}})^{\preceq \yel{z}}$, 
so the equation $\Phi_{i+1}^{\preceq \cont{z}}=(\Psi_{i+1}^{\preceq\blu{z}})^{\preceq \yel{z}}$ is true by the induction hypothesis;
and if $w\preceq\blu{z}$ and $u\preceq\yel{z}$, then  $\Phi_{i+1}^{\preceq \cont{z}}=\seq{\Phi_{i}^{\preceq\cont{z}},\pp\alpha}$ and $(\Psi_{i+1}^{\preceq\blu{z}})^{\preceq\yel{z}}=\seq{(\Psi_{i}^{\preceq\blu{z}})^{\preceq\yel{z}},\pp\alpha}$. But, by the induction hypothesis, $\Phi_{i}^{\preceq\cont{z}}=(\Psi_{i}^{\preceq\blu{z}})^{\preceq\yel{z}}$. Hence 
$\Phi_{i+1}^{\preceq \cont{z}}=(\Psi_{i+1}^{\preceq\blu{z}})^{\preceq\yel{z}}$.
A similar argument applies when the $i$th iteration of LOOP deals with Case (iv).

{\em Clause (g):} Below we implicitly rely on the induction hypothesis, according to which 
$\Theta_i\in\legal{\costi\costi E^*\mli \costi E^*}{}$ and hence $\Phi_i\in\legal{\costi E^*}{}$ and 
$\Psi_i\in\legal{\costi\costi E^*}{}$. 
Note that, with the clean environment assumption in mind, all of the moves made in any of Cases (i)-(iv) of LOOP have the prefix `$1.$' or `$2.$', i.e. are made either in the antecedent or the consequent of $\cost\cost E^*\mli\cost E^*$. Hence, in order to show that $\Theta_{i+1}$ is a legal position of 
 $\cost\cost E^*\mli \cost E^*$, it would suffice to verify that $\Phi_{i+1}\in\legal{\costi E^*}{}$ and 
$\Psi_{i+1}\in\legal{\costi\costi E^*}{}$. 

Suppose the $i$th iteration of LOOP deals with Case (i) or (ii). The clean environment assumption guarantees that  $\Psi_{i+1}\in\legal{\costi\costi E^*}{}$. In the consequent of $\cost \cost E^*\mli\cost E^*$ only replicative moves are made. 
Replicative moves can yield an illegal position ($\Phi_{i+1}$ in our case) of a $\cost$-game only if they yield a non-prelegal 
position.  But, by clause (a), $\Phi_{i+1}$ is a prelegal position of $\cost E^*$. Hence it is also a legal position of $\cost E^*$. 

Suppose now the $i$th iteration of LOOP deals with Case (iii). Again, that  
$\Psi_{i+1}\in\legal{\costi\costi E^*}{}$ is guaranteed by the clean environment assumption. So, we only need to verify that $\Phi_{i+1}\in\legal{\costi E^*}{}$. By clause (a), this position is a prelegal position of $\cost E^*$. So, it remains to see that, for any leaf 
$y$ of $\tree{\costi E^*}{\seq{\Phi_{i+1}}}$, $\Phi_{i+1}^{\preceq y}\in\legal{E^*}{}$.  
Pick an arbitrary leaf $y$ of $\tree{\costi E^*}{\seq{\Phi_{i+1}}}$  --- i.e., by clause (a), of $\cont{T}_{i+1}$. Let $z$ be the leaf of $T_{i+1}$ with $y=\cont{z}$. We already know that $\Psi_{i+1}\in\legal{\costi\costi E^*}{}$. By clause (d), we also know that $\blu{z}$ is a leaf of $\tree{\costi \costi E^*}{\seq{\Psi_{i+1}}}$. Consequently, 
$\Psi_{i+1}^{\preceq \blu{z}}\in\legal{\costi E^*}{}$. Again by clause (d), $\yel{z}$ is a leaf of 
$\tree{\costi E^*}{\seq{\Psi_{i+1}^{\preceq \blu{z}}}}$. Hence, $(\Psi_{i+1}^{\preceq \blu{z}})^{\preceq \yel{z}}$ should be a legal position of $E^*$.
But, by clause (f), $\Phi_{i+1}^{\preceq \cont{z}}=(\Psi_{i+1}^{\preceq \blu{z}})^{\preceq \yel{z}}$. Thus,
$\Phi_{i+1}^{\preceq \cont{z}}\in\legal{E^*}{}$, i.e. $\Phi_{i+1}^{\preceq y}\in\legal{E^*}{}$.

Finally, suppose the $i$th iteration of LOOP deals with Case (iv). By the clean environment assumption, $\Phi_{i+1}\in\legal{\costi E^*}{}$. Now consider $\Psi_{i+1}$. This position 
is a prelegal position of $\cost\cost E^*$ by clause (b).  
So, in order for $\Psi_{i+1}$  to be a legal position 
of $\cost \cost E^*$, for every leaf $x$ of 
$\tree{\costi\costi E^*}{\seq{\Psi_{i+1}}}$ we should have $\Psi_{i+1}^{\preceq x}\in\legal{\costi E^*}{}$. Consider an arbitrary such leaf $x$. By clause (c), $\Psi_{i+1}^{\preceq x}$ is a prelegal position of $\cost E^*$. Hence, a sufficient condition for 
$\Psi_{i+1}^{\preceq x}\in\legal{\costi E^*}{}$ is that, for every leaf $y$ of $\tree{\costi E^*}{\seq{\Psi_{i+1}^{\preceq x}}}$,  $(\Psi_{i+1}^{\preceq x})^{\preceq y}\in\legal{E^*}{}$. So, let $y$ be an arbitrary such leaf.
 By clause (e), there is a leaf $z$ of $T_{i+1}$ such that $\blu{z}=x$ and $\yel{z}=y$.
Therefore, by clause (f), $\Phi_{i+1}^{\preceq \cont{z}}=(\Psi_{i+1}^{\preceq x})^{\preceq y}$. But we know that $\Phi_{i+1}\in\legal{\costi E^*}{}$ and hence (with clause (a) in mind) $\Phi_{i+1}^{\preceq \cont{z}}\in\legal{E^*}{}$. Consequently,
$(\Psi_{i+1}^{\preceq x})^{\preceq y}\in\legal{E^*}{}$.
\end{proof}

\begin{lemma}\label{sep21c} 
For every finite initial segment $\Upsilon$ of $\Theta$, there is $i\in N$ such that $\Upsilon$ is a $($not necessarily proper$)$ initial segment of 
$\Theta_i$ and hence of every $\Theta_j$ with $i\leq j\in N$. 
\end{lemma}

\begin{proof} 
The statement of the lemma is straightforward when there are infinitely many iterations of LOOP, for each iteration 
adds a nonzero number of new moves to the run and hence there are arbitrarily long $\Theta_i$s, each of them  being an initial segment of $\Theta$. Suppose now 
 LOOP is iterated a finite number $m$ of times. It would be (necessary and) sufficient to verify that in this case $\Theta=\Theta_m$, i.e. no moves are made during the last iteration of LOOP. But this is indeed so. From the description of LOOP we see that the machine does not make any moves during a given iteration unless the environment makes a move $\gamma$ first. So, assume $\oo$ makes move $\gamma$ during the $m$th iteration of LOOP. By the clean environment assumption, 
we should have $\seq{\Theta_m,\oo\gamma}\in\legal{\costi\costi E^*\mli\costi E^*}{}$. It is 
easy to see that such a $\gamma$ would have to satisfy the conditions of one of the Cases (i)-(iv) of LOOP. But then there would be an $(m+1)$th iteration of LOOP, contradicting out assumption that there are only $m$ iterations. \end{proof}

Let us use $\Psi$ and $\Phi$ to denote $\rneg\Theta^{1.}$ and $\Theta^{2.}$, respectively. Of course, the statement of 
Lemma \ref{sep21c} is true for $\Phi$ and $\Psi$ (instead of $\Theta$) as well.
Taking into account that, by definition, a given run is legal iff all of its finite initial segments are legal, the following fact is an immediate corollary of Lemmas \ref{sep21c} and \ref{sep21b}(g):

\begin{equation}\label{sep21d}
\mbox{\em $\Theta\in\legal{\costi\costi E^*\mli \costi E^*}{}$. Hence, $\Psi\in\legal{\costi\costi E^*}{}$ and $\Phi\in\legal{\costi E^*}{}$.}
\end{equation}

To complete our proof of Lemma \ref{l5}, we need to show that  
\[\win{\costi\costi E^*\mli \costi E^*}{}\seq{\Theta}=\pp.\]
With (\ref{sep21d}) in mind, if $\win{\costi\costi E^*}{}\seq{\Psi}=\oo$, by the definition of $\mli$, we are done. Assume now  
$\win{\costi\costi E^*}{}\seq{\Psi}=\pp$. Then, by the definition of $\cost$, 
there is a complete branch $x$ of $\tree{\costi\costi E^*}{\seq{\Psi}}$ such that   
$\win{\costi E^*}{}\seq{\Psi^{\preceq x}}=\pp$. This, in turn, means that, for some complete branch 
$y$ of $\tree{\sti E^*}{\seq{\Psi^{\preceq x}}}$, 
\begin{equation}\label{sp19}
\win{E^*}{}\seq{(\Psi^{\preceq x})^{\preceq y}}=\pp.
\end{equation}
 Fix these $x$ and $y$. For each $i\in N$, let $x_i$ denote the (obviously unique) leaf of $\tree{\costi \costi E^*}{\seq{\Psi_{i}}}$ such that $x_i\preceq x$; and let $y_i$ denote the (again unique) leaf of $\tree{\costi E^*}{\seq{\Psi_{i}^{\preceq x_i}}}$ such that $y_i\preceq y$. 
Next, let 
$z_i$ denote the leaf of $T_i$ with $\blu{z}_i=x_i$ and $\yel{z}_i=y_i$. According to Lemma \ref{sep21b}(e), 
such a $z_i$ exists and is unique. 

Consider any $i$ with $i+1\in N$. Clearly $x_i\preceq x_{i+1}$ and $y_i\preceq y_{i+1}$. By our choice of the $z_j$,
we then have $\blu{z}_i\preceq\blu{z}_{i+1}$ and $\yel{z}_i\preceq\yel{z}_{i+1}$. Hence, by Lemma \ref{sep19}, 
$z_i\preceq z_{i+1}$. Let us fix $z$ as the shortest (perhaps infinite if $N$ is infinite) colored bitstring such that for every $i\in N$, $z_i\preceq z$. Based on the 
just-made observation that we always have  $z_i\preceq z_{i+1}$,  such a $z$ exists. And, in view of Lemma \ref{sep21b}(a), it is not hard to see that 
$\cont{z}$ is a complete branch of $\tree{\costi E^*}{\seq{\Phi}}$.

With Lemma \ref{sep21c} in mind, Lemma \ref{sep21b}(f) easily allows us to find that $\Phi^{\preceq \cont{z}}=(\Psi^{\preceq x})^{\preceq y}$. Therefore, by (\ref{sp19}), $\win{E^*}{}\seq{\Phi^{\preceq \cont{z}}}=\pp$. By the definition of $\cost$, this means that $\win{\costi E^*}{}\seq{\Phi}=\pp$. Hence, by the definition of $\mli$ and with (\ref{sep21d}) in mind, 
$\win{\costi\costi E^*\mli\costi E^*}{}\seq{\Theta}=\pp$. Done. 

\subsection{Finishing the soundness proof for affine logic}\label{smain}

Now we are ready to prove Theorem \ref{main}. Consider an arbitrary sequent $S$ provable in $\al$. 
By induction on the $\al$-derivation of $S$, we are going to show that $S$ has a uniform solution $\cal E$. 
This is sufficient to conclude that $\al$ is `uniformly sound'. The theorem also claims `constructive soundness', i.e. that such an $\cal E$ can be effectively built from a given $\al$-derivation of $S$. This claim of the theorem will be automatically taken care of by the fact that our proof of the existence of $\cal E$ is constructive: all of the uniform-validity and closure lemmas on which
we rely provide a way for actually constructing a corresponding uniform solution. With this remark in mind and for 
the considerations of readability, in what follows we only talk about uniform validity without explicitly mentioning 
uniform solutions for the corresponding formulas/sequents and without explicitly showing how to construct such solutions. 

 There are 16 cases to consider, corresponding to the 16 possible rules that might have been used at the last step of an $\al$-derivation of $S$, with $S$ being the conclusion of the rule. In each non-axiom case below, ``induction hypothesis" means the assumption that the premise(s) of the corresponding rule is (are) uniformly valid. The goal in each case is to show that the conclusion of the rule is also uniformly valid. ``Modus ponens" should be understood as Lemma \ref{l1a}, 
``generalized modus ponens'' as Lemma \ref{l1}, and ``transitivity" as Lemma \ref{l1c}. Also, clauses (f) and (g) of Lemma \ref{l8}, in combination with modus ponens, always allow us to rewrite a statement $\uvalid \s{G_1}\mld\s{H}\mld\s{G_2}$ as $\uvalid \s{G_1}\mld(\s{H})\mld\s{G_2}$, and vice versa. We will often explicitly or implicitly rely on this fact, which we call {\bf associativity} (of $\mld$).{\vspace{10pt}

{\bf Identity Axiom:} By Lemma \ref{l8}(a).\vspace{10pt}

{\bf $\twg$-Axiom:} Of course, $\uvalid \twg$.  \vspace{10pt}
 
{\bf Exchange:} By Lemma \ref{l8}(b), $\uvalid E\mld F\mli F\mld E$. And, by Lemma \ref{l8}(e), 
\[\uvalid (E\mld F\mli F\mld E)\mli (\s{G}\mld E\mld F\mld\s{H}\mli\s{G}\mld F\mld E\mld\s{H}).\] Hence, by modus 
ponens, \[\uvalid \s{G}\mld E\mld F\mld\s{H}\mli\s{G}\mld F\mld E\mld\s{H}.\] But, 
by the induction hypothesis, $\uvalid \s{G}\mld E\mld F\mld\s{H}$. Hence, by modus ponens,  
$\uvalid \s{G}\mld F\mld E\mld\s{H}$.\vspace{10pt}
 
{\bf Weakening:} Similar to the previous case, using Lemma \ref{l8}(i) instead of Lemma \ref{l8}(b).\vspace{10pt} 

{\bf $\pcost$-Contraction}: By Lemma \ref{l8}(e) (with  
empty $\s{S}$),   
\[\uvalid (\pcost E\mld\pcost E\mli\pcost E)\mli (\s{G}\mld \pcost E \mld \pcost E\mli \s{G}\mld \pcost E).\]
And, by Lemma \ref{pl6c}, $\uvalid \pcost E\mld\pcost E\mli\pcost E$. Hence, by modus ponens, 
\(\uvalid \s{G}\mld \pcost E\mld\pcost E \mli \s{G}\mld \pcost E.\)
But, by the induction hypothesis, $\uvalid\s{G}\mld \pcost E\mld \pcost E$. Hence, by modus ponens, 
$\uvalid \s{G}\mld \pcost E$.\vspace{10pt}  

{\bf $\cost$-Contraction}: Similar to $\pcost$-contraction, using Lemma \ref{l6c} instead of Lemma \ref{pl6c}.\vspace{10pt}

{\bf $\add$-Introduction:}  By Lemma \ref{l8}(j), $\uvalid E_i\mli E_1\add\ldots\add E_n$; and, by Lemma \ref{l8}(e), 
\[\uvalid (E_i\mli E_1\add\ldots\add E_n)\mli\bigl(\s{G}\mld E_i\mli\s{G}\mld (E_1\add\ldots\add E_n)\bigr).\]  Modus ponens yields 
$\uvalid \s{G}\mld E_i\mli\s{G}\mld (E_1\add\ldots\add E_n)$.
But, by the induction hypothesis, $\uvalid \s{G}\mld E_i$. So, by modus ponens, 
$\uvalid \s{G}\mld (E_1\add\ldots\add E_n)$.\vspace{10pt}

{\bf  $\adc$-Introduction:} By the induction hypothesis,  \[\uvalid \s{G}\mld E_1, \ \ \ldots, \ \ \uvalid \s{G}\mld E_n.\] 
And, from Lemma \ref{l8}(k), 
\[\uvalid (\s{G}\mld E_1)\mlc\ldots\mlc (\s{G}\mld E_n)\mli\s{G}\mld (E_1\adc\ldots\adc E_n).\]
Generalized modus ponens yields 
$\uvalid \s{G}\mld (E_1\adc\ldots\adc E_n)$.\vspace{10pt}

{\bf $\mld$-Introduction:} In view of associativity, this rule is trivial.\vspace{10pt}

{\bf $\mlc$-Introduction:} By the induction hypothesis,  \[\uvalid \s{G_1}\mld E_1, \ \ \ldots, \ \ \uvalid \s{G_n}\mld E_n.\] And, 
from Lemma \ref{l8}(d), 
\[\uvalid (\s{G_1}\mld E_1)\mlc\ldots\mlc (\s{G_n}\mld E_n)\mli\s{G_1}\mld\ldots\mld\s{G_n}\mld (E_1\mlc\ldots\mlc E_n).\]
Generalized modus ponens yields 
$\uvalid \s{G_1}\mld\ldots\mld\s{G_n}\mld (E_1\mlc\ldots\mlc E_n)$.\vspace{10pt}

{\bf $\pcost$-Introduction:}  By the induction hypothesis, $\uvalid \s{G}\mld E$.  And, by Lemma \ref{pl6a}, 
$\uvalid E\mli \pcost E$. So, by Lemma \ref{l8}(e) and modus ponens applied twice, 
$\uvalid \s{G}\mld \pcost E$.\vspace{10pt}

{\bf $\cost$-Introduction:} Similar to $\pcost$-introduction, using Lemma \ref{l6a} instead of Lemma 
\ref{pl6a}.\vspace{10pt}

{\bf $\pst$-Introduction:} By the induction hypothesis, $\uvalid \s{\pcost G}\mld E$. 
If $\s{\pcost G}$ is empty, then  $\s{\pcost G}\mld E = E$ and thus $\uvalid E$. Hence, by Lemma \ref{pl10}, 
$\uvalid \pst E$, i.e. $\uvalid \s{\pcost G}\mld \pst E$. Suppose now $\s{\pcost G}$ is not empty. Associativity 
allows us to rewrite $\uvalid \s{\pcost G}\mld E$ as \ ($\uvalid(\s{\pcost G})\mld E$ and thus) \ $\uvalid \gneg \s{\pcost G}\mli E$. Then, by Lemma \ref{pl10}, 
$\uvalid \pst(\gneg \s{\pcost\ G}\mli E)$. From here, by Lemma \ref{pl4} and modus ponens, we get 
$\uvalid \pst\gneg \s{\pcost G}\mli \pst E$, which can be rewritten as $\uvalid \pcost\s{\pcost G}\mld \pst E$. But, by Lemma \ref{pl4a},
$\uvalid \pcost\s{\pcost G}\mli \s{\pcost\pcost G}$ and, by Lemma \ref{l8}(e), 
\[\uvalid (\pcost\s{\pcost G}\mli \s{\pcost\pcost G})\mli (\pcost\s{\pcost G}\mld \pst E\mli \s{\pcost\pcost G}\mld \pst E).\] 
Applying modus ponens twice yields     $\uvalid \s{\pcost\pcost G}\mld \pst E$. From here, using Lemmas \ref{pl5}, \ref{l8}(e) and modus ponens as many times as 
the number of disjuncts in $\s{\pcost\pcost G}$, we get $\uvalid \s{\pcost G}\mld \pst E$.\vspace{10pt}

{\bf $\st$-Introduction:} Similar to $\pst$-introduction, using Lemmas \ref{l10}, \ref{l4}, \ref{l4a} and \ref{l5} instead of Lemmas 
\ref{pl10}, \ref{pl4}, \ref{pl4a} and \ref{pl5}, respectively.\vspace{10pt}

{\bf $\ade$-Introduction:} By Lemma \ref{oct5b}, 
$\uvalid E(t)\mli \ade xE(x)$. And, by Lemma \ref{l8}(e), \[\uvalid (E(t)\mli \ade xE(x))\mli(\s{G}\mld E(t)\mli
\s{G}\mld \ade xE(x)).\] Modus ponens yields $\uvalid \s{G}\mld E(t)\mli\s{G}\mld\ade xE(x)$. But, by the induction hypothesis,  $\uvalid \s{G}\mld E(t)$. Hence, by modus ponens, $\uvalid \s{G}\mld \ade x E(x)$.\vspace{10pt}

{\bf $\ada$-Introduction:} First, consider the case when $\s{G}$ is nonempty. By the induction hypothesis, 
we have $\uvalid \s{G}\mld E(y)$, which can be rewritten as   $\uvalid \gneg\s{G}\mli E(y)$. Therefore, by Lemma \ref{l10a},  
$\uvalid \ada y\bigl(\gneg \s{G}\mli E(y)\bigr)$ and, by Lemma \ref{oct5a} and modus ponens, 
$\uvalid \ada y\vspace{1pt}\gneg \s{G}\mli \ada yE(y)$. At the same time, by Lemma \ref{oct5c}, $\uvalid  \gneg\s{G} \mli
\ada y \gneg \s{G}$. By transitivity, we then get $\uvalid \gneg \s{G}\mli \ada yE(y)$. But, by Lemma \ref{oct99}, $\uvalid \ada yE(y)\mli\ada x E(x)$. Transitivity yields $\uvalid \gneg \s{G}\mli \ada xE(x)$, which can be rewritten as  
$\uvalid \s{G}\mld \ada xE(x)$.
The case when $\s{G}$ is empty is simpler, for then $\uvalid \s{G}\mld \ada xE(x)$, i.e. $\uvalid\ada xE(x)$, can be obtained directly from the induction hypothesis by Lemmas \ref{l10a}, \ref{oct99} and modus ponens.

\section{What could be next?}\label{ssconcl}
As a conclusive remark, the author wants to point out that the story told in this chapter was, in fact, only about the tip of the iceberg. Even though the phrase ``{\em the} language of CL" was used in some semiformal contexts, such a language has no official boundaries and, depending on particular needs or taste, remains open to various interesting extensions. In a broad sense, CL is not a particular syntactic system or a particular semantics for a particular collection of operators, but rather a platform and ambitious program for redeveloping logic as a formal theory of computability, as opposed to the formal theory of truth which it has more traditionally been. 

The general framework of CL is also ready to accommodate any reasonable weakening modifications of its absolute-strength  computation model HPM,
thus keeping a way open for 
studying logics of sub-Turing computability and developing a systematic theory of interactive complexity. Among modifications of this sort, for example,  might be depriving the HPM of its infinite work tape, leaving in its place just a write-only buffer where the machine constructs its moves. In such a modification the exact type of read access to the run and valuation tapes becomes relevant, and a reasonable restriction would apparently be to allow --- perhaps now multiple --- read heads to move only in one direction. An approach favoring 
this sort of machines would try to model Turing (unlimited) or sub-Turing (limited) computational resources such as memory, time, etc. as games, and then understand computing a problem $A$ with resources represented by $R$ as 
computing  $R\mli A$, thus making explicit not only trans-Turing (incomputable) resources 
as we have been doing in this paper, but also all of the Turing/sub-Turing resources needed or allowed for computing $A$ 
--- the resources that the ordinary HPM or Turing machine models take for granted. 
So, with $T$ representing the infinite read/write tape as a computational resource, 
computability of $A$ in the old sense would mean nothing but computability of $T\mli A$ in the new sense: 
having $T$ in the antecedent would amount to having infinite memory, only this time provided externally (by the environment)
via the run tape rather than internally via the work tape.} 

Complexity and sub-Turing computability aside, there are also good philosophical reasons for questioning the legitimacy of the presence of an infinite work tape, whether it be in our HPM model or in the ordinary Turing machine (TM) model. The point is that neither HPMs nor TMs can be implemented --- even in principle --- as actual physical beings. This is so for the simple reason that     
no real mechanical device will ever have an infinite (even if only potentially so) internal memory. The reason why this fact 
does not cause much frustration and usually remains unnoticed   
is that the tape can be easily thought of as an {\em external resource}, and thus TMs or HPMs can be identified only with their finite control parts; then and only then, they indeed become implementable devices. Yet, the standard formal treatment of TMs or our treatment of
HPMs does not account for this implicit intuition, and views the infinite work tape as a part of the machine.  Computability logic, with its flexibility and ability to keep an accurate and explicit count of all resources, makes it possible to painlessly switch from TMs or HPMs to truly finite devices, and make things really what they were meant to be.

An alternative or parallel direction for CL to evolve with a focus shift from computability to complexity, could be 
extending its vocabulary with {\em complexity-conscious} operators. For example, winning a complexity-conscious version $\ada^p x\ade^p y A(x,y)$ of $\ada x\ade y A(x,y)$ could mean existence of a polynomial-time function $f$ such that, to any move $m$ by the environment, the machine responds with a move $n$ within time $f(m)$, and then wins the game $A(m,n)$. 

Time has not yet matured for seriously addressing complexity or sub-Turing computability issues though, and in the nearest future CL will probably remain focused on just computability: as it happens, there are still too many unanswered questions here. The most important and immediate task is finding axiomatizations for incrementally expressive fragments of CL --- first of all, fragments that involve recurrence operators, for which practically no progress has been made so far (with the intuitionistic fragment of CL being one modest exception). It is also highly desirable to fully translate {\bf CL4}-style ad hoc axiomatizations into some systematic and ``nice'' proof theory, such as cirquent calculus.  So far this has only been done (in \cite{Japcirq,Japjlc2}) for the $\gneg,\mlc,\mld$-fragment.

\begin{theindex}
\item {Abramsky} \pageref{0Abramsky}
\item {admissible interpretation} \pageref{0admissible interpretation}
\item {affine logic} \pageref{0affine logic},\pageref{0affine logic2}
\item {$\al$} \pageref{0al}
\item {$\al$-formula} \pageref{0al-formula}
\item {algorithmically solvable} \pageref{0algorithmically solvable}
\item {arity of a game} \pageref{0arity of a game}
\item {arity of a letter} \pageref{0arity of a letter}
\item {arity of an atom} \pageref{0arity of an atom}
\item {atom} \pageref{0atom}
\indexspace

\item {bitstring} \pageref{0bitstring}
\item {bitstring tree (BT)} \pageref{0bitstring tree (BT)}
\item {Blass} \pageref{0Blass0},\pageref{0Blass}
\item {blind conjunction} \pageref{0blind conjunction and disjunction}
\item {blind disjunction} \pageref{0blind conjunction and disjunction}
\item {blind existential quantification} \pageref{0blind existential quantification}
\item {blind operations} \pageref{0blind operations}
\item {blind universal quantification} \pageref{0blind universal quantification}
\item {blindly bound} \pageref{0blindly bound}
\item {blue content} \pageref{0blue content}
\item {branch of a BT} \pageref{0branch of a BT}
\item {branching corecurrence} \pageref{0branching corecurrence0},\pageref{0branching corecurrence}
\item {branching recurrence} \pageref{0branching recurrence0},\pageref{0branching recurrence}
\item {breadth (of a game)} \pageref{0breadth (of a game)}
\item {BT (bitstring tree)}  \pageref{0bitstring tree (BT)}
\item {BT-structure} \pageref{0treestructure of a DBT}
\indexspace

\item {canonical tuple} \pageref{0canonical tuple}
\item {capital `S' semantics} \pageref{0capital `S' semantics}
\item {CBT (colored bitstring tree)} \pageref{0colored bitstring tree (CBT)}
\item {$\cal CCS$} \pageref{0ccs}
\item {$\clt$} \pageref{0clt}
\item {$\clt$-formula} \pageref{0clt-formula}
\item {$\predell$} \pageref{0predell}
\item {$\predell$-formula} \pageref{0predell-formula}
\item {choice conjunction} \pageref{0choice conjunction}
\item {choice disjunction} \pageref{0choice disjunction}
\item {choice existential quantification} \pageref{0choice existential quantification}
\item {choice operations} \pageref{0choice operations}
\item {choice universal quantification} \pageref{0choice universal quantification}
\item {Church-Turing thesis} \pageref{0Church-Turing thesis}
\item {cirquent calculus} \pageref{0cirquent calculus}
\item {clean environment assumption} \pageref{0clean environment assumption}
\item {color of a bit} \pageref{0color (of a bit)}
\item {colored bit} \pageref{0colored bit}
\item {colored bitstring} \pageref{0colored bitstring}
\item {colored bitstring tree (CBT)} \pageref{0colored bitstring tree (CBT)}
\item {complete branch of a BT} \pageref{0complete branch of a BT}
\item {computability logic (CL)} \pageref{0CL (Computability Logic)}
\item {computable} \pageref{0computable}
\item {computation branch} \pageref{0computation branch}
\item {computational problem} \pageref{0computational problem},\pageref{0computational problem3}
\item {computational resource} \pageref{0computational resource},\pageref{0cr2}
\item {configuration} \pageref{0configuration}
\item {constant} \pageref{0constant1},\pageref{0constant2}
\item {constant DBT} \pageref{0constant DBT}
\item {constant game} \pageref{0constant game}
\item {content (of a colored bit)} \pageref{0content (of a colored bit)}
\item {content (of a colored bitstring)} \pageref{0content (of a colored bitstring)}
\item {content (of a game)} \pageref{0content (of a game)}
\item {copy-cat strategy} \pageref{0copy-cat strategy}
\indexspace

\item {decorated bitstring tree (DBT)} \pageref{0decorated bitstring tree (DBT)}
\item {decoration} \pageref{0decoration}
\item {delay} \pageref{0delay}
\item {depend (a game on a variable)} \pageref{0depend (a game on a variable)}
\item {depth (of a game)} \pageref{0depth (of a game)}
\indexspace

\item {$\epsilon$} \pageref{0epsilon}
\item {elementarization} \pageref{0elementarization}
\item {elementary atom} \pageref{0elementary atom}
\item {elementary game} \pageref{0elementary game},\pageref{0elementary game2}
\item {elementary letter} \pageref{0elementary letter}
\item {elementary formula} \pageref{0elementary formula}
\item {empty string} \pageref{0empty string}
\item {empty run} \pageref{0empty run}
\item {environment} \pageref{0environment}
\item {EPM} \pageref{0EPM}
\subitem {\{EPMs\}} \pageref{0epms}
\item {equivalent games} \pageref{0equivalent games}
\item {excluded middle} \pageref{0excluded middle}
\indexspace

\item {fair computation branch} \pageref{0fair computation branch}
\item {finitary game} \pageref{0finitary game}
\item {finite game} \pageref{0finite game}
\item {finite-breadth game} \pageref{0finite-breadth game}
\item {finite-depth game} \pageref{0finite-depth game}
\item {free game} \pageref{0free game}
\indexspace

\item {game} \pageref{0game}
\item {general atom} \pageref{0general atom} 
\item {general letter} \pageref{0general letter}
\item {general-base formula} \pageref{0general-base formula}
\item {Gentzen} \pageref{0Gentzen}
\item {G\"{o}del's Dialectica interpretation} \pageref{0Dialectica interpretation}
\item {granting permission} \pageref{0granting permission}
\indexspace

\item {heterogeneous position} \pageref{0heterogeneous position}
\item {Heyting's intuitionistic calculus} \pageref{0Heyting's intuitionistic calculus}
\item {HPM} \pageref{0HPM}
\subitem {\{HPMs\}} \pageref{0hpms}
\indexspace

\item {illegal move} \pageref{0illegal move}
\item {illegal run} \pageref{0illegal run}
\subitem {$\xx$-illegal} \pageref{0xx-illegal run}
\item {initial configuration} \pageref{0initial configuration}
\item {instable formula} \pageref{0instable formula}
\item {instance of a game} \pageref{0instance of a game}
\item {internal informational resource} \pageref{0internal informational resource}
\item {interpret} \pageref{0interpret} 
\item {interpretation (as a function)} \pageref{0interpretation (as a function)}
\item {interpretation (as a game)} \pageref{0interpretation (as a game)}
\item {intuitionistic logic} \pageref{0intuitionistic logic}
\item {iteration principle} \pageref{0iteration principle}
\indexspace

\item {Jagadeesan} \pageref{0Jagadeesan}
\indexspace

\item {Kleene's realizability} \pageref{0Kleene's realizability}
\item {knowledgebase} \pageref{0knowledgebase}
\item {Kolmogorov's thesis} \pageref{0Kolmogorov's thesis}
\item {Kripke model} \pageref{0Kripke model}
\indexspace

\item {label} \pageref{0label}
\item {labeled move (labmove)} \pageref{0labeled move (labmove)}
\item {labmove (labeled move)} \pageref{0labeled move (labmove)}
\item {leaf} \pageref{0leaf}
\item {legal move} \pageref{0legal move}
\item {legal run} \pageref{0legal run}
\item {linear logic} \pageref{0linear logic}
\item {logical atom} \pageref{0logical atom}
\item {Lorenzen} \pageref{0Lorenzen}
\item {Lorenzen's game semantics} \pageref{0Lorenzen's game semantics}
\item {lost run} \pageref{0lost run}
\item {lowercase `s' semantics} \pageref{0lowercase `s' semantics}
\item {$\legal{A}{}$} \pageref{0legalruns}
\item {$\legal{A}{e}$} \pageref{0lr}
\item {$\Legal{A}$} \pageref{0lll}
\indexspace

\item {machine} \pageref{0machine}
\item {mapping reducibility} \pageref{0mapping reducibility}
\item {maximal run} \pageref{0maximal run}
\item {move} \pageref{0move}
\item {move state} \pageref{0move state}
\indexspace

\item {negation} \pageref{0negation}
\item {negative occurrence} \pageref{0negative occurrence}
\item {node of a BT} \pageref{0node of a BT}
\item {nonlogical atom} \pageref{0non-logical atom}
\item {$\mov$} \pageref{0mov}
\item {nonreplicative (lab)move} \pageref{0nonreplicative (lab)move}
\indexspace

\item {$\xx$} \pageref{0xx}
\item {parallel conjunction} \pageref{0parallel conjunction}
\item {parallel corecurrence} \pageref{0parallel corecurrence}
\item {parallel disjunction} \pageref{0parallel disjunction}
\item {parallel existential quantification} \pageref{0parallel existential quantification}
\item {parallel operations} \pageref{0parallel operations}
\item {parallel recurrence} \pageref{0parallel recurrence}
\item {parallel universal quantification} \pageref{0parallel universal quantification}
\item {perifinite-depth game} \pageref{0perifinite-depth game}
\item {permission state} \pageref{0permission state}
\item {position} \pageref{0position}
\item {positive occurrence} \pageref{0positive occurrence}
\item {predicate} \pageref{0predicate}
\item {predicate letter} \pageref{0predicate letter}
\item {prefixation} \pageref{0prefixation}
\item {prelegal position} \pageref{0prelegal position}
\item {prelegal run} \pageref{0prelegal run}
\item {procedural rule} \pageref{0Procdural rule}
\item {provider} \pageref{0provider}
\indexspace

\item {quantifier} \pageref{0quantifier}
\indexspace

\item {recurrence operations} \pageref{0recurrence operations}
\item {reducible} \pageref{0reducible}
\item {reduction (as an EPM)} \pageref{0reduction (as an EPM)}
\item {reduction (as an HPM)} \pageref{0reduction (as an HPM)}
\item {reduction (as a game operation)} \pageref{0reduction (as a game operation)}
\item {$\rep$} \pageref{0rep}
\item {relative Kolmogorov complexity} \pageref{0kc}
\item {replicative (lab)move} \pageref{0replicative (lab)move}
\item {run} \pageref{0run}
\item {run spelled by a computation branch} \pageref{0run spelled by a computation branch}
\item {run tape} \pageref{0run tape}
\indexspace

\item {sequent} \pageref{0sequent}
\item {sequent calculus} \pageref{0sequent calculus}
\item {sequential conjunction} \pageref{0sequential conjunction and disjunction}
\item {sequential corecurrence} \pageref{0sequential coreccurrence}
\item {sequential disjunction} \pageref{0sequential conjunction and disjunction}
\item {sequential quantifiers} \pageref{0sequential quantifiers}
\item {sequential recurrence} \pageref{0sequential recurrence}
\item {singleton DBT} \pageref{0singleton DBT}
\item {solution (as an EPM)} \pageref{0solution2}
\item {solution (as an HPM)} \pageref{0solution}
\item {stable formula} \pageref{0stable formula}
\item {static game} \pageref{0static game1},\pageref{0static game}
\item {strategy} \pageref{0strategy}
\item {strict game} \pageref{0strict game}
\item {strong completeness} \pageref{0strong completeness}
\item {structure (of a game)} \pageref{0structure (of a game)}
\item {substitution} \pageref{0substitution}
\item {substitution of variables} \pageref{0substitution of variables}
\item {substitutional instance} \pageref{0substitutional instance}
\item {successor configuration} \pageref{0successor configuration}
\item {surface occurrence} \pageref{0surface occurrence}
\indexspace

\item {term} \pageref{0term1},\pageref{0term2}
\item {$\tree{\ldots}{\seq{\ldots}}$} \pageref{0trtr}
\item {Turing} \pageref{0Turing}
\item {Turing reducibility} \pageref{0Turing reducibility}
\indexspace

\item {underlying BT-structure} \pageref{0underlying BT-structure}
\item {uniform solution} \pageref{0uniform solution}
\item {uniform-constructive soundness} \pageref{0uniform-constructive soundness}
\item {uniformly valid} \pageref{0uniformly valid}
\item {unilegal run} \pageref{0unilegal run}
\item {unistructural game} \pageref{0unistructural game}
\indexspace

\item {valid} \pageref{0valid}
\item {valuation} \pageref{0valuation}
\item {valuation tape} \pageref{0valuation tape}
\item {variable} \pageref{0variable1},\pageref{0variable2}
\indexspace

\item {win} \pageref{0win}
\item {winnable} \pageref{0winnable}
\item {winnability} \pageref{0winnability}
\item {$\win{A}{}$} \pageref{0winA}
\item {$\win{A}{e}$} \pageref{0wn}
\item {won run} \pageref{0won run}
\item {work tape} \pageref{0work tape}
\indexspace

\item {yellow content} \pageref{0yellow content}
\indexspace

\indexspace

\item {$\twg$ (as a game)} \pageref{0twg (as a game)}\vspace{3pt}
\item {$\twg$ (as a player)} \pageref{0twg (as a player)}\vspace{3pt}
\item {$\tlg$ (as a game)} \pageref{0tlg (as a game)}\vspace{3pt}
\item {$\tlg$ (as a player)} \pageref{0tlg (as a player)}\vspace{3pt}
\item {$\gneg$ (as an operation on games)} \pageref{0gneg (as an operation on games)}\vspace{3pt}
\item {$\pneg$ (as an operation on players)} \pageref{0pneg (as an operation on players)}\vspace{3pt}
\item {$\rneg$ (as an operation on runs)} \pageref{0rneg (as an operation on runs)}\vspace{3pt}
\item {$\mlc$} \pageref{0mlc}\vspace{3pt}
\item {$\mld$} \pageref{0mld}\vspace{3pt}
\item {$\mla$} \pageref{0mla}\vspace{3pt}
\item {$\mle$} \pageref{0mle}\vspace{3pt} 
\item {$\mli$} \pageref{0mli}\vspace{3pt}
\item {$\cla$} \pageref{0cla}\vspace{3pt}
\item {$\cle$} \pageref{0cle}\vspace{3pt}
\item {$\adc$} \pageref{0adc}\vspace{3pt}
\item {$\add$} \pageref{0add}\vspace{3pt}
\item {$\ada$} \pageref{0ada}\vspace{3pt}
\item {$\ade$} \pageref{0ade}\vspace{3pt}
\item {$\sqc$} \pageref{0sqc}\vspace{3pt}
\item {$\sqd$} \pageref{0sqd}\vspace{3pt}
\item {$\pst$} \pageref{0pst}\vspace{3pt}
\item {$\pcost$} \pageref{0pcost}\vspace{3pt}
\item \hspace{-2pt}{$\pintimpl$} \pageref{0pintimpl}\vspace{3pt}
\item {$\st$} \pageref{0st}\vspace{3pt}
\item {$\cost$} \pageref{0cost}\vspace{3pt}
\item \hspace{-2pt}{$\intimpl$} \pageref{0intimpl}\vspace{3pt}
\item {$\sst$} \pageref{0sst}\vspace{3pt}
\item {$\scost$} \pageref{0scost}\vspace{3pt}
\item \hspace{-2pt}{$\sintimpl$} \pageref{0sintimpl}\vspace{3pt} 
\item {$\circ$} \pageref{0circ}\vspace{3pt}
\item {$\spadesuit$} \pageref{0spadesuit}\vspace{3pt}
\item {$\emptyrun$} \pageref{0emptyrun}\vspace{3pt}

\item {$\preceq$} \pageref{0preceq}\vspace{3pt}
\item {$\models$} \pageref{0models}\vspace{3pt}
\item {\mbox{$\valid$}} \pageref{0validd}\vspace{3pt}
\item {{\mbox{$\vdash\hspace{-4pt}\vdash\hspace{-4pt}\vdash$}}} \pageref{0uvalid}\vspace{3pt}
\item {$\mapsto$} \pageref{0mapsto}\vspace{3pt}
\item {$\cont{v}$} \pageref{0contv}\vspace{3pt}
\item {$\blu{v}$} \pageref{0bluv}\vspace{3pt}
\item {$\yel{v}$} \pageref{0yelv}\vspace{3pt}
\item {$\cont{T}$} \pageref{0ttt}\vspace{3pt}
\item {$e[A]$} \pageref{0eee}\vspace{3pt}
\item {$\seq{\Phi}A$} \pageref{0seqPhiA}\vspace{3pt}
\item {$A(x_1/t_1,\ldots,x_n/t_n)$} \pageref{0suv}\vspace{3pt}
\item {$\Gamma^\alpha$} \pageref{apr2}\vspace{3pt}
\item {\(\Gamma^{\preceq u}\)} \pageref{susu}\vspace{3pt}

\end{theindex}


\newpage

\begin{thebibliography}{99}


\bibitem{Abr94} S. Abramsky and R. Jagadeesan. {\em Games and full completeness for multiplicative linear logic}. {\bf Journal of Symbolic Logic} 59 (2) (1994), pp. 543-574.

\bibitem{Ben01} J. van Benthem. {\bf Logic in Games}. ILLC preprint, University of Amsterdam, Amsterdam, 2001.

\bibitem{Bla72} A. Blass. {\em Degrees of indeterminacy of games}. {\bf Fundamenta Mathematicae} 77 (1972), pp. 151-166. 

\bibitem{Bla92} A. Blass. {\em A game semantics for linear logic}. {\bf Annals of Pure and Applied Logic} 56 (1992), pp 183-220.

\bibitem{Fel85} W. Felscher. {\em Dialogues, strategies, and intuitionistic provability}. {\bf Annals of Pure and Applied Logic} 28 (1985), pp. 217-254.

\bibitem{Gir87} J.Y. Girard. {\em Linear logic}. {\bf Theoretical Computer Science} 50 (1) (1987), pp.  1-102.


\bibitem{God58} K. G\"{o}del. {\em \"{U}ber eine bisher noch nicht ben\"{u}tzte Erweiterung des finiten Standpunktes}. {\bf Dialectica} 12 (1958), pp. 280-287.



\bibitem{Gol04}
D.Q. Goldin, S.A. Smolka, P.C. Attie and E.L. Sonderegger. {\em Turing machines, transition systems, and interaction}.
{\bf Information and Computation} 194 (2) (2004), pp. 101-128.

\bibitem{Jap97} G. Japaridze. {\em A  constructive game semantics for the language of linear logic}. {\bf Annals of Pure and Applied Logic} 85 (1997), pp. 87-156.

\bibitem{Jap00} G. Japaridze. {\em The propositional logic of elementary tasks}. {\bf Notre Dame Journal of Formal Logic} 41 (2) (2000), pp. 171-183.

\bibitem{Jap02a} G. Japaridze. {\em The logic of tasks}. {\bf Annals of Pure and Applied  Logic} 117 (2002), pp. 263-295.

\bibitem{Jap03} G. Japaridze. {\em Introduction to computability logic}. {\bf Annals of Pure and Applied Logic} 123 (2003), pp. 1-99.


\bibitem{CL1} G. Japaridze. {\em Propositional computability logic I}. {\bf ACM Transactions on Computational Logic} 7 (2) 
(2006), pp. 302-330.

\bibitem{CL2} G. Japaridze. {\em Propositional computability logic II}. {\bf ACM Transactions on Computational Logic} 
7 (2) (2006), pp. 331-362. 

\bibitem{Japic} G. Japaridze. {\em Computability logic: a formal theory of interaction}. In: {\bf Interactive Computation: The New Paradigm}. D.Goldin, S.Smolka and P.Wegner, eds. Springer Verlag,  Berlin, 2006, pp. 183-223. 

\bibitem{Japcirq} G. Japaridze. {\em Introduction to cirquent calculus and abstract resource semantics}. {\bf Journal of Logic and Computation} 16 (4) (2006), pp. 489-532.

\bibitem{CL3} G. Japaridze. {\em From truth to computability I}. {\bf Theoretical Computer Science} 357 (2006), pp. 100-135.


\bibitem{CL4} G. Japaridze. {\em From truth to computability II}. {\bf Theoretical Computer Science} 379 (2007), pp. 20-52.
   

\bibitem{int1} G. Japaridze. {\em Intuitionistic computability logic}. {\bf Acta Cybernetica} 18 (1) (2007), pp. 77-113. 


\bibitem{Japjsl1} G. Japaridze. {\em The logic of interactive Turing reduction}. {\bf Journal of Symbolic Logic} 72 (1) (2007), pp. 243-276. 

\bibitem{Japapal} G. Japaridze. {\em The intuitionistic fragment of computability logic at the propositional level}. {\bf Annals of Pure and Applied Logic} 
147 (3) (2007), pp.187-227.

\bibitem{Japjsl2} G. Japaridze. {\em Many concepts and two logics of algorithmic reduction}. {\bf Studia Logica} (to appear). 

\bibitem{Japjlc2} G. Japaridze. {\em Cirquent calculus deepened}. {\bf Journal of Logic and Computation} (to appear). 

\bibitem{Japseq} G. Japaridze. {\em Sequential operators in computability logic}. {\bf Information and Computation} (to appear). 

\bibitem{Kle52} S.C. Kleene. {\bf Introduction to Metamathematics}. D. van Nostrand Company, New York / Toronto, 1952. 


\bibitem{Kol32} A. N. Kolmogorov. {\em Zur Deutung der intuitionistischen Logik}. {\bf Mathematische Zeitschrift} 35 (1932), pp. 58-65. 

\bibitem{Kon89} K. Konolige.  {\em On the relation between default and autoepistemic logic}. {\bf Artificial Intelligence} 35 (3) (1988), pp. 343--382. 

\bibitem{Lev00} H. Levesque and G. Lakemeyer. {\bf  The Logic of Knowledge Bases}. The MIT Press,
Cambridge, MA, 2000.

\bibitem{Lor59} P. Lorenzen. {\em Ein dialogisches Konstruktivit\"{a}tskriterium}. In: {\bf Infinitistic Methods}. In: PWN,  Proc. Symp. Foundations of Mathematics, Warsaw, 1961, pp. 193-200.

\bibitem{Mil93} R. Milner. {\em Elements of interaction}. {\bf Communications of the ACM} 36 (1) 
(January 1993), pp. 79-89.

\bibitem{Moo85} R. Moore.  {\em A formal theory of knowledge and action}. In:   {\bf Formal Theories of Commonsense Worlds}. J. Hobbs and R. Moore, eds. Ablex, Norwood, NJ, 1985. 

\bibitem{Pie01} A. Pietarinen. {\bf Semantic Games in Logic and Language}. Academic dissertation. University of Helsinki, Helsinki, 2002.

\bibitem{Sip06} M. Sipser. {\bf Introduction to the Theory of Computation}, 2nd edition. Thomson Course Technology, USA, 
2006.

\bibitem{Tur36} A. Turing. {\em On Computable numbers with an application to the entsheidungsproblem}. {\bf Proceedings of the London Mathematical Society} 2:42 (1936), pp. 230-265.

\bibitem{Ver} N. Vereshchagin. {\em Japaridze's computability logic and intuitionistic propositional calculus}. Moscow State University Preprint, 2006. http://lpcs.math.msu.su/$\sim$ver/papers/japaridze.ps
 

\bibitem{Weg98} P. Wegner. {\em Interactive foundations of computing}. {\bf Theoretical Computer Science} 192 (1998), pp.  315-351.


\end{thebibliography}
\end{document}